\documentclass[aps,prx,showpacs,preprintnumbers,twocolumn,superscriptaddress,nofootinbib]{revtex4-2}
\usepackage{amsmath,amssymb}
\usepackage{bm}
\usepackage{tipa}
\usepackage{upgreek}
\usepackage{comment}
\usepackage{mathrsfs}
\usepackage{graphicx}
\usepackage{lipsum}
\usepackage{braket}
\usepackage{enumitem}
\usepackage{mathbbol}
\usepackage{booktabs}
\usepackage{gensymb}
\usepackage[normalem]{ulem}
\usepackage{color}
\usepackage[colorlinks,bookmarks=true,citecolor=blue,linkcolor=red,urlcolor=blue]{hyperref}
\usepackage{hyperref}
\renewcommand{\vec}[1]{\mathbf{#1}}
\usepackage{pifont}

\def \k{{\mathbf k}}
\def \d{{\mathbf d}}
\def \q{{\mathbf q}}
\def \r{{\mathbf r}}
\def \n{{\mathbf n}}
\def \R{{\mathbf R}}

\makeatletter 
    
\renewcommand\onecolumngrid{
\do@columngrid{one}{\@ne}%
\def\set@footnotewidth{\onecolumngrid}
\def\footnoterule{\kern-6pt\hrule width 1.5in\kern6pt}%
}

\renewcommand\twocolumngrid{
        \def\footnoterule{
        \dimen@\skip\footins\divide\dimen@\thr@@
        \kern-\dimen@\hrule width.5in\kern\dimen@}
        \do@columngrid{mlt}{\tw@}
}%

\makeatother    

\begin{document}

\title{Textured Exciton Insulators} 
 \author{Yves H. Kwan}\thanks{These authors contributed equally.} 
	\affiliation{Princeton Center for Theoretical Science, Princeton University, Princeton NJ 08544, USA}

 \author{Ziwei Wang}\thanks{These authors contributed equally.} 
	\affiliation{Rudolf Peierls Centre for Theoretical Physics, Parks Road, Oxford, OX1 3PU, UK}

 	\author{Glenn Wagner}
	\affiliation{Department of Physics, University of Zurich, Winterthurerstrasse 190, 8057 Zurich, Switzerland}
	
	\author{Steven H. Simon}
	\affiliation{Rudolf Peierls Centre for Theoretical Physics, Parks Road, Oxford, OX1 3PU, UK}
	\author{S.A. Parameswaran}
	\affiliation{Rudolf Peierls Centre for Theoretical Physics, Parks Road, Oxford, OX1 3PU, UK}
  \affiliation{Max Planck Institute for the Physics of Complex Systems, Nöthnitzer Str. 38, 01187 Dresden, Germany}
\author{Nick Bultinck}
	\affiliation{Department of Physics, Ghent University, Krijgslaan 281, 9000 Gent, Belgium}

\begin{abstract}
We introduce and study new interacting topological states that arise
in  time-reversal symmetric bands with an underlying obstruction to forming localized states. If the $U(1)$ valley symmetry linked to independent charge conservation in each time-reversal sector is spontaneously broken, the corresponding `excitonic' order parameter is forced to form a topologically non-trivial texture across the Brillouin zone. We show that the resulting phase, which we dub a \textit{textured exciton insulator}, cannot be given a local-moment description due to a form of delicate topology. Using  toy models of bands with Chern or Euler obstructions to localization we construct explicit examples of the  Chern or Euler texture insulators (CTIs or ETIs) they support, and demonstrate that these are  generically  competitive ground states at intermediate coupling. We  construct field theories that capture the response properties of these new states. Finally, we identify the  incommensurate Kekulé spiral phase observed in magic-angle bi- and trilayer graphene as a concrete realization of an ETI. 
\end{abstract}

\maketitle

\section{Introduction}
Understanding the insulating state remains an enduring challenge in condensed matter physics. An early distinction was drawn between band insulators --- where the insulating behaviour is driven by interactions of electrons with the periodic lattice of positive ions --- and Mott or correlated insulators, which require electron-electron interactions. This has since been greatly enriched by the injection of ideas from both symmetry and topology, most famously with the prediction~\cite{kane_quantum_2005,Kane2005,Bernevig2006, fu_topological_2007, fu_topological_2007-1, moore_topological_2007, roy_topological_2009} and subsequent discovery~\cite{Koenig2007, hsieh_topological_2008} of topological insulators and their myriad generalizations, leading to a finer classification of band insulators. An equally important yet perhaps less widely known theme is the recognition that in many cases correlated insulators must either break symmetries or host emergent  fractionalized excitations~\cite{Lieb1961, Hastings2004, Oshikawa2000, parameswaran_topological_2013, watanabe_filling_2015}, features that arise from the interplay of crystalline  and time reversal symmetries with charge or spin conservation. Together, these ideas serve as key organizing principles for classifying different insulating phases of matter. 

A traditional starting point for understanding interacting insulators takes an atomic limit wherein the electron charge is frozen on lattice sites, so that the ultimate  phase structure is decided by the behaviour of other on-site degrees of freedom, most commonly the electron spin. However, such a description in terms of `local moments' can be obstructed in situations where the combination of symmetry and band topology forbids the construction of localized orbitals for the low-energy electrons~\cite{Bradlyn2017,Khalaf2018,Slager2013}. 

\begin{figure}[t!]
    \centering
    \includegraphics[width = \linewidth]{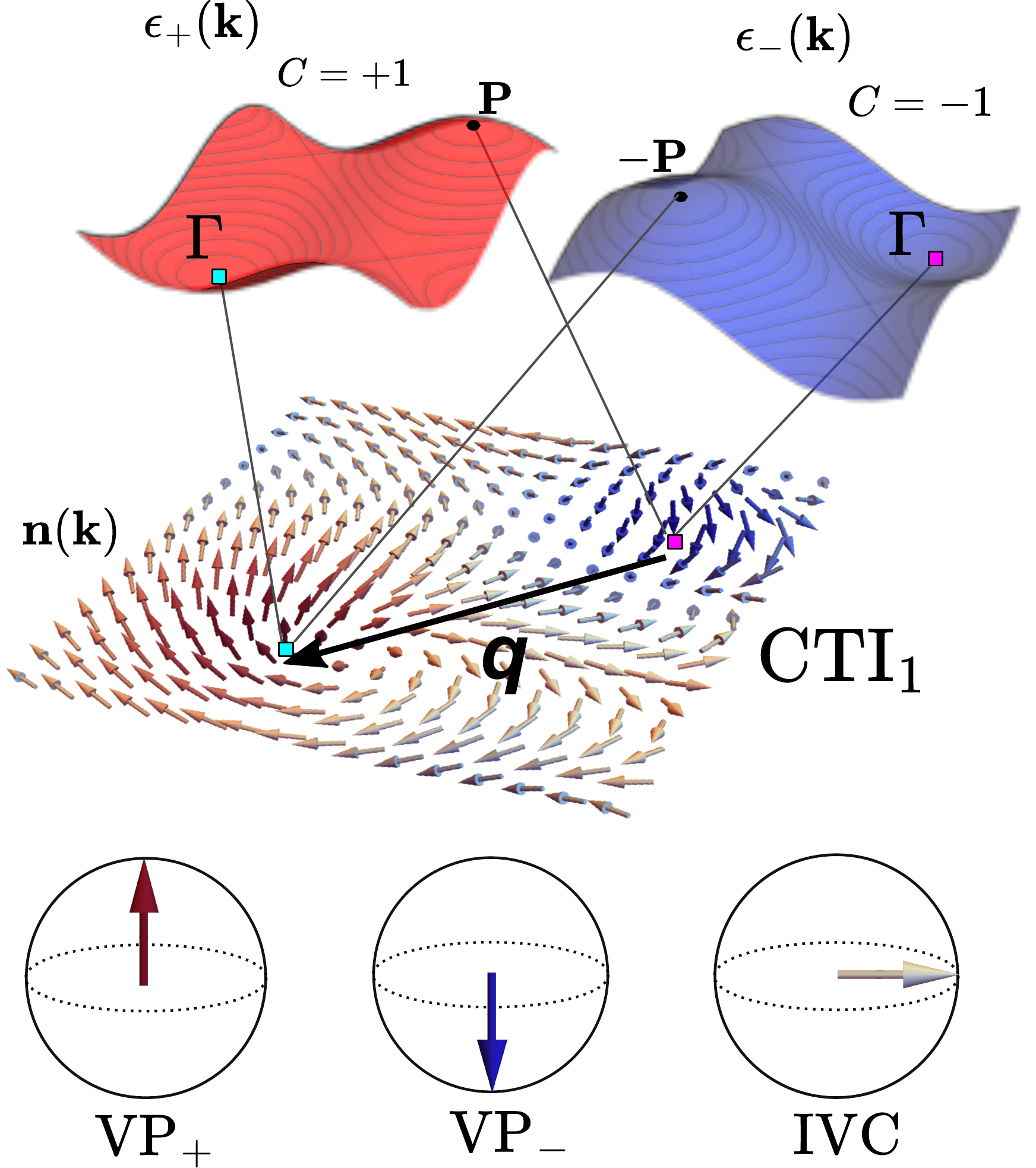}
    \caption{\textbf{Schematic representation of a Chern texture insulator from $C=\pm1$ bands.} Top plots show the non-interacting band structures $\epsilon_\tau(\bm{k})$ in valley $\tau=\pm$ with Chern numbers $C_\tau=\pm 1$. The $\Gamma$-point, where the dispersion has its minimum, is indicated with a cyan (magenta) square in $\tau=+$ ($-$). The maximum of the dispersion is at $\tau \mathbf{P}$. Middle plot shows the momentum-dependent valley pseudospin order $\bm{n}(\bm{k})$ in the CTI$_1$ state at half-filling in a smooth gauge. Intervalley coherence hybridizes $\bm{k}$ in $\tau=+$ with $\bm{k}+\bm{q}$ in $\tau=-$. The spiral wavevector $\bm{q}$ is influenced by kinetic energy considerations, e.g.~aligning the band maximum at $\bm{P}$ in $\tau=+$ with the band minimum at $\Gamma$ in $\tau=-$. Color shows the $\tau_z$ component, which is also related to the Berry curvature. $\bm{n}(\bm{k})$ is defined using a smooth gauge, which implies that the order parameter winds by $4\pi$ around the Brillouin zone due to topological considerations. Bottom plots illustrate Bloch spheres with the pseudospin $\bm{n}(\bm{k})$ corresponding to valley polarization along $\tau=+$, valley polarization along $\tau=-$, and intervalley coherence respectively.}
    \label{fig:splash}
\end{figure}

In this work, we introduce new examples of broken-symmetry correlated insulators that cannot be understood from a local-moment picture. The absence of such a description ultimately stems from the  obstruction to a  symmetry-preserving localized Wannier representation as dictated by the non-zero Chern or Euler invariants of the underlying symmetry-unbroken bands. This  forces any `excitonic' order parameter describing the {\it breaking} of one of the protecting global $U(1)$ symmetries within these bands to form a topologically non-trivial texture across the Brillouin zone. 
Although this texturing is typically  disfavored at strong coupling, we demonstrate that it characterizes the generic non-fractionalized and translation symmetry-preserving insulating state at intermediate coupling. 
We explore properties of these new insulating states via a series of toy models and effective field theories. (We use the terms `Chern texture insulator' (CTI) or `Euler texture insulator' (ETI)  --- omitting `exciton' for brevity --- to reflect that these arise from distinct obstructions, and reserve the umbrella term `textured exciton insulator' to describe both collectively.)
We then show that the previously-predicted  `incommensurate Kekulé spiral' (IKS)~\cite{kwan_kekule_2021} state of matter in magic-angle graphene bilayers and trilayers  is an example of an Euler texture insulator, and explain the hitherto-mysterious fact that it only emerges in the presence of strain. Given the recent observation of IKS order via scanning tunneling microscopy (STM)~\cite{nuckolls2023quantum,kim2023imaging}, this connection immediately roots the ETI in experimental reality. In a companion paper~\cite{companion}, we use numerical Hartree-Fock studies to identify several  Chern-obstructed moiré materials as candidates for realizing CTIs.
\subsection{Motivation}

We motivate our considerations by introducing 
the simplest setting in which a textured excitonic insulator can arise, consisting of a pair of degenerate Chern bands with $C_\tau =  \tau C$,  distinguished by a time-reversal-odd `valley' index $\tau =\pm 1$.
Such a valley degree of freedom is relevant to several semiconductors, graphene~\cite{Guinea2010,Guinea2010Generating,Levy2010}, and moiré materials.
We will assume that this is a purely internal degree of freedom\footnote{In the moiré setting this is an excellent approximation  facilitated by the separation of atomic and superlattice scales. However, the spatial structure linked to valley degrees of freedom is important when considering experimental probes, especially STM.}
that realises a U(1)$_V$  symmetry, equivalent to the independent conservation of charges in each valley. Working at an electron density sufficient to fill one of the two bands sets the stage for one of the basic questions we explore in this work: how do repulsive interactions open a gap in this time-reversal invariant system? 

Opening an insulating gap in the absence of translational symmetry breaking or fractionalization requires the spontaneous breaking of (at least) one of the two symmetries: time-reversal or U(1)$_V$~\cite{Lieb1961,Oshikawa2000,Hastings2004}. The first case is easy to understand: electrons  can form a fully valley-polarized (VP) phase termed an {\it orbital Chern insulator}~\cite{Bultinck2019mechanism,Zhang2019anomalous,Zhang2019,zhu2020voltage}; for $|C|=1$ this has the same underlying topology as a Landau level, and is hence non-localizable.
Breaking  U(1)$_V$ symmetry is more subtle, since this implies inter-valley coherence (IVC), i.e.~a non-zero expectation value of the form $\Delta_{\mathbf{k}\mathbf{k}'}=\langle c^\dagger_{\tau,\mathbf{k}} c^{\phantom\dagger}_{-\tau,\mathbf{k}'}\rangle$, which is equivalent to an `exciton condensate' where the pairing is between a particle and a hole taken from bands with equal and opposite Chern numbers. For topological reasons, such an order parameter cannot be constant throughout the Brillouin zone of allowed momenta $\mathbf{k},\mathbf{k}'$, and is forced to vanish within it, making it energetically unfavorable at strong coupling unless interactions are tuned into a regime that is physically difficult to achieve~\cite{Bultinck2019mechanism}. Therefore the resulting  inter-Chern coherent U(1)$_V$-breaking states were considered unlikely to emerge in the flat-band, strong coupling limit~\cite{Bultinck2019mechanism,Zhang2019,Lee2019}. Similar arguments can also be applied to `Euler-obstructed' bands\footnote{Previous work has considered the related but distinct question of Euler-obstructed \emph{Cooper pairing} in bands with non-trivial Euler number~\cite{Yu2023EulerTBG} (see also Ref.~\cite{Yu2022EulerMajorana}).}, characterized by a $\hat{C}_{2z} \hat{\mathcal{T}}$ symmetry (a composition of time-reversal and a $\pi$-spatial rotation), which can in a sense be viewed as a `doubled' version of the $C=\pm1$ problem described above.

Remarkably,
upon introducing a finite bandwidth that moves the problem away from strong coupling, a new type of U(1)$_V$-breaking state, that we term the `Chern texture insulator' (CTI), emerges under generic conditions.  The structure of the CTI is best understood by considering how  the valley order parameter evolves across the BZ: it lies at the poles of the valley Bloch sphere (indicating maximal VP) at a pair of time-reversal conjugate points (`nodes') in the BZ, but rotates gradually to lie in the plane (indicating maximal IVC order) away from these points (see Fig.~\ref{fig:splash}). The mandates of the underlying band topology are satisfied by a $2\pi$ winding of the IVC order about each of the nodes. (Formally, the pattern near each node is that of a `meron'~\cite{CALLAN1977375,Gross1978,Moon1995}.) If the two nodes have the same VP, the texture breaks time reversal, leading to a $|C|\neq 0$ state with coexisting time reversal breaking and IVC. This state,  that we dub the tilted-valley-polarized (TVP) insulator, cannot be localized for the same reason as the orbital Chern insulator (although we show below that its IVC properties mark it as a subtly distinct toplogical phase). In contrast, if the nodes have opposite VP the texture can preserve time reversal. The resulting phase, the  CTI, has $C=0$ so it is not obvious whether a similar order could arise in a model of local moments. We address this question next, by describing the topological properties of this state.

\subsection{Topology of the Chern Texture Insulator}\label{subsec:intro_topology_CTI}

In order to understand the  topology manifested by the CTI, it is useful to  first distinguish three types of topology encountered in band theory: namely, `strong', `fragile', and `delicate'. The distinction between trivial and strong topological phases is robust against the addition of any number of topologically trivial filled or empty bands\footnote{This is linked to the notion of 
 `stable equivalence' in $K$-theory.}. In contrast, fragile topological phases can be trivialized upon the addition of {\it filled} trivial  bands, i.e. valence bands  that lie below the Fermi energy~\cite{Bouhon2020,Po2018Fragile,Yang2019}. Finally, the recently-introduced notion of {\it delicate} topological phases refers to those unstable to adding additional filled or unfilled bands either above or below the Fermi energy (i.e.~both in the conduction and valence subspaces)~\cite{Nelson_Multicellularity,Nelson2022}. The distinction between the forms of topology can depend on the symmetries imposed: for example, the flat bands of TBG exhibit strong topology if the (approximate) particle-hole symmetry is elevated to an exact symmetry~\cite{Song2019,Song2021}, but  otherwise only exhibit fragile topology~\cite{Po2019,Ahn2019,Kang2018,Koshino2018,Zou2018,Liu2019}. Which of these notions applies to the Chern texture  insulator?

To address this question conceptually, we now introduce a simplified four-band model that captures the essential topological features while abstracting away detailed energetics. (We simply sketch the arguments here and defer detailed computations to Appendix~\ref{secapp:CTI_localization}.) Consider the Bloch Hamiltonian
\begin{equation}\label{eq:HtoyCTIintro}
H = \left[\d(\k)\cdot\boldsymbol{\sigma}\right]\tau^z \,,
\end{equation}
where $\tau^z$ labels the U$(1)$ charge, and the Pauli matrices $\sigma^i$ act on an orbital index labelled by $A$ and $B$. We take the time-reversal symmetry (TRS) to act as $\mathcal{T}=\tau^x\mathcal{K}$ ($\mathcal{K}: i\rightarrow -i$ is complex conjugation),  which implies that $d_x(\k)$ and $d_z(\k)$ are odd functions of $\k$ and $d_y(\k)$ is even. We can introduce a topological gap while preserving both U$(1)$ and $\mathcal{T}$ by choosing  $\d(\k)$ such that the unit vector $\d({\k})/|\d({\k})|$ covers the unit sphere once, resulting in a lower (upper) band with valley-dependent Chern number $C = \tau_z$ ($-\tau_z$). 
Half-filling the lower band  subspace  furnishes a setting for the CTI:  a pair of $|C|=1$ bands that are exchanged by $\hat{\mathcal{T}}$,  a global U$(1)_V$ symmetry corresponding to the valley charge conservation, and enough electrons to fully fill exactly one band. (If we take charge neutrality to coincide with the gap between the lower and upper subspaces, this corresponds to filling $\nu=-1$.)

Now, we observe that since $C\neq 0$, it is impossible to choose a smooth and periodic gauge for the Bloch functions for any single band, and this forces slower-than-exponential decay of their Fourier transform, i.e.~the Wannier functions~\cite{Brouder2007exponential}. Hence it is impossible to construct exponentially localized Wannier states while preserving U$(1)_V$ (unless, of course, we  hybridize the lower and upper bands).  We now argue that upon introducing intervalley coherence  and hence spontaneously breaking U$(1)_V$, it is possible to remove this obstruction  by hybridizing only the lower bands in the two valleys. This is especially transparent for the model in Eq.~\eqref{eq:HtoyCTIintro}. To see why, first note that it is possible to build a perfectly localized valley-polarized $\nu=-1$ state by hybridizing upper and lower bands within a single valley (say, the one with $\tau_z = 1$) --- for example, we could simply place a single electron in the $A$ orbital in each unit cell. Evidently, such a state has inter-Chern but not inter-valley coherence. Next, we note that by construction the orbital wavefunction of the $C=-1$, $\tau_z = -1$  lower band  is exactly equivalent to that of the $C=-1$, $\tau_z = +1$ upper band. Therefore, by starting with the wavefunction of inter-Chern valley-polarized state described above written in the  band-and-valley basis and simply swapping labels in the $C=-1$ sector (simultaneously from $\tau_z =+1$ to $\tau_z=-1$ and upper to lower), we obtain an inter-Chern and inter-valley coherent state built solely from the lower bands in each valley. 
The resulting occupied inter-valley coherent band has $C=0$ and hence admits an exponentially-localized Wannierization~\cite{alexandradinata2018nogo}.

Nevertheless, a vestige of the underlying  obstruction remains if we consider correlation functions of operators with a definite U$(1)_V$ charge: such correlators cannot be made strictly on-site. To see this, we argue by contradiction. 
For a fermion bilinear to have a purely on-site expectation value, its Fourier transform must be constant across the Brillouin zone.
However, the non-zero Chern number implies that each of the two orbital components of the valence band Bloch states  in a {\it single} valley vanish at least once in the Brillouin zone. Therefore, in the state we have constructed, any momentum-space fermion bilinear with a definite valley charge will pick up such zeroes, obstructing its adiabatic continuity to a constant non-zero value. Hence we conclude that such expectation values cannot be made fully onsite.
Thus, we find that  such operators retain a memory of the non-trivial topology of the U$(1)_V$-symmetric bands, which manifests itself as an obstruction to a trivial atomic limit with strictly on-site correlations. Perhaps more intuitively, we argue (see Appendix~\ref{secapp:CTI_localization}) that as a corollary, the charge variance within a single unit cell is always non-vanishing. In a true ‘local moment’ picture we would expect that such charge fluctuations vanish. This is similar to the notion of `multicellularity' introduced in the context of delicate topological phases~\cite{Nelson_Multicellularity}. This analogy suggests  that the  obstruction to making operators with definite valley charge onsite can be removed if we mix in another set of trivial orbitals either above or below the Fermi energy. We confirm explicitly that this is indeed the case in Appendix~\ref{secapp:CTI_localization}.

We conclude from this that the CTI exhibits an obstruction to an atomic limit that is a form of delicate topology inherited from the fragile/strong topology of the  parent bands. 
 While mixing with other bands can remove this obstruction, this is typically penalized energetically in many settings, making the delicate topological limit a good starting approximation. In fact, we will show that the CTI generically (i.e., except possibly for a measure-zero set of parameters) belongs to the subset of delicate topological phases that exhibit a stronger form of `non-compact' topology~\cite{SchindlerBernevigNOnCompact2021}\footnote{Note that it is not known if being delicate is a sufficient condition for non-compactness~\cite{SchindlerBernevigNOnCompact2021}, hence this requires a separate analysis.}.  This means that 
 there exist no Wannier functions which are strictly compact,
 i.e.~they can be exponentially localized but with tails extending to arbitrarily distant unit cells.
 
While the model we have introduced is highly simplified, we will show that its important features  apply more generally, to systems with either non-trivial strong or fragile topology in the symmetric bands.

\subsection{From Chern  to Euler Textures}
The Euler texture insulator (ETI) poses a conceptually distinct setting from the CTI. Here, the minimal model involves {\it two} bands in each valley --- four in total --- as required in order to define a  nontrivial Euler class~\cite{Ahn2019,Song2019}. This is associated with a symmetry that acts {\it within} a single valley --- for example the $\hat{\mathcal{C}}_{2z}\hat{\mathcal{T}}$ symmetry in TBG --- and forces the two bands in a valley to be linked by a pair of Dirac points with the same winding number (which is reversed in the other valley). Our goal is to find a $\hat{\mathcal{C}}_{2z}$- and $\hat{\mathcal{T}}$-preserving insulator at $\nu=\pm1$, such that we have enough electrons to fill exactly one of these four bands. 
Attempting to replicate the CTI construction in the `Chern bases' within the four-band subspace would break $\hat{\mathcal{C}}_{2z}$ symmetry, necessitating a different approach.

Recall that in the CTI, gaplessness due to the topologically mandated zeros in the U$(1)_V$ order parameter was evaded by valley polarizing in the vicinity of these nodes. Imagine now that we are instead in the ETI setting, where even before imposing any IVC  we have gapless $\hat{\mathcal{C}}_{2z}\hat{\mathcal{T}}$-mandated Dirac nodes in each valley, which we can view as a local singularity in the Bloch states. We can then bring the Dirac points in one valley, say $\tau=+$, into coincidence with non-singular patches of the BZ in the opposite valley with $\tau=-1$ (if necesssary, by boosting the momentum in one valley)  and then choose to locally polarize into this non-singular valley. TRS then forces us to do the opposite for the singular points in the $\tau=-$ valley, and we are then free to impose IVC everywhere else (one can show that this IVC would have to involve both inter- and intra-Chern components). The resulting valley texture is similar to that of the CTI, but plays a subtly different role: instead of directly gapping out nodes in an IVC order parameter, here the  valley texture instead `hides' the Euler-enforced band singularity in one valley by forcing the electrons into the other valley where the singularity is at a different point of the BZ, and vice-versa. It is because of this distinct topological role played by a superficially similar valley texture that we distinguish the CTI and ETI. 

The preservation of $\hat{\mathcal{C}}_{2z}$ has further important implications for the topological response of the ETI, making it  distinct from the CTI; one can instead view the ETI as symmetry-tuned to a boundary between two distinct CTIs with opposite choice of valley Chern number. The ETI is also a delicate topological phase in that it can be trivialized by hybridizing with additional trivial bands.

\subsection{Plan of Attack and Outline of This Paper}
 Itineracy, interactions, and topology each play an important role in the CTI and ETI: the finite bandwidth favors an equal occupation of the two valleys; Coulomb interactions seed exchange ferromagnetism; and topology forces the order parameter to wind. A final axis of complexity arises from the  freedom of the IVC to involve electrons at $\mathbf{k}-\tau \mathbf{q}/2$ in the  valley $\tau$; when $\mathbf{q}\neq 0$ the order parameter is a `spiral' in valley space. This allows the system to lower the energy cost of the textured exciton insulator  by adjusting the points where the electron is polarized in valley $\tau$ to coincide with minima of the dispersion in that valley. The problem of optimizing $\mathbf{q}$ to achieve the lowest energy can be framed in terms of a heuristic `lobe principle' that appropriately aligns the dispersion minimum in one valley with the dispersion maximum in the other valley~\cite{kwan_kekule_2021}. The combination of this freedom as well as the ability of the valley order parameter to smoothly evolve from VP to IVC across the Brillouin zone --- thereby minimizing the loss of exchange energy --- together explain why the textured exciton insulator emerges as a competitive ground state for non-zero dispersion (and also why it was missed in previous variational studies). The additional variational freedom of the state to adjust its choice of $\mathbf{q}$ in response to changes in twist angle, strain, etc.~also explain why it is a relatively robust energetic state that survives in experimentally realistic systems~\cite{nuckolls2023quantum,kim2023imaging}.

 Understanding the influence of each of these ingredients is a challenging task,  that we address by deploying  a series of  models and approximations.  In outlining these now, we also sketch the organization of this paper.

We begin in Sec.~\ref{sec:1BLLL} by introducing the CTI within the context of the so-called `Lowest Landau level' (LLL) model, which comprises a pair of LLLs with valley-contrasting magnetic fields (and hence opposite Chern numbers) subject to a periodic potential. The LLL basis states are  caricatures of the strong-coupling bands that are used as a starting point for study of several graphene-based moir\'e materials, while the finite dispersion  allows us to move away from the flat-band limit and into the intermediate coupling regime. This setup, first studied in Ref.~\cite{Bultinck2019mechanism}, is relevant to $\hat{C}_{2z}$-violating moir\'e platforms which can have isolated valley Chern bands in the single-particle band structure. As discussed above, the IVC order parameter cannot be uniform due to the Chern obstruction. To address this, Ref.~\cite{Bultinck2019mechanism} introduced the exciton vortex lattice (EVL), a candidate IVC phase that is gapless due to the presence of vortices  in the IVC order parameter. This phase, among others, was predicted to compete with the valley-polarized Chern insulator in the presence of finite dispersion and/or valley interaction anisotropy. In our self-consistent Hartree-Fock (HF) calculations, we find that the phase diagram is dominated for a large  parameter regime by a previously overlooked phase: the Chern-texture insulator, which we identify as the CTI$_1$\footnote{We use the subscript `$n$' in CTI$_n$ to indicate that the underlying valley-symmetric bands have Chern number $C=+n\tau_z$.}. We demonstrate that the CTI$_1$ can be viewed as a gapped version of the EVL, where the vortex cores are converted into merons whose positions are dictated energetically by the lobe principle. We also identify a second $|C|=1$ insulating phase distinct from the fully valley-polarized phase positioned between the CTI$_1$ and the fully valley-polarized phase, that we dub the tilted valley polarized (TVP) insulator.

To generalize the problem to capture  `Euler-obstructed bands' --- as relevant, for instance, to TBG ---  we must include $\hat{C}_{2z}$ symmetry. This is done, in Sec.~\ref{sec:2BLLL}, by considering a setup with two bands in each valley with net non-zero windings at Dirac point band touchings (equivalent to non-zero Euler number). We argue that at half-filling of the conduction or valence bands, a valley $U(1)$ breaking insulating mean-field state with $\hat{C}_{2z}\hat{\mathcal{T}}$ symmetry must be an ETI, which is a $\hat{C}_{2z}$-symmetric generalization of the CTI discussed previously, where the role of the valley texture is to `hide' the Dirac nodes --- which are mandated by $\hat{C}_{2z}\hat{\mathcal{T}}$ --- by locally valley-polarizing in the BZ so as to empty the  valley in which the nodes occur.

In Sec.~\ref{sec:field_theory}, we further investigate various properties of the CTI$_1$ and TVP via low-energy effective field theories, which we use to show that a real-space IVC vortex in the CTI$_1$ induces a charge density wave (CDW) halo at $2\bm{k}^*$, with $\pm \bm{k}^*$ the momenta of the meron cores in the ground state.
We also construct the appropriate response actions for the two phases. The CTI$_1$ exhibits a mixed Chern-Simons term between the gauge fields for charge and translation, implying that a real-space vortex of a translational-symmetry breaking charge density-wave order binds an electric charge of $\pm e/2$. Interestingly, we find that the TVP realizes the second entry of Kitaev's 16-fold way~\cite{Kitaev2006}. We comment on the connection between the CTI and the ETI in light of the topological response.

As we have emphasized at the outset, moir\'e materials provide natural candidate platforms for the textured exciton insulators introduced in this work, and indeed this work was motivated in part by the STM observation of IKS order near $\nu=-2$ in twisted bilayer~\cite{nuckolls2023quantum} and mirror-symmetric trilayer graphene~\cite{kim2023imaging}. In Sec.~\ref{sec:IKSasETI} we argue that the IKS state in both these systems is an example of an ETI, thereby demonstrating the experimental reality of the ETI. In a companion paper~\cite{companion}, we use microscopic HF simulations to predict the emergence of CTIs associated with spiral IVC order in an array of other moiré materials that lack $\hat{\mathcal{C}}_{2z}$ symmetry.

We close with a discussion and outlook in Sec.~\ref{sec:discussion}. Various  topological aspects of the CTI/ETI are discussed in a pair of appendices while additional analytical derivations,  model Hamiltonians, and numerical results are provided in the Supplementary Material (SM).

\tableofcontents

\section{LLL model and Chern texture insulators}\label{sec:1BLLL}

In this section, we study a lowest Landau level (LLL) based Hamiltonian designed to capture the essential physics of CTIs in systems that lack $\hat{C}_{2z}$ symmetry. We begin by explaining the basic picture of CTIs in Sec.~\ref{subsec:intro_picture}. The Hamiltonian of the LLL model is described in Sec.~\ref{subsec:1BLLL_model}, while its phase diagram is presented in Sec.~\ref{subsec:1BLLL_phase}. In Sec.~\ref{subsec:1BLLL_EVL}, we describe the exciton vortex lattice (EVL) order parameter, which provides physical intuition behind the various phases. Finally, we discuss some generalizations in Sec.~\ref{subsec:1BLLL_extensions}.

\subsection{Basic picture}\label{subsec:intro_picture}

A key aspect of the CTI is the interplay between energetic considerations and topological constraints, which we now briefly exemplify with the simplest $C_{2z}$-breaking scenario of two TRS-related bands with opposite Chern number $C=\tau$ in the two valleys $\tau=\pm$ (see Fig.~\ref{fig:splash}). 

For one electron per unit cell, an insulating mean-field state that preserves (generalized) translation symmetry can be described by a normalized valley pseudospin $\bm{n}(\bm{k})$ defined over the Brillouin zone (BZ). The ground state at strong interaction $U$ is expected to be a TRS-breaking valley-polarized Chern insulator ($\bm{n}\parallel \hat{z}$) in order to minimize exchange energy. If the single-particle bandwidth $W$ is comparable to $U$, then the dispersion $\epsilon_\tau(\bm{k})$ needs to be taken into account. An insulating mean-field state with good kinetic and exchange energy can be obtained by inducing IVC as follows. In the example of Fig.~\ref{fig:splash}, the band dispersions in the two valleys both have their global minimum in the BZ at $\Gamma$, while their band maxima are at $\tau\mathbf{P}$. As a result, hybridizing the valleys at relative wavevector $\bm{q}=0$ is not desirable since only one of these minima can be occupied. Similarly, there is no guarantee that such IVC will effectively avoid occupation of the band maxima at $\tau\mathbf{P}$. The \emph{lobe principle}~\cite{kwan_kekule_2021} resolves these concerns by considering IVC at a finite $\bm{q}$. In particular, the `boost' momentum $\bm{q}$ is chosen such that the band maximum (which is associated with a high-energy `lobe') in $\tau=+$ aligns with the band minimum (a low-energy lobe) in $\tau=-$, and vice versa. For Fig.~\ref{fig:splash}, this is achieved with $\bm{q}=-\mathbf{P}$. An energetically competitive TRS-invariant valley spiral is then constructed by inducing IVC across most of the BZ, except at the lobes where $\bm{n}(\bm{k})$ tilts towards the valley with lower kinetic energy. Favorable valley exchange is ensured by enforcing a smooth modulation of $\bm{n}(\bm{k})$ across the BZ.

At this point, our discussion of IVC has not addressed the electronic band topology. Therefore, it  could equally well be applied to topologically trivial models, such as the Hubbard model with spin $U(1)$-preserving spin-orbit coupling. For the latter in the strong-coupling regime, the IVC spiral is connected to kinetically-driven (superexchange) magnetism of localized valley moments~\cite{sahebsara2008spiral,zhu2018intervalley,Pan2020WSe2,wolf2021valleyspiral}. Such a picture evidently cannot hold if the underlying bands are topological. For the valley Chern bands in Fig.~\ref{fig:splash}, the key topological constraint is that the IVC order parameter $\langle c^\dagger_{\bm{k},+}c_{\bm{k}+\bm{q},-}\rangle$ [i.e.~the in-plane part of $\bm{n}(\bm{k})$] must possess at least two vortices in the BZ. This arises because in a smooth gauge, $c^\dagger_{\bm{k},\tau}$ winds by $2\pi \tau$ around the BZ, and hence the IVC order parameter has a net $4\pi$ winding~\cite{Bultinck2019mechanism}. These vortices would na\"ively just lead to gapless points (see the discussion of the exciton vortex lattice in Sec.~\ref{subsec:1BLLL_EVL}) and an energy penalty for the IVC spiral. However, the lobe principle naturally admits patches in momentum space
where the valley pseudospin is compelled by the kinetic energy to orient towards the poles. Hence, the IVC vortices can be accommodated as smooth merons in $\bm{n}(\bm{k})$ that maintain good exchange energy, as illustrated in Fig.~\ref{fig:splash}, leading to a fully gapped insulator.

We call the resulting state a Chern texture insulator (CTI$_1$). It is an example of a textured exciton insulator, because condensation of intervalley excitons of the valley-polarized state, which can restore $\hat{\mathcal{T}}$-symmetry, is topologically frustrated due to the opposite Chern numbers carried by the bands. This mandates a complex texture in the valley pseudospin $\bm{n}(\bm{k})$. In fact, at this filling, the CTI$_1$ is the unique time reversal invariant  insulating mean-field phase that preserves translation symmetry for all $U(1)_V$-symmetric observables. The subscript `1' in CTI$_1$ indicates that the constituent bands have $C=\pm1$. This notation alludes to the existence of related CTIs distinguished by the type and degree of the underlying topological frustration, but which share common characteristics such as time-reversal symmetry and kinetically-driven IVC between bands related by $\hat{\mathcal{T}}$.  Our analysis clarifies that the lobe principle and the triggering of $\bm{k}$-dependent intervalley order, possibly at finite spiral wavevector $\bm{q}$, are general \emph{energetic} considerations that frequently arise in the intermediate-coupling regime where $U\simeq W$. Note that the commensurability of $\bm{q}$ is considered a subsidiary issue that depends on the spatial symmetries and details of the model. However, there can be additional \emph{topological} constraints on IVC depending on the topology of the participating bands.
The complex interplay between these energetic and topological factors yields the CTI, which forms a novel family of electronic states.

\subsection{Model}\label{subsec:1BLLL_model}

\subsubsection{Magnetic Bloch basis}

We now turn to a concrete interacting model to explicate the properties of the CTI and related competing states. The description of the LLL model below closely follows that of Ref.~\cite{Bultinck2019mechanism}. The single-particle Hilbert space consists of two LLLs that carry a valley
index $\tau=\pm$ and experience opposite magnetic fields $\bm{B}=-\tau B\hat{z}$ leading to opposite Chern numbers $C=\tau$. With the magnetic length $\ell_B$ set to 1, the LLL wavefunctions in Landau gauge $\bm{A}=-\tau B x \hat{y}$ are
\begin{equation}
    u_{k\tau}(\bm{r})=\frac{1}{\sqrt{\pi^{\frac{1}{2}}L_y}}e^{iky}e^{-\frac{(x-\tau k)^2}{2}}
\end{equation}
with corresponding creation operators $c^\dagger_{k\tau}$. Note that the position-momentum locking goes in opposite directions in the two valleys. This setup clearly breaks $\hat{C}_{2z}$ symmetry, which interchanges the valleys and would require the Chern numbers to be identical. We note that related setups of opposite-field Landau levels have previously been considered in Refs.~\cite{Chen2012FQHtorusFTI,Mukherjee2019FQHsphereFTI,Bultinck2019mechanism,furukawa2014global,zhang2018composite,kwan2021exciton,kwan2022hierarchy,eugenio2020DMRG,stefanidis2020excitonic,Chatterjee2022dmrg,shi2024excitonic,yang2023phase}.

We consider a square lattice with lattice constant \mbox{$a=\sqrt{2\pi}$} corresponding to one flux per unit cell. We define the magnitude of the primitive reciprocal lattice vector $Q=\frac{2\pi}{a}=a$. We consider a cylinder geometry with $L_x=N_xa\rightarrow\infty$ and  periodic direction $L_y=N_ya$. The first magnetic BZ contains momenta $\bm{k}$ such that $k_x\in[0,Q),k_y\in[0,Q)$, for both valleys. The magnetic Bloch functions
\begin{align}
\begin{split}
    \phi_{\bm{k}\tau}(\bm{r})&=\frac{1}{\sqrt{N_x}}\sum_{n}^{}e^{i\tau k_x(k_y+nQ)}u_{k_y+nQ,\tau}(\bm{r})\\
    k_x&=\frac{2\pi n}{N_x a},\quad n=0,\ldots,N_x-1\\
    k_y&=\frac{2\pi n}{N_y a},\quad n=0,\ldots,N_y-1,
    \end{split}
\end{align}
are eigenfunctions of the magnetic translation operators 
\begin{equation}
\begin{gathered}\label{eq:MTO}
    \tilde{T}_{\tau,x}=e^{-i\tau Qy}T_x,\quad \tilde{T}_{\tau,y}=T_y\\
    \phi_{\bm{k},\tau}(\bm{r}+a\hat{x})=e^{i\tau Qy}e^{ik_x a}\phi_{\bm{k},\tau}(\bm{r})\\\phi_{\bm{k},\tau}(\bm{r}+a\hat{y})=e^{ik_y a}\phi_{\bm{k},\tau}(\bm{r})
\end{gathered}
\end{equation}
where $T_x,T_y$ are ordinary translation operators (i.e. $T_x=e^{iap_x}$). The magnetic Bloch functions are defined with a \emph{smooth} gauge that is \emph{not} periodic in the momentum argument $\bm{k}$
\begin{equation}\label{eq:Bloch_kperiodicity}
    \phi_{\bm{k}+Q\hat{x},\tau}(\bm{r})=e^{i\tau k_y Q}\phi_{\bm{k},\tau}(\bm{r}),\quad\phi_{\bm{k}+Q\hat{y},\tau}(\bm{r})=\phi_{\bm{k},\tau}(\bm{r}).
\end{equation}
The gauge choice above is inherited by the associated magnetic Bloch basis creation operators $d^\dagger_{\bm{k},\tau}$. We impose an anti-unitary time-reversal symmetry $\hat{\mathcal{T}}$ that relates the valleys 
\begin{equation}
    \hat{\mathcal{T}}d^\dagger_{\bm{k},\tau}\hat{\mathcal{T}}^{-1}=d^\dagger_{-\bm{k},-\tau}.
\end{equation}

\subsubsection{Single-particle dispersion}

To model a single-particle dispersion, we project a periodic potential $V_\text{SP}(\bm{r})$ onto the LLLs
\begin{equation}
    \hat{H}_\tau^\text{SP}=\int d\bm{r}\,V_\text{SP}(\bm{r})\psi^\dagger_\tau(\bm{r})\psi_\tau(\bm{r}),
\end{equation}
where $\psi^\dagger_\tau(\bm{r})$ is the LLL-projected fermion creation operator. The harmonics of the potential, which are compatible with the choice of magnetic unit cell, are mapped to the dispersion $\epsilon_{\tau}(\bm{k})$ according to~\cite{Bultinck2019mechanism}
\begin{align}
\begin{split}\label{eq:SP_general}
	\cos\left(\frac{2\pi n_xx}{a}+\phi_x\right)&\rightarrow \epsilon_{\tau}(\bm{k})=e^{-\frac{n_x^2\pi}{2}}\cos(\tau n_xk_ya+\phi_x)\\
	\cos\left(\frac{2\pi n_yy}{a}+\phi_y\right)&\rightarrow \epsilon_{\tau}(\bm{k})=e^{-\frac{n_y^2\pi}{2}}\cos(\tau n_yk_xa+\phi_y).
 \end{split}
\end{align}
For a lowest-harmonic square cosine potential with $\phi_x=\phi_y=0$, we obtain
\begin{equation}
\begin{gathered}\label{eq:SP}
    V_\text{SP}(\bm{r})=-w\left(\cos\frac{2\pi x}{a}+\cos\frac{2\pi y}{a}\right)\\
    \epsilon_{\tau}(\bm{k})=-\frac{W}{4}(\cos k_xa + \cos k_y a),\quad W= 4we^{-\frac{\pi}{2}},
\end{gathered}
\end{equation}
which will be used in this section unless otherwise stated. 

\subsubsection{Interactions}
For the interaction Hamiltonian, we consider the $U(1)_\text{v}$-symmetric density-density interaction
\begin{align}\label{eq:1BLLL_densityint}
\hat{H}_\text{int}=\frac{1}{2A}\sum_{\bm{q}\in\text{all},\tau,\tau'}\tilde{U}_{\tau\tau'}(\bm{q}):\rho_\tau(\bm{q})\rho_{\tau'}(-\bm{q}):
\end{align}
where $A=L_xL_y$ is the system area, and $\rho_{\tau}(\bm{q})$ is the projected density operator
\begin{align}
\rho_\tau(\bm{q})
&=e^{-\frac{\bm{q}^2}{4}}\sum_{\bm{k}\in\text{BZ}}e^{i\tau(k_x+\frac{q_x}{2})q_y}d^\dagger_{\bm{k}\tau}d_{\bm{k}+\bm{q},\tau}\\
&\equiv \sum_{\bm{k}\in\text{BZ}}\Lambda_\tau (\bm{k},\bm{q})d^\dagger_{\bm{k}\tau}d_{\bm{k}+\bm{q},\tau}
\end{align}
which implicitly defines the form factors $\Lambda_\tau (\bm{k},\bm{q})$ (see also Ref.~\cite{Wu2013pseudopotentials} and Supplementary Material, Sec.~\ref{secsupmat:1BLLL_additional}). 

The bare interaction potential in valley space is
\begin{equation}
    \tilde{U}_{\tau\tau'}(\bm{q})=\begin{pmatrix}
        u_0(\bm{q})+u_1(\bm{q}) & u_0(\bm{q})-u_1(\bm{q}) \\
        u_0(\bm{q})-u_1(\bm{q}) & u_0(\bm{q})+u_1(\bm{q})
    \end{pmatrix}_{\tau\tau'}.
\end{equation}
For the isotropic component $u_0(\bm{q})$, we choose the dual gate screened interaction $u_0(\bm{q})=2\pi U\frac{\tanh qd_\text{sc}}{q}$ where $d_\text{sc}$ is the distance between the gates and the sample plane. We also introduce a valley-anisotropic component $u_1(\bm{q})$. For simplicity, we consider an overall scaling
\begin{equation}
    u_1(\bm{q})=\alpha_\text{v}u_0(\bm{q})
\end{equation}
which effectively corresponds to different dielectric constants for the intravalley and intervalley interactions for $\alpha_\text{v}\neq 0$. Unless otherwise stated, we set $U=1$. 

\subsubsection{Winding of order parameters and boosting}\label{subsec:1BLLL_winding}
In a smooth gauge, certain order parameters are expected to wind around the BZ. Consider the IVC (inter-Chern) order parameter $\Delta(\bm{k})$ for a translation-invariant state, and its value after shifting in momentum space by $Q\hat{x}$
\begin{equation}\label{eq:IVC_OP}
    \Delta(\bm{k})\equiv\langle d^\dagger_{\bm{k},+}d_{\bm{k},-}\rangle\rightarrow \Delta(\bm{k}+Q\hat{x})=\Delta(\bm{k}) e^{2ik_y Q},
\end{equation}
where we have used Eq.~\ref{eq:Bloch_kperiodicity}. On the other hand, $\Delta(\bm{k})$ is periodic in the $k_y$ direction. This implies that $\Delta(\bm{k})$ must wind by $4\pi$ when traversing the boundary of the BZ counter-clockwise. In particular, this means that any $\Delta(\bm{k})$ that corresponds to a physically sensible state (i.e.~no discontinuities) must have at least one zero in the BZ in order to accommodate the winding, which could be achieved by e.g.~two vortices each with winding $2\pi$. 

In our HF calculations, we will sometimes `boost' the momentum in valley $\tau=-$ by $\bm{q}$ in order to access valley spiral orders. This amounts to allowing hybridization between $d^\dagger_{\bm{k},+}$ and $d^\dagger_{\bm{k}+\bm{q},-}$, in which case the analogous (smooth) order parameter is
\begin{equation}\label{eq:IVC_OP_q}
    \Delta_{\bm{q}}(\bm{k})\equiv\langle d^\dagger_{\bm{k},+}d_{\bm{k}+\bm{q},-} \rangle
\end{equation}
which similarly must have vortices in the BZ to accommodate the $4\pi$ winding.

\subsection{Phase diagram}\label{subsec:1BLLL_phase}

\begin{figure}
    \centering
    \includegraphics[width = \linewidth]{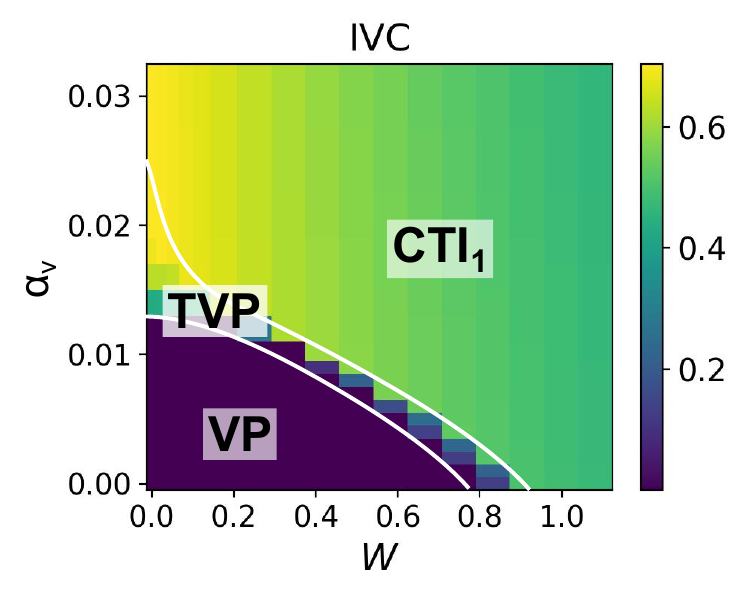}
    \caption{\textbf{HF phase diagram of the LLL model at  half-filling.} $\alpha_\text{v}$ is the valley interaction anisotropy, and $W$ is the kinetic bandwidth arising from a square cosine potential. The gate distance $d_\text{sc}=6a$, and the interaction scale $U=1$. Color indicates the magnitude of intervalley coherence (IVC). White lines indicate approximate phase boundaries. System size is $24\times 24$, and the valley boost is fixed to either $\bm{q}=(0,0)$ or $(Q/2,Q/2)$. [VP: valley-polarized; TVP: tilted valley-polarized; CTI$_1$: Chern texture insulator]}
    \label{fig:1BLLL_phase_24x24}
\end{figure}

We perform self-consistent HF calculations on an $N_x\times N_y$ momentum mesh at half-filling, i.e.~when there are enough electrons to fully occupy one of the two bands. The one-body density matrix $P_{\bm{k},\tau;\bm{k}',\tau'}=\langle d^\dagger_{\bm{k},\tau}d_{\bm{k}',\tau'}\rangle$ is constrained to satisfy a generalized translation invariance parameterized by a wavevector $\bm{q}$, which will be referred to as the `boost'.
This means that only $P_{\bm{k},\tau;\bm{k},\tau}$, $P_{\bm{k},+;\bm{k}+\bm{q},-}$, and $P_{\bm{k}+\bm{q},-;\bm{k},+}$ are allowed to be non-vanishing, i.e.~IVC is only permitted at wavevector $\bm{q}$:
\begin{equation}
    P_{\bm{k},+;\bm{k}',-}\propto \Delta_{\bm{q}}(\bm{k})\delta_{\bm{k}',\bm{k}+\bm{q}}.
\end{equation}
A non-zero IVC at $\bm{q}\neq\bm{0}$ implies valley spiral order. On the other hand, valley-diagonal observables yield translation-invariant expectation values since $\langle d^\dagger_{\bm{k},\tau}d_{\bm{k}',\tau}\rangle\sim \delta_{\bm{k},\bm{k}'}$. 

Fig.~\ref{fig:1BLLL_phase_24x24} presents a HF phase diagram as a function of kinetic bandwidth $W$ and valley interaction anisotropy $\alpha_\text{v}$.  [See SM, Sec.~\ref{secsupmat:1BLLL_additional} for analogous results with an additional second harmonic component to the dispersion $\epsilon_{\tau}(\bm{k})$]. For reasons to be made apparent later, we restrict the boost to $\bm{q}=(0,0)$ or $(Q/2,Q/2)$. All data points shown have a positive indirect gap. We first summarize the broad features of the phase diagram. Around the lower left corner $(\alpha_\text{v}=W=0)$, we find the valley-polarized (VP) phase where all electrons occupy one of the two valleys. This is an exact eigenstate across the phase diagram, since it is the unique many-body state in its $U(1)_\text{v}$ sector. The VP phase is surrounded by a narrow sliver of states with finite IVC, but which still have a non-zero valley polarization. We dub this the tilted valley-polarized (TVP) phase, whose boost is $\bm{q}=(Q/2,Q/2)$, except for $W=0$ where we find that $\bm{q}=(0,0)$ and $(Q/2,Q/2)$ are degenerate\footnote{In fact, all $\bm{q}$ are degenerate in the $W=0$ limit.}. For larger $W$ and $\alpha_\text{v}$, the phase diagram is dominated by the CTI$_1$. The CTI$_1$ has $\hat{\mathcal{T}}$ symmetry, and its boost $\bm{q}$ follows the same rules as for the TVP. 
Below, we discuss these phases and their transitions in more detail by zooming in on the $\alpha_\text{v}$ and $W$ axes of Fig.~\ref{fig:1BLLL_phase_24x24}.

\subsubsection{$\alpha_\mathrm{v}$-axis}\label{subsubsec:1BLLL_alpha_axis}

\begin{figure*}
    \centering
    \includegraphics[width = \linewidth]{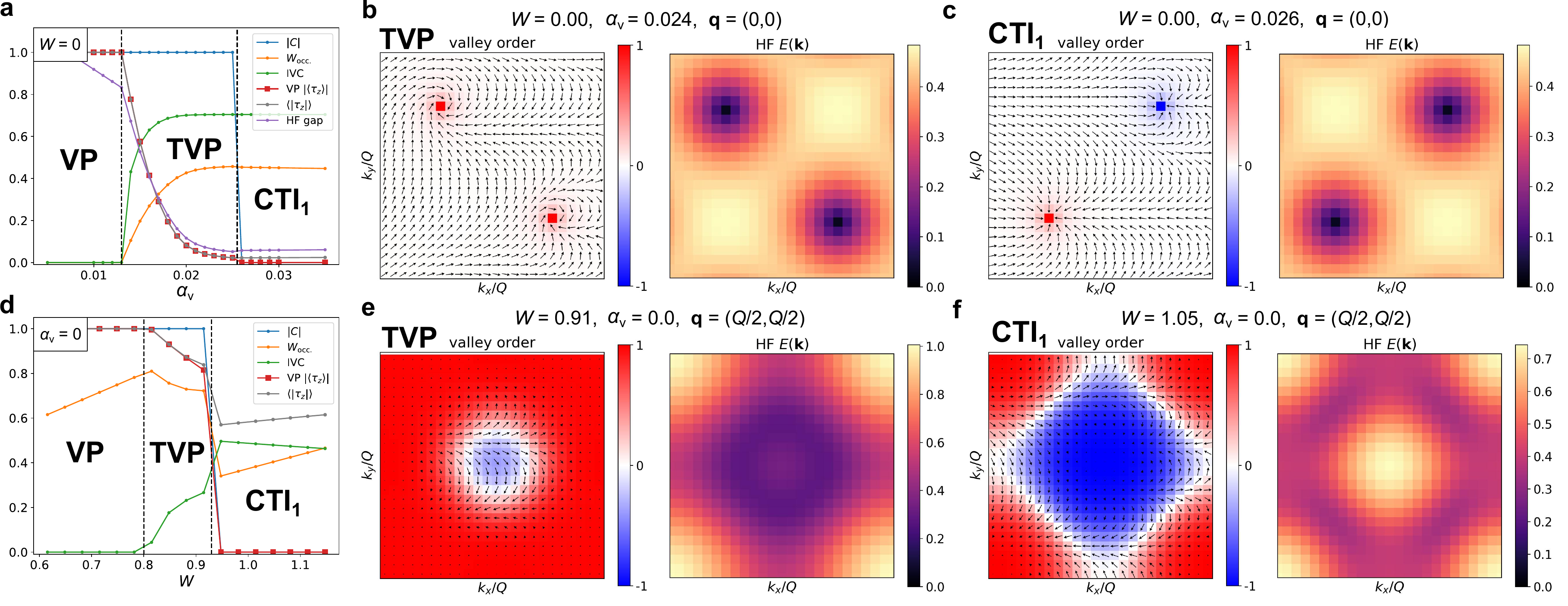}
    \caption{\textbf{HF phase diagram of the LLL model at half-filling along the $\alpha_\text{v}$ and $W$ axes.} The gate distance $d_\text{sc}=6a$, and the interaction scale $U=1$. a) Phase diagram as a function of anistropy $\alpha_\text{v}$ for zero dispersion $(W=0)$. $C$ is the Chern number, $W_\text{occ.}$ is the bandwidth of the occupied HF valence band, IVC is intervalley coherence, and VP is valley polarization. Boost is fixed at $\bm{q}=(0,0)$. b) Properties of a representative TVP state. Left: momentum-dependent valley order parameter. Color indicates $\langle\tau_z\rangle$, while arrows indicate in-plane order. Right: HF conduction band dispersion. The valence band has $-E(\bm{k})$ due to particle-hole symmetry. c,e,f) Same as b) except for different parameters and states. d) Same as a) except as a function of dispersion $W$ for zero anistropy $(\alpha_\text{v}=0)$. Boost is fixed at $\bm{q}=(Q/2,Q/2)$. System size for all plots is $24\times 24$.}
    \label{fig:1BLLL_axes}
\end{figure*}

Fig.~\ref{fig:1BLLL_axes}a focuses around the phase transitions in the flat band limit ($W=0$), with boost $\bm{q}=(0,0)$. The VP phase for small $\alpha_\text{v}$ has uniform Berry curvature, and its HF band structure is also flat for $W=0$.
As proposed in Ref.~\cite{Bultinck2019mechanism}, the VP is the candidate state for the quantized anomalous Hall (QAH) insulator observed in hBN-aligned TBG at $\nu=+3$~\cite{Serlin900,Sharpe_2019}, and is expected to have a finite threshhold of stability against $\alpha_\text{v}$ and $W$. The fact that VP survives a non-zero $\alpha_\text{v}>0$ (i.e.~weaker intervalley vs.~intravalley interactions) is at first glance surprising --- in the usual quantum Hall bilayer at $\nu=1$, a non-zero interlayer distance $d$ weakens the interlayer interaction and drives interlayer excitonic condensation to minimize the charging energy~\cite{Moon1995,Eisenstein2014exciton}. However the situation in the LLL model is different, as the two valleys have opposite Chern numbers, and there is no $SU(2)_\text{v}$ symmetry even for $\alpha_\text{v}=0$. Moreover, Ref.~\cite{Bultinck2019mechanism} argued that IVC is energetically penalized because the corresponding order parameter $\Delta(\bm{k})$ must admit vortices due to topological constraints (see Sec.~\ref{subsec:1BLLL_winding}). By mapping to an effective problem of a superconductor in a magnetic field, Ref.~\cite{Bultinck2019mechanism} derived a trial time-reversal invariant IVC state, dubbed the exciton vortex lattice (EVL), and showed that its energy was higher than the VP for small $\alpha_\text{v}$. In the BZ, the EVL possesses two $2\pi$ vortices where the band structure forms Dirac points. The EVL and its order parameter will be discussed in more detail in Sec.~\ref{subsec:1BLLL_EVL}.

At $\alpha_\text{v}\simeq 0.013$, the VP begins to develop IVC through what appears to be a continuous phase transition. In the resulting TVP phase, which is still a QAH state with $|C|=1$, the HF gap and valley polarization remain finite. In fact, the valley pseudospin remains within one hemisphere throughout the BZ. For the boost $\bm{q}=(0,0)$ enforced here, the IVC develops across the entire BZ except at two fully valley-polarized nodes which can either be at $\pm(Q/4,Q/4)$ or $\pm(Q/4,3Q/4)$. Just below the phase transition at higher $\alpha_\text{v}\simeq0.025$, the momentum-resolved valley pseudospin resembles two merons in momentum space (Fig.~\ref{fig:1BLLL_axes}b left). Note that the `core polarization' and the in-plane winding of $2\pi$ are identical for the merons, the latter being consistent with the net $4\pi$ winding around the BZ that is topologically required for the IVC order parameter $\Delta(\bm{k})$. While the HF dispersion $E(\bm{k})$ of the conduction band (the HF valence band dispersion is $-E(\bm{k})$ due to particle-hole symmetry) has a minimum at the meron cores, it remains gapped throughout the TVP phase. As will be explained in Sec.~\ref{subsec:1BLLL_EVL}, the TVP in this limit resembles the EVL with small valley masses of the same sign at the two vortices.

A  weakly first-order topological transition occurs at $\alpha_\text{v}\simeq0.025$ where one of the merons changes its core valley polarization (see Fig.~\ref{fig:1BLLL_axes}c). The resulting CTI$_1$ preserves $\hat{\mathcal{T}}$ and hence has $C=0$. While the net valley polarization $|\langle\tau_z\rangle|$ vanishes, its average magnitude across the BZ $\langle|\tau_z|\rangle$ is non-zero due to the valley polarization around the meron cores. At these positions, the HF gap is smallest, and the Berry curvature has pronounced peaks with opposite signs. The CTI$_1$ resembles the TVP except with opposite valley masses at the two EVL vortices\footnote{We emphasize that the  momentum-space merons have identical vorticities in the TVP and CTI$_1$, which is compatible with the $4\pi$ winding of the pseudospin around the BZ. Since the pseudospin is not BZ-periodic we cannot define a topological skyrmion number.}.

\subsubsection{$W$-axis}\label{subsubsec:Waxis}

For valley isotropic interactions ($\alpha_\text{v}=0$), Fig.~\ref{fig:1BLLL_axes}d shows the phase diagram with boost $\bm{q}=(Q/2,Q/2)$ in the vicinity of the phase transitions. We remark that the physics of the LLL model along the $W$-axis is more relevant to the moir\'e materials in Sec.~\ref{sec:IKSasETI} and Ref.~\onlinecite{companion}, since the long-range density-density interaction naturally has $\alpha_\text{v}=0$, and the bandwidth $W$ is usually comparable to the interaction strength and can often be tuned with a displacement field. Similar to Sec.~\ref{subsubsec:1BLLL_alpha_axis}, the transition at $W\simeq 0.8$ between the VP and TVP appears to be continuous or very weakly first-order. However, the TVP has some qualitative differences to that along the $\alpha_\text{v}$ axis. As shown in Fig.~\ref{fig:1BLLL_axes}e for a TVP that has a net polarization in $\tau=+$, the valley pseudospin tilts into the opposite hemisphere around $\bm{k}=(Q/2,Q/2)$, but maintains finite IVC and does not fully polarize along $\tau=-$. The zeros in $\Delta_{\bm{q}}(\bm{k})$, required by the net $4\pi$ pseudospin winding, are at $\bm{k}=(0,Q/2)$ and $(Q/2,0)$. 

As can be inferred from the small peak in the HF dispersion of the TVP at $(Q/2,Q/2)$ (Fig.~\ref{fig:1BLLL_axes}e), the non-zero valley boost and tilting into the opposite valley is driven by the details of the non-interacting band structure, which we now explain with the lobe principle outlined in Sec.~\ref{subsec:intro_picture}. For the square cosine potential used here, the dispersion $\epsilon(\bm{k})$ has a minimum at $(0,0)$ and a maximum at $(Q/2,Q/2)$ in the unboosted `lab frame'. Consider initially a VP state that is fully polarized in $\tau=+$. This has good exchange energetics, but poor kinetic energy around the maximum of $\epsilon(\bm{k})$ at $(Q/2,Q/2)$. The system can lower its kinetic energy while remaining insulating (with uniform particle occupation of 1 across the BZ) in the following way. We boost valley $\tau=-$ by $\bm{q}=(Q/2,Q/2)$, which aligns the dispersion minimum in that valley with the maximum in $\tau=+$. Then, we tilt the VP state towards $\tau=-$ at $\bm{k}=(Q/2,Q/2)$ (where $\bm{k}$ is defined relative to the unboosted BZ of $\tau=+$). This tilting, which introduces IVC, is done smoothly in momentum space in order to reduce the loss of exchange gain. For sufficiently large $W$, the lowering of kinetic energy is able to overcome the exchange penalty.

At larger $W\simeq 0.92$, there is a strongly first-order transition to the $\hat{\mathcal{T}}$-invariant CTI$_1$ phase. As shown in Fig.~\ref{fig:1BLLL_axes}f, the valley pseudospin is fully polarized along $\tau=+$ ($\tau=-$) at $\bm{k}=(0,0)$ [$\bm{k}=(Q/2,Q/2)$]. The HF band structure clearly shows the influence of the kinetic energy. Unlike in the $W=0$ case in Sec.~\ref{subsubsec:1BLLL_alpha_axis}, the HF gap is maximum at the valley-polarized meron cores, which are now significantly more spread out in momentum space to take advantage of the band extrema. The way that exchange and the dispersion $\epsilon(\bm{k})$ combine to shape the properties of the CTI$_1$ is reminiscent of the lobe principle used to explain the stabilization of the IKS in TBG~\cite{kwan_kekule_2021}. The precise relationship between the CTI  and the IVC phases in moir\'e graphene will be explained in more detail in Sec.~\ref{sec:IKSasETI} and Ref.~\onlinecite{companion}. 
As we will see, while the IKS in TBG requires a $\hat{C}_{2z}$-symmetric model for a proper description (see Sec.~\ref{sec:2BLLL}), the CTI in the LLL model has closely related cousins in other materials~\cite{companion}.

Since the lowest-harmonic dispersion of Eq.~\ref{eq:SP} has intervalley nesting at half-filling, the CTI$_1$ phase persists as $W\rightarrow\infty$. However, we emphasize that the CTI$_1$ is fundamentally an intermediate-coupling order which does not rely on weak-coupling nesting instabilities. In Sec.~\ref{secsupmat:1BLLL_additional} of the SM, we show that the CTI$_1$ phase survives for a large range of parameters at finite $W\sim1$ in the presence of higher harmonics in the dispersion, which eliminates the nesting.

\subsection{Exciton vortex lattice}\label{subsec:1BLLL_EVL}

In this section, we discuss exciton vortex lattice (EVL), which was first introduced in Ref.~\cite{Bultinck2019mechanism}, and its relation to the CTI.

\subsubsection{Construction of order parameter}
We first briefly recap the motivation behind the EVL~\cite{Bultinck2019mechanism}. 
Our goal is to write down an order parameter corresponding to strong IVC between the two valleys, which in real space involves the fermion bilinear $\Delta(\bm{r})\propto\langle{\psi}^\dagger_{+}(\bm{r}){\psi}_{-}(\bm{r})\rangle$, with ${\psi}^\dagger_{\tau}(\bm{r})$ the position creation operator in valley $\tau$. 
By performing a particle-hole transform in valley $\tau=-$, we see that $\Delta(\bm{r})$ effectively describes a superconducting order parameter in a fictitious system that experiences a net magnetic field. As there is no Meissner effect in 2d, the system is forced to accommodate the field through vortices in the order parameter, which will arrange to form a lattice\footnote{Since the lattice vectors of the LLL model form a square lattice due to the external potential, the vortices will not arrange in the usual triangular lattice.}. The resulting phase is dubbed the exciton vortex lattice (EVL).

In momentum space, we can derive the order parameter of the EVL (see SM, Sec.~\ref{secsupmat:1BLLL_additional} for details)
\begin{align}
    \begin{split}\label{eq:1BLLL_EVL_OP}
        \Delta_{\text{EVL},\bm{q}}(\bm{k}-\frac{\bm{q}}{2})
        &=\Delta_0 e^{2ik_xk_y-k_y^2}\theta_3\left(\frac{k_x+ik_y}{Q},\frac{s+i}{2}\right)\\
        &=\Delta_{\text{EVL},\bm{q}}(-\bm{k}-\frac{\bm{q}}{2})
    \end{split}
\end{align}
where $\Delta_0$ is some overall normalization and $\bm{q}$ denotes the intervalley boost of $\tau=-$ relative to $\tau=+$. The Jacobi theta function of the third kind $\theta_3(z,\alpha)$ has nodes at $z=m\alpha + n + \frac{\alpha + 1}{2}$ where $m,n\in\mathbb{Z}$. Hence $\Delta_{\text{EVL},\bm{q}}(\bm{k}-\frac{\bm{q}}{2})$ either has vortices at $\pm(Q/4,3Q/4)$ for $s=1$, or $\pm(Q/4,Q/4)$ for $s=-1$. 

\begin{figure}
    \centering
    \includegraphics[width = \linewidth]{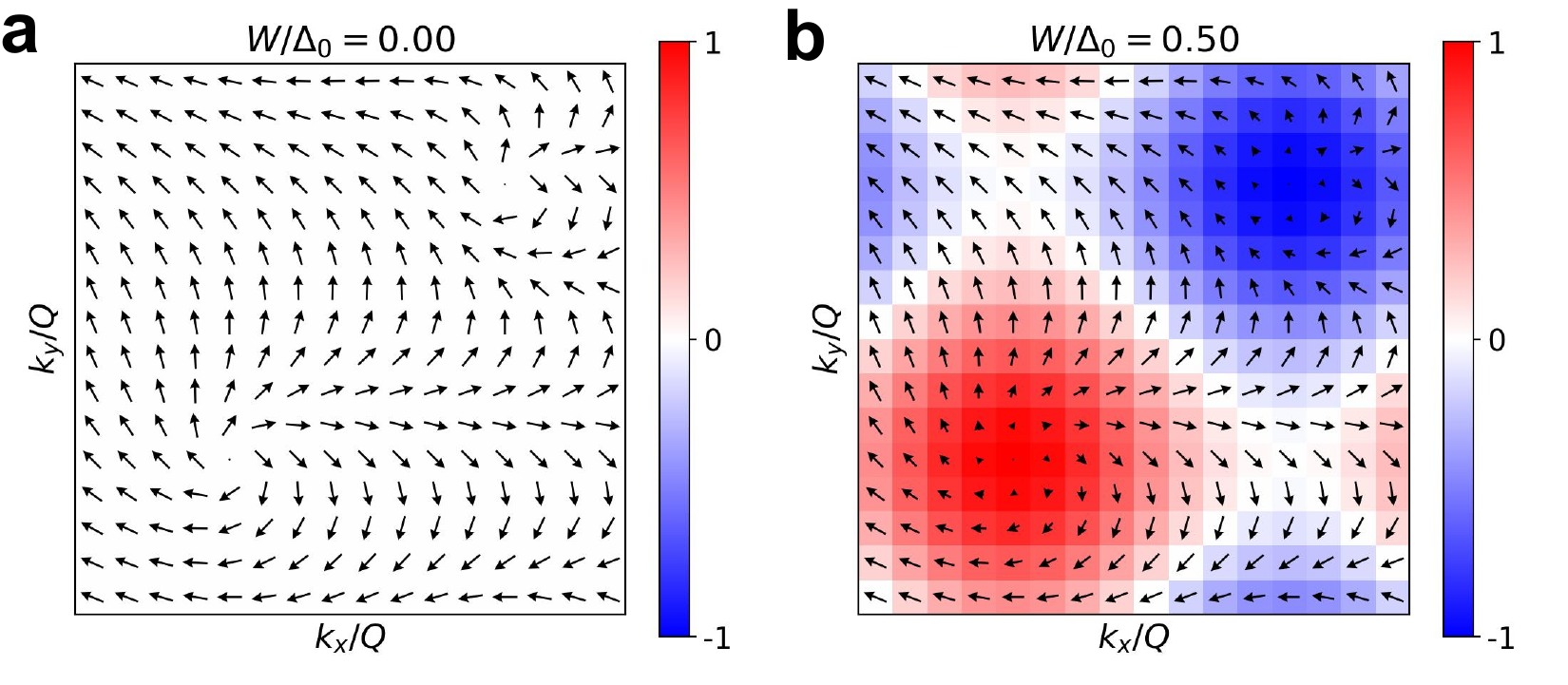}
    \caption{\textbf{Properties of the CTI$_1$ constructed from the EVL ansatz (Eq.~\ref{eq:trial_BFEI}) with $\bm{q}=0$ and $s=-1$ in the LLL model.} Plots show the in-plane pseudospin (arrows) and out-of-plane valley polarization (color) in the BZ. a) Results for $W/\Delta_0=0$. b) Same as a) but for $W/\Delta_0=0.5$.} 
    \label{fig:EVL_BML_ansatz}
\end{figure}

Fig.~\ref{fig:EVL_BML_ansatz}a plots the in-plane order of the ground state projector of the mean-field Hamiltonian
\begin{equation}
h(\bm{k})=\begin{pmatrix}
        0 & [\Delta_{\text{EVL}}(\bm{k})]^*\\
        \Delta_{\text{EVL}}(\bm{k}) & 0
    \end{pmatrix}
\end{equation}
constructed using the EVL order parameter with $\bm{q}=0$. Thus, $\Delta_{\text{EVL}}(\bm{k})$ controls the direction of the in-plane component of the valley pseudospin $\bm{n}(\bm{k})$. As we have chosen $s=-1$, the IVC exhibits nodes at $\pm(Q/4,Q/4)$. Consistent with the discussion in Sec.~\ref{subsec:1BLLL_winding}, the pseudospin winds by $4\pi$ when going around the BZ counter-clockwise. 
Since the magnetic Bloch operators are periodic in the $k_y$ direction, it is straightforward to detect the vorticity of each vortex by tracking the rotation of the pseudospin along $k_y$ as a function of $k_x$. For instance, the pseudospin does not wind along $k_y$ at $k_x=0$, but after passing the vortex at $k_x=Q/4$, the pseudospin winds by $2\pi$ along $k_y$.

\subsubsection{Trial state for CTI}
The quasiparticle spectrum of the EVL state is gapless due to the Dirac points at the two vortices. However, there is no symmetry that protects these Dirac points. Hence the EVL requires fine-tuning, and it is natural to ask what proximate phases are obtained by gapping the Dirac points, which would likely reduce the total energy. In fact, we already know the answer from the discussion in Sec.~\ref{subsubsec:1BLLL_alpha_axis}. As shown in Fig.~\ref{fig:1BLLL_axes}b,c, the TVP can be reached from the EVL by generating identical interaction-induced valley masses at the two Dirac points, while the CTI$_1$ is obtained by generating opposite valley masses. By smoothly canting towards the poles of the valley Bloch sphere at the vortex cores and forming pseudospin merons in momentum space, the system can lower the exchange energy, while respecting the topological constraints imposed on the order parameter (i.e.~vanishing of IVC at the meron cores). This possibility of gapping the Dirac points of the EVL by tilting in valley space was missed in the analysis of Ref.~\cite{Bultinck2019mechanism}. Along the $\alpha_\text{v}$-axis of the HF phase diagram (Fig.~\ref{fig:1BLLL_axes}a), the transition between the TVP and CTI$_1$ appears very weakly first-order, such that the EVL is skipped entirely. 

The CTI$_1$ at finite bandwidth $W$ (Fig.~\ref{fig:1BLLL_axes}f) can also be obtained by orienting the pseudospin of the EVL (with appropriate $\bm{q}$) along opposite poles at the vortex cores. This case is conceptually simpler because the single-particle dispersion $\epsilon_\tau(\bm{k})$ is an obvious candidate for the origin of the valley masses\footnote{The general form of the ansatz in Eq.~\ref{eq:trial_BFEI} can also be used to model the TVP and CTI$_1$ for $W=0$, though one would need to specify some other functions to lie along the diagonals of $h(\bm{k})$.}. To see this, consider for simplicity the $\bm{q}=0,s=-1$ EVL whose vortices are at $\pm(Q/4,Q/4)$ in the BZ (Fig.~\ref{fig:EVL_BML_ansatz}a). We then apply a lowest-harmonic cosine potential (Eq.~\ref{eq:SP_general}) with phases $\phi_x=-\phi_y=-\pi/2$ chosen so that the minima and maxima of the dispersions in the two valleys anti-align, and coincide with the vortex cores\footnote{If we chose different phases $\phi_x,\phi_y$, then we would need to tune the intervalley boost $\bm{q}$ in order to anti-align the valley band extrema, and have them coincide with the vortices. The construction of the CTI$_1$ trial state follows similarly.}. We now solve the $\hat{\mathcal{T}}$-symmetric mean-field Hamiltonian 
\begin{equation}
\begin{gathered}
    H^\text{MF}=\sum_{\bm{k},\tau,\tau'}d^\dagger_{\bm{k},\tau}h_{\tau,\tau'}(\bm{k})d_{\bm{k},\tau'}\\
    h(\bm{k})=\begin{pmatrix}
        \epsilon_+(\bm{k}) & [\Delta_{\text{EVL}}(\bm{k})]^*\\
        \Delta_{\text{EVL}}(\bm{k}) & \epsilon_-(\bm{k})
    \end{pmatrix}\label{eq:trial_BFEI}
\end{gathered}
\end{equation}
to obtain the density matrix $P_{\tau,\tau'}(\bm{k})=\langle d^\dagger_{\bm{k},\tau}d_{\bm{k},\tau'}\rangle$. As shown in Fig.~\ref{fig:EVL_BML_ansatz}b, the resulting pseudospin texture forms two merons with opposite core polarities in momentum space, resembling the CTI$_1$ obtained within self-consistent HF calculations. Apart from the $U(1)_\text{v}$ phase, the only non-trivial free parameter in Eq.~\ref{eq:trial_BFEI} is the ratio $W/\Delta_0$ between the bandwidth and the IVC mean-field strength. By tuning this ratio for $\bm{q}=(Q/2,Q/2)$ and $\phi_x=\phi_y=0$, we can recover the CTI$_1$ state in Fig.~\ref{fig:1BLLL_axes}f with errors in the local pseudospin of less than $1\%$, reinforcing the notion that the CTI$_1$ is a gapped version of the EVL.

\subsection{Extensions}\label{subsec:1BLLL_extensions}

So far, the calculations in this section have used a first-harmonic single-particle dispersion whose `lobe' spacing $(Q/2,Q/2)$ coincides with the vortex spacing in the EVL ansatz of Eq.~\ref{eq:1BLLL_EVL_OP}. As a result, the CTI$_1$ can be intuitively understood as simply applying opposite valley masses around the two Dirac points (see Eq.~\ref{eq:trial_BFEI} and Fig.~\ref{fig:EVL_BML_ansatz}). However, the band structures in realistic materials are complex and prone to interaction-induced renormalization. External effects like strain can also break rotational symmetries. In the Supplementary Material (Sec.~\ref{secsupmat:1BLLL_additional}), we show that the CTI$_1$ survives for more general single-particle dispersions $\epsilon_\tau(\bm{k})$. This underscores the fact that {\it i}) the topological frustration of the CTI$_1$ is a generic feature of coherence between opposite Chern bands, and {\it ii}) the flexibility of the lobe principle provides a broad energetic basis for the stabilization of CTIs in a variety of conditions.

The LLL model does not straightforwardly accommodate bands with higher Chern numbers $|C|>1$, but we can anticipate the general properties of the CTI$_n$ for a one-band model with $C=n\tau$ in the two valleys. In this case, the IVC order parameter in a smooth gauge must wind by $4n\pi$ around the BZ. Hence, the CTI$_n$ must have vortices/merons in the BZ to accommodate this winding. Their positions and multiplicities will be controlled by the interplay between exchange physics and the details of $\epsilon_\tau(\bm{k})$. An example of a CTI$_n$ with $n=2$ in a moir\'e system is presented in Ref.~\onlinecite{companion}.

Finally, we comment on the effects of introducing additional bands. As discussed in Sec.~\ref{subsec:intro_topology_CTI} and App.~\ref{secapp:toy_local}, the CTI has $C=0$ and possesses exponentially-localized Wannier orbitals, but yet does not admit an atomically localized description. However, this obstruction is delicate in that it is not robust to adding trivial bands. In App.~\ref{secapp:CTI_remote}, we show that a CTI can in principle relax its non-trivial valley texture by mixing into these bands, eventually connecting to the atomically trivial limit, though such mixing may be energetically unfavorable especially if the additional bands are sufficiently remote in energy.

\section{Euler-textured insulators}\label{sec:2BLLL}

In this section we discuss the physics of the ETI in $\hat{C}_{2z}$- and $\hat{\mathcal{T}}$-symmetric systems where the single-particle Hamiltonian in each valley has a non-trivial Euler topology. The focus is on the topological considerations underlying the ETI, though we briefly comment on energetic considerations where relevant. We first summarize the properties of the Euler class that will be relevant for our purposes. The Euler class $e_2$ is an integer-valued topological invariant that characterizes the topology of two-band systems with a space-time inversion symmetry that squares to one, which we will take to be $\hat{C}_{2z}\hat{\mathcal{T}}$ in this work. A non-zero $e_2$ presents an Euler obstruction to a symmetric exponentially-localized Wannier representation, and enforces gapless band crossings with a net winding number of $-2e_2$ between the two bands~\cite{Ahn2019}. A single $\hat{C}_{2z}\hat{\mathcal{T}}$-symmetric band from this two-band subspace therefore has singularities. It is always possible to construct a non-symmetric basis consisting of bands with Chern numbers $C=\pm e_2$, such that $\hat{C}_{2z}\hat{\mathcal{T}}$ maps the Chern bands onto each other~\cite{Xie2020bounded,Unal2020Euler,Bouhon2020Weyl,Yu2023EulerTBG}. An example of $|e_2|=1$ is given by the central bands of TBG within a single spin-valley flavor. This can be seen from the existence of a sublattice-polarized `Chern basis' in TBG with $C=\tau\sigma$, where $\hat{C}_{2z}\hat{\mathcal{T}}$ flips the sublattice index but leaves the valley invariant. In Sec.~\ref{secsupmat:additional_ETI} of the SM, we introduce another two-band model with $e_2=1$ that is based on LLLs and generalizes the model of Sec.~\ref{sec:1BLLL}.

In Sec.~\ref{sec:2BLLL_ETI}, we first demonstrate that a gapped $U(1)_\text{v}$-breaking mean-field insulator at $\nu=\pm1$ in a two-valley model with an Euler obstruction in each valley
that otherwise preserves all other symmetries must be an ETI. 
In Sec.~\ref{sec:C2z_effective}, we discuss an effective toy model that further elucidates the physics of the ETI. Further variants of the toy model are studied in the Supplementary Material (Sec.~\ref{secsupmat:additional_ETI}).

\subsection{General arguments}\label{sec:2BLLL_ETI}

We consider a model with two bands per valley\footnote{Spin can be incorporated as a spectator degree of freedom.} which preserves $\hat{C}_{2z}$, $\hat{\mathcal{T}}$ and $U(1)_\text{v}$. As usual, $\hat{C}_{2z}$ and $\hat{\mathcal{T}}$ flip the valley index. We require that the two single-particle bands within a valley carry a non-zero Euler class $e_2$. Importantly, this means that within a valley, any single $\hat{C}_{2z}\mathcal{T}$-symmetric band selected from the two-band space must have singularities in the BZ.
We add electron interactions that preserve all the symmetries. The filling $\nu=0$ corresponds to charge neutrality, so that the system is fully occupied at $\nu=+2$ and fully unoccupied at $\nu=-2$.

We are interested in constructing a gapped $\hat{C}_{2z}$- and $\hat{\mathcal{T}}$-symmetric insulator at $\nu=\pm1$. Ruling out fractionalization and both time-reversal and translation symmetry breaking means that such an insulator must be intervalley coherent. Since there are multiple bands and no net Chern number in each valley, the CTI discussion in Sec.~\ref{sec:1BLLL} does not straightforwardly apply here, and it is a priori not clear that there is an obstruction to having a uniform non-vanishing IVC across the BZ. 

We now demonstrate that such an insulating IVC state is still frustrated at $\nu=\pm1$ owing to the non-trivial Euler topology. To see this, we focus on $\nu=-1$ for concreteness, and consider the occupied intervalley-coherent band $\ket{\text{HF},\bm{k}}$ at $\nu=-1$ that is separated by a charge gap to other bands. This band can be expressed generally as
\begin{equation}\label{eq:IR_basis}
    \ket{\text{HF},\bm{k}}=\alpha(\bm{k})\ket{\text{HF}_+,\bm{k}}+\beta(\bm{k})\ket{\text{HF}_-,\bm{k}}
\end{equation}
which implicitly defines a pair of time-reversal-related, valley-diagonal, and normalized bands $\ket{\text{HF}_+,\bm{k}}$ and $\ket{\text{HF}_-,\bm{k}}$, dubbed the `valley-filtered' bands\footnote{For a given $\ket{\text{HF},\bm{k}}$, the valley-filtered bands are unique up to a momentum-dependent phase.}. The coefficients $\alpha(\bm{k})$ and $\beta(\bm{k})$, whose magnitudes are gauge-independent, parameterize the momentum-dependent valley structure of the IVC insulator. Given our assumptions, each valley-filtered band individually obeys $\hat{C}_{2z}\hat{\mathcal{T}}$-symmetry. However, due to the non-trivial Euler class of the two-band Hilbert space in each valley, the valley-filtered bands must contain singularities in the BZ as pointed out above. In order to construct a physically smooth insulating state, the valley pseudospin cannot solely lie in-plane, but must fully cant towards the poles for some momenta in order to hide the effects of these singularities. In other words, $\alpha(\bm{k})$ and $\beta(\bm{k})$ each must vanish (i.e.~the system is locally valley polarized) at different points in the BZ, manifesting a non-trivial valley texture. We dub this type of insulating order an ETI, an example of which is the IKS in TBG and TSTG (see Sec.~\ref{sec:IKSasETI}).

Our analysis highlights that a $\hat{C}_{2z}$- and $\hat{\mathcal{T}}$-symmetric ETI at $|\nu|=1$ can be insulating despite the intrinsic two-band Euler topology within each valley, because the $U(1)_v$ symmetry-breaking mixes the valleys into a combined four-band problem. An alternative $\hat{C}_{2z}\hat{\mathcal{T}}$- and $U(1)_v$-symmetric stripe insulator at half-filling of a single spin-valley flavor of TBG, which furnishes a single two-band space with $|e_2|=1$, was proposed in Ref.~\cite{Kang2020} (see also Ref.~\cite{Xie2023nu3}). There, the spontaneous doubling of the moir\'e unit cell leads to a four-band problem in the reconstructed BZ, enabling a gap to be opened at the Fermi level via non-Abelian braiding of Dirac points~\cite{Ahn2019,Wu2019nonabelian}. 

We now compare and contrast the construction of the CTI and ETI with reference to the general decomposition of a gapped IVC band in Eq.~\ref{eq:IR_basis}. In the CTI$_n$ setting, there is only one band in each valley, such that the valley-filtered bands $\ket{\text{HF}_\tau,\bm{k}}$ are fixed, and are generic Chern bands. Owing to the mandatory $4\pi n$ winding of the order parameter, introducing IVC as in Eq.~\ref{eq:IR_basis} leads to vortices, whose positions can be anywhere in the BZ, with associated Dirac points in the electronic spectrum. The vortex positions are ultimately fixed by energetic considerations, such as the lobe principle, and the CTI$_n$ is realized upon gapping these Dirac points and forming a smooth valley pseudospin texture. 

On the other hand, in the ETI setting where each valley contains two bands with non-trivial Euler class, the valley-filtered bands are not uniquely specified at the outset. Indeed, we could pick any set of $\ket{\text{HF}_\tau,\bm{k}}$ as long as they satisfy $\hat{C}_{2z}$ and $\hat{\mathcal{T}}$. Crucially, different choices lead to different positions of singularities which are required due to the Euler topology. To create an IVC insulator, these singularities 
would need to be hidden by locally valley-polarizing in momentum space, leading to an ETI with a non-trivial valley texture. 
Again, the precise choice of ETI that is stabilized (if at all) in some Hamiltonian depends on energetic details that are influenced by many factors such as band dispersion and quantum geometry.

\subsection{Toy model for ETI}\label{sec:C2z_effective}

\begin{figure}[t]
    \centering
    \includegraphics[width = \linewidth]{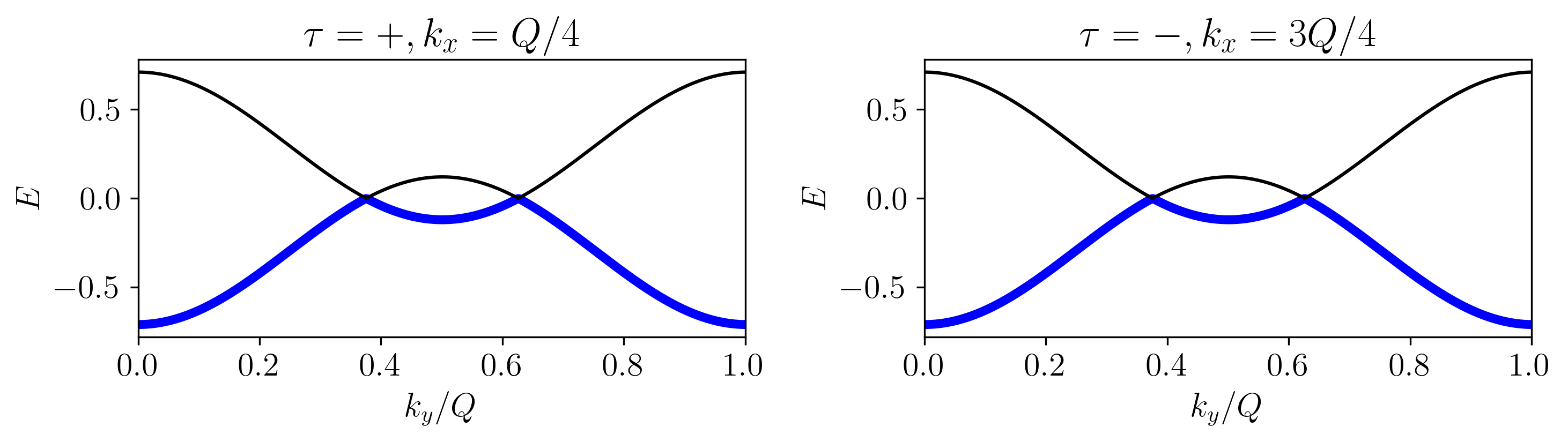}
    \caption{\textbf{Non-interacting band structure of the ETI toy model.} In each valley, we take a linecut in $k_y$ that intersects the non-interacting Dirac points. We take $\bm{k}_1=(Q/4,3Q/8)$ and $\bm{k}_2=(Q/4,5Q/8)$ in Eq.~\ref{eq:fk_H+}. When constructing the ETI in Sec.~\ref{sec:C2z_effective}, we project onto the lower bands (highlighted blue).}
    \label{fig:toy_model_valley}
\end{figure}

The previous subsection highlighted the choice of valley-filtered bands $\ket{\text{HF}_\tau,\bm{k}}$ in Eq.~\ref{eq:IR_basis} as an important aspect of the ETI, i.e.~which band from each valley is chosen to participate in intervalley coherence in the occupied band? To further illuminate the physics of the ETI, we focus on a regime where the appropriate valley-filtered bands can be approximated in a physically intuitive manner. In particular, we imagine that the band structure in each valley is strongly dispersive, with the lower and upper bands connected with two non-interacting Dirac points with identical winding consistent with $|e_2|=1$ (see Fig.~\ref{fig:toy_model_valley}). At $\nu=-1$, the non-interacting Fermi level intersects the lower band in each valley. If the interaction is relatively weak compared to the bandwidth, then a putative ETI at $\nu=-1$ would primarily be built from the lower bands, and a reasonable assumption is to take these to be $\ket{\text{HF}_\tau,\bm{k}}$ (highlighted in blue). This fixes at the outset the singularities of the valley-filtered bands to where they would have connected to the upper bands in the non-interacting band structure.

\begin{figure}
    \centering
    \includegraphics[width = \linewidth]{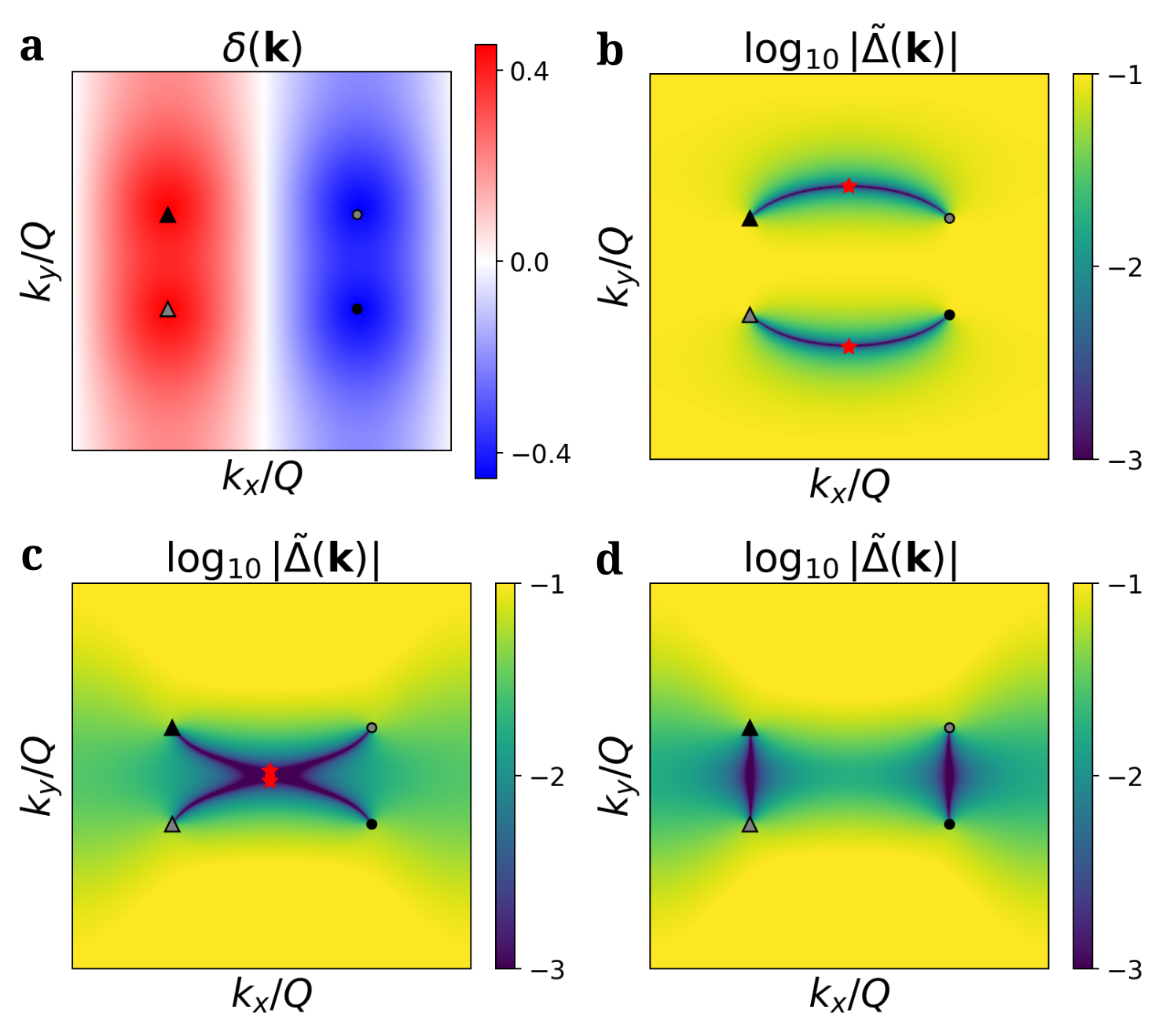}
\caption{\textbf{Toy model for ETI in the BZ, projected to the lower bands.} 
The unprojected valley-diagonal kinetic term is given in Eq.~\ref{eq:fk_H+}. Any boost $\bm{q}$ has been absorbed into the momentum origin such that TRS takes $\bm{k}\rightarrow -\bm{k}$.
a) Difference $\delta(\bm{k})$ between the dispersions of the lower bands in the two valleys. Non-interacting Dirac points in valley $\tau=+$ are at $\bm{k}_1=(Q/4,3Q/8)$ [grey triangle] and $\bm{k}_2=(Q/4,5Q/8)$ [black triangle].
TRS-related non-interacting Dirac points in valley $\tau=-$ are indicated with dots.
b,c,d) Projected IVC matrix element $\tilde{\Delta}(\bm{k})$ [Eq.~\ref{eq:toy_model_Htilde}] for a fixed TIVC coupling strength $\Delta_{\text{TIVC}}=0.1$ (Eq.~\ref{eq:TIVCmass}) and variable inter-Chern coupling strength (Eq.~\ref{eq:HCTI}). b,c,d) corresponds to $\Delta_{\text{inter}}=0, -0.085, -0.1$ respectively. Note that the lower limit of the color scale has been clamped. Red stars indicate residual Dirac points in the projected Hamiltonian of Eq.~\ref{eq:toy_model_Htilde}. The residual Dirac points have annihilated in d), leading to the formation of an ETI.
}
    \label{fig:2BLLL_fk_model}
\end{figure}

We now introduce a phenomenological toy model that captures the formation of an ETI in the regime described above. We choose the BZ to be the same as that of the LLL in Sec.~\ref{sec:1BLLL}, so that $k_x\in[0,Q),k_y\in[0,Q)$ for both valleys. 
The two bands in each valley are described by $2\times2$ matrices $H_+(\bm{k})$ and $H_-(\bm{k})$ in sublattice space\footnote{We assume that any requisite boosting has been absorbed into the momentum coordinates so that any IVC will be induced at $\bm{q}=0$, and TRS takes $\bm{k}\rightarrow -\bm{k}$.}:
\begin{align}\label{eq:fk_H+}
        H_+(\bm{k})&= \begin{pmatrix}
        0 & f(\bm{k}; \bm{k}_1)^*f(\bm{k}; \bm{k}_2)^*\\
        f(\bm{k}; \bm{k}_1)f(\bm{k}; \bm{k}_2) & 0
    \end{pmatrix},\\
\label{eq:fk_H-}
H_-(\bm{k})&= H^*_+(-\bm{k}).
\end{align}
The system is invariant under TRS and two-fold rotation which act on valley and sublattice as $\hat{\mathcal{T}}=\tau^x\mathcal{K}$ and $\hat{C}_{2z}=\tau^x\sigma^x$ respectively. TRS leads to the constraint \eqref{eq:fk_H-} that links the Hamiltonians in the two valleys. 
The sublattice basis carries Chern number $C=\tau^z\sigma^z$ so that the Euler class $|e_2|=1$ in each valley. 

The inter-sublattice tunneling function $f(\bm{k}; \bm{k}^\prime)$ has a vortex at $\bm{k}^\prime$ in the BZ and satisfies the appropriate (quasi)-periodicity in momentum space for Chern bands. In particular, we choose here
\begin{multline}
    {f}(\bm{k}; \bm{k}^\prime) = e^{i(k_x - k^\prime_x + Q/2)k_y}e^{-(k_y - k^\prime_y + Q/2)^2} \\ \times \theta_3((k - k^\prime)/Q + (1 + i)/2, i),
\end{multline}
which satisfies $f(\bm{k}+Q\hat{x}; \bm{k}^\prime)=f(\bm{k}; \bm{k}^\prime)e^{ik_y Q}$ and $f(\bm{k}+Q\hat{y}; \bm{k}^\prime)=f(\bm{k}; \bm{k}^\prime)$, meaning that $H_\tau(\bm{k})$ obeys boundary conditions consistent with the LLL Bloch wavefunctions (see Eq.~\ref{eq:Bloch_kperiodicity}).  The non-interacting Dirac points that connect the lower and upper bands in valley $\tau$ are located at $\tau\bm{k}_1$ and $\tau\bm{k}_2$ and have opposite winding in the two valleys\footnote{For physically natural IVC insulators where the intervalley physics is driven by a kinetic lobe principle, $\bm{k}_1$ and $\bm{k}_2$ are not expected to be at time-reversal invariant momenta.}. Since we are interested in filling $\nu=-1$ where the upper bands will be remote in energy, and are assuming that the ETI will be constructed solely from the lower bands, we can safely project into the lower bands:
\begin{gather}
    \ket{\text{lower},+,\bm{k}}=\ket{+}\otimes\frac{1}{\sqrt{2}}\begin{pmatrix}
        1\\-e^{i\theta(\bm{k})}
    \end{pmatrix}\\
    \ket{\text{lower},-,\bm{k}}=\ket{-}\otimes\frac{1}{\sqrt{2}}\begin{pmatrix}
        1\\-e^{-i\varphi(\bm{k})}
    \end{pmatrix},
\end{gather}
where
\begin{gather}
    \theta(\bm{k})=\text{arg}[f(\bm{k}; \bm{k}_1)f(\bm{k}; \bm{k}_2)]\\
    \varphi(\bm{k})=\text{arg}[f(-\bm{k}; \bm{k}_1)f(-\bm{k}; \bm{k}_2)].
\end{gather}

The goal is to induce a gapped $\hat{\mathcal{T}}$- and $\hat{C}_{2z}$-symmetric IVC state at half-filling of this projected subspace (equivalent to $\nu=-1$ in the full four-band problem). The projected Hamiltonian can be generally parameterized as
\begin{gather}\label{eq:toy_model_Htilde}
    \tilde{H}(\bm{k})=\begin{pmatrix}
        \epsilon(\bm{k})+{\delta(\bm{k})} & \tilde{\Delta}(\bm{k}) \\
        [\tilde{\Delta}(\bm{k})]^*& \epsilon(\bm{k})-{\delta(\bm{k})}
    \end{pmatrix},
\end{gather}
where $\epsilon(\bm{k})+\tau\delta(\bm{k})$ is the dispersion of the lower band in valley $\tau$, and $\tilde{\Delta}(\bm{k})$ describes the projected intervalley coherence, to be discussed below.
In order to open a direct gap at $\nu=-1$, $\sqrt{|\tilde{\Delta}(\bm{k})|^2+\delta(\bm{k})^2}$ must be non-vanishing for all momenta $\bm{k}$. In Fig.~\ref{fig:2BLLL_fk_model}a, we show a representative plot of the valley dispersion difference $\delta(\bm{k})$. This takes values of opposite signs near the non-interacting Dirac points within each valley\footnote{Despite the fact that the upper bands have been projected out, we continue to use the term `non-interacting Dirac point' to mean the momentum points where the lower bands would have connected to the upper bands in the original Hamiltonian, and hence where $\ket{\text{lower},\tau,\bm{k}}$ has singularities.}, since they are high-energy features of the band dispersion of the lower bands. The non-interacting Dirac points in one valley are therefore separated by lines of $\delta(\bm{k})=0$ from those of the other valley. The formation of an ETI thus requires a projected IVC mean-field $\tilde{\Delta}(\bm{k})$ that does not vanish at momenta where $\delta(\bm{k})=0$.

As a first attempt to obtain an ETI, we add following $\hat{C}_{2z}\hat{\mathcal{T}}$-symmetric uniform IVC term\footnote{The name TIVC refers to a particular symmetry-breaking order within the low-energy $U(4)\times U(4)$ strong-coupling hierarchy in TBG~\cite{bultinck_ground_2020,TBG4}. 
This can lead to a fully gapped state in spinful TBG (which contains eight flat bands) at even integer fillings, which in the spinless 4-band ETI toy model translates to $\nu=0,\pm2$.
However as we argue here, uniform $\Delta_{\text{TIVC}}$ leads to a gapless state at $\nu=\pm1$ of the spinless 4-band setting.}
\begin{equation}\label{eq:TIVCmass}
    H_\text{TIVC}=\Delta_{\text{TIVC}}\tau^x\sigma^x\,,
\end{equation}
which could arise from interaction effects. Note that since $\tau^x\sigma^x$ only hybridizes Chern bands with the same Chern number, there is no topological obstruction to having a uniform $\Delta_\text{TIVC}$. 
 In Fig.~\ref{fig:2BLLL_fk_model}b, we plot the corresponding projected matrix element $|\tilde{\Delta}(\bm{k})|$, which exhibits two nodal lines that connect the non-interacting Dirac points in opposite valleys at $\bm{k}_1$ and $-\bm{k}_2$, as well as $\bm{k}_2$ and $-\bm{k}_1$. 
 The bands of $\tilde{H}(\bm{k})$ are now connected by two residual Dirac points
 (red stars in Fig.~\ref{fig:2BLLL_fk_model}b) where $\delta(\bm{k})=\tilde{\Delta}(\bm{k})=0$ with opposite winding number. In the Supplemental Material (Sec.~\ref{secsupmat:additional_ETI}), we show that this happens generally for any $\bm{k}_1,\bm{k}_2$. 

To bring the residual Dirac points together and annihilate them, we add an additional inter-Chern IVC term 
that preserves the $\hat{C}_{2z}$ and $\hat{\mathcal{T}}$ symmetries but now couples bands of {\it opposite} Chern numbers
\begin{equation}
    H_{\text{inter}} = \text{Re}[\Delta_{\text{inter}}(\bm{k})]\tau^x + \text{Im}[\Delta_{\text{inter}}(\bm{k})]\tau^y\sigma^z.\label{eq:HCTI}
\end{equation}
We choose $\Delta_{\text{inter}}(\bm{k}) = \Delta_{\text{inter}}f(\bm{k}; \bm{k}_{\text{inter}})f(\bm{k}; (\hat{x} + \hat{y})Q-\bm{k}_{\text{inter}})$ with $\bm{k}_{\text{inter}}=0$. This functional form satisfies the topological winding required for hybridization between bands of different Chern number. $H_{\text{inter}}$ is also projected to the lower bands, which yields a contribution to $\tilde{\Delta}(\bm{k})$.

As shown in Figs.~\ref{fig:2BLLL_fk_model}c,d, a sufficiently strong $\Delta_\text{inter}$ `rewires' the nodal lines of the projected IVC matrix element $|\tilde{\Delta}(\bm{k})|$, so that they now connect the non-interacting Dirac points within each valley. As a result, the residual Dirac points between the lower two IVC bands annihilate and a direct gap opens at $\nu=-1$, leading to an ETI\footnote{Additional interaction effects that do not affect the topological properties of the state may be required to open a full indirect gap.}. This illustrates that a simple uniform IVC term does not suffice to generate a gapped ETI, and a more complicated combination of IVC terms is required.

We relegate more detailed aspects of the toy model and related Hamiltonians to the Supplementary Material (Secs.~\ref{secsupmat:additional_ETI} and \ref{secsupmat:2BLLL_additional}). There, we also 
discuss energetic considerations regarding the positions $\bm{k}_1,\bm{k}_2$ of the non-interacting Dirac points. Having them close together within one valley, and far apart from those in the other valley, is conducive towards the formation of a gapped ETI. 
Finally, we also 
repeat the calculation without projecting onto the lower bands.

\section{Field theory}\label{sec:field_theory}
In the LLL model introduced and analyzed in Sec.~\ref{sec:1BLLL}, we found two unconventional exciton condensates, the TVP and CTI$_1$ states, which spontaneously break the valley $U(1)_\text{v}$ symmetry. The TVP state also breaks time-reversal symmetry and is a quantum anomalous Hall state, whereas the CTI$_1$ state is time-reversal symmetric. In this section we analyse the physical consequences of the delicate and strong topology of these two states via a low-energy effective field theory. At the end of this section we comment on the generalization to the ETI.

To identify the relevant continuum model, let us start by summarizing the main ingredients of the LLL model. We start with two Chern bands with valley quantum number $\tau^z = C$, where $C=\pm 1$ is the Chern number. The relevant symmetries, besides charge conservation and translation, are
\begin{eqnarray}
U(1)_\text{v} &: e^{i\theta \tau^z} & \text{(valley $U(1)_\text{v}$)}\\
\hat{\mathcal{T}} & : \tau^x \mathcal{K} &\text{(time reversal,  $\mathcal{K}: i\rightarrow -i$)}.
\end{eqnarray}
Both the TVP and CTI$_1$ states can be described at the mean-field level by the following two-band Hamiltonian in the valley ($\tau^z$) basis (c.f.~Eq.~\ref{eq:trial_BFEI})
\begin{equation}
H(\k) = \left(\begin{matrix}\varepsilon_+(\k) & \Delta(\k) \\ \Delta^*(\k) & \varepsilon_-(\k) \end{matrix}\right),
\end{equation}
where $\Delta(\k)$ is the valley symmetry-breaking order parameter. If time-reversal is preserved, then $\varepsilon_-(\k) = \varepsilon_+(-\k)$. As explained previously, because the valley coherence is between two bands with opposite Chern number with $|C|=1$, the order parameter must have two vortices with the same winding. In the vicinity of a vortex located at $\k^*$ we can write
\begin{equation}\label{Deltavortex}
\Delta(\k) = \Delta\left[ (k^*_x - k_x)+ i(k^*_y-k_y)\right] + \mathcal{O}((\k^*-\k)^2)\,.
\end{equation}
If time-reversal is preserved, then the two vortices are at $\pm \k^*$. Crucially, we see that near the vortex, the mean-field Hamiltonian takes the form of a Dirac Hamiltonian, and $M = (\varepsilon_+(\k^*) - \varepsilon_-(\k^*))/2$ is the Dirac mass.

We can now write down a low-energy Hamiltonian assuming that the Dirac mass is much smaller than $\Delta$. If time-reversal symmetry is preserved, corresponding to the CTI$_1$ state, it is given by\footnote{The generalization of CTI$_1$ to CTI$_n$ can be done in a straightforward way by taking $|n|$ copies of the Dirac Hamiltonian in Eq. \eqref{HDCTI}. If sgn$(n)$ is negative, one simply changes the chirality of all the Dirac fermions (e.g. by taking $k_y \rightarrow - k_y$).}
\begin{equation}\label{HDCTI}
H_D(\k) = (k_x\tau^x + k_y\tau^y + M\tau^z)\mu^z\,,
\end{equation} 
where we have adopted units such that $\Delta = 1$, and $\mu^z$ labels the two mini-valleys, i.e.~the two different gapped Dirac fermions. Note that time-reversal acts as $\hat{\mathcal{T}} = \mu^x\tau^x \mathcal{K}$, and translation becomes a mini-valley $U(1)$ symmetry. The time-reversal breaking case, corresponding to the TVP state, can be described with
\begin{equation}
H_D(\k) = (k_x\tau^x + k_y\tau^y)\mu^z + M \tau^z\,,
\end{equation}
i.e.~now the Dirac mass has the same sign in the two mini-valleys. Below we will use these two different Dirac Hamiltonians as the starting point for our analysis of the topological properties of the TVP and CTI$_1$ states\footnote{See Supplementary Material, Sec.~\ref{secsupmat:edgephysics} for a discussion of the edge modes of these states based on the Dirac Hamiltonians.}.

\subsection{IVC vortex-core states}
Let us first introduce a \emph{real-space} vortex in the IVC order parameter. The corresponding low-energy Hamiltonian in a single mini-valley is
\begin{equation}\label{vortexH}
H = \left[R(\theta_\r)\right]_{mn} \tau^m ( i\partial_n+ \frac{1}{2}\partial_n\theta_\r \tau^z ) + M\tau^z\,,
\end{equation}
where
\begin{equation}
R(\theta_\r) = \left(\begin{matrix} \cos \theta_\r & \sin \theta_\r \\ -\sin \theta_\r & \cos\theta_\r \end{matrix}\right)
\end{equation}
is the SO(2) matrix which rotates over an angle $\theta_\r$. For a vortex configuration, we take $\theta_\r$ to be the polar angle, i.e.~$\theta_\r = \tan^{-1}(y/x)$. The Hamiltonian in Eq. \eqref{vortexH} is the same as the Read-Green Hamiltonian which describes a vortex in the order parameter of a $p+ip$ superconductor~\cite{ReadGreen}, with the important difference that in our case $H$ is defined in valley space, and not in Nambu space. Despite this difference, we can immediately conclude that there is a midgap state which is exponentially localized at the vortex core. In particular, it holds that $H|\psi\rangle = 0$ for
\begin{equation}
|\psi\rangle = \left(\begin{matrix} 1 \\ -i \end{matrix}\right)\times \frac{e^{-Mr/2}}{\sqrt{r}}\,,
\end{equation}
where $r$ is the radial distance to the vortex core at the origin, and we have assumed that $M>0$. Note that generically $|\psi\rangle$ is a midgap state and not a zero-mode. For a superconductor it would be a true zero mode because of the particle-hole symmetry. In our case, however, the Hamiltonian can contain a term $\mathbb{1}(\varepsilon_+(\k^*) +\varepsilon_-(\k^*))/2$ proportional to the identity, which we have ignored in our low-energy Dirac Hamiltonian. This term shifts the entire single-particle spectrum and hence shifts $|\psi\rangle$ away from zero energy. Despite not being a zero-mode, $|\psi\rangle$ nevertheless remains a midgap state (in the two-band model).

\subsubsection{CTI$_1$ case}\label{vortexCTI}
In the time-reversal symmetric (CTI$_1$) case the vortex core will bind the following two midgap states, coming from the two different mini-valleys:
\begin{eqnarray}
|\psi_{\k^*}\rangle & = &  e^{i\k^*\cdot \r}\left(\begin{matrix} 1 \\ -i \end{matrix}\right)\times \frac{e^{-Mr/2}}{\sqrt{r}} \\
|\psi_{-\k^*}\rangle & = &  e^{-i\k^*\cdot \r}\left(\begin{matrix} 1 \\ -i \end{matrix}\right)\times \frac{e^{-Mr/2}}{\sqrt{r}}.
\end{eqnarray}
These two states are degenerate as they are interchanged by time reversal.  We can energetically split the midgap states in a time-reversal symmetric way by raising/lowering the energy of the states
\begin{eqnarray}
|\psi_{+}\rangle & = &  \cos(\k^*\cdot \r + \alpha)\left(\begin{matrix} 1 \\ -i \end{matrix}\right)\times \frac{e^{-Mr/2}}{\sqrt{r}} \\
|\psi_{-}\rangle & = &  \sin(\k^*\cdot \r + \alpha)\left(\begin{matrix} 1 \\ -i \end{matrix}\right)\times \frac{e^{-Mr/2}}{\sqrt{r}}.
\end{eqnarray}
These states will induce a halo of charge density oscillations with wavevector $2\k^*$. So we can lower the energy of the vortices of the CTI$_1$ state by coupling to a charge density wave (CDW) order parameter.

\subsubsection{TVP case}
In the time-reversal broken (TVP) case the two vortex-core states coming from the two mini-valleys are
\begin{eqnarray}
|\psi_{\k^*}\rangle & = &  e^{i\k^*\cdot \r}\left(\begin{matrix} 1 \\ -i \end{matrix}\right)\times \frac{e^{-Mr/2}}{\sqrt{r}} \nonumber\\
|\psi_{-\k^*}\rangle & = &  e^{-i\k^*\cdot \r}\left(\begin{matrix} 1 \\ i \end{matrix}\right)\times \frac{e^{-Mr/2}}{\sqrt{r}}.\label{VSTB}
\end{eqnarray}
In this case taking linear combinations does not induce charge oscillations.

\subsection{Response action}
Let us now couple the low-energy theory to background gauge fields in order to identify potential quantized response coefficients. In the presence of background gauge fields, the Lagrangian (without mass term) is
\begin{equation}\label{Lagr}
\mathcal{L} = \psi^\dagger iD_t\psi - \psi^\dagger \mu^z e^n_m \tau^m iD_n \psi\,,
\end{equation}
where $\psi$ contains four complex fermions $\psi_{\tau\mu}$ labeled by valley and mini-valley, and
\begin{equation}
D_\mu = \partial_\mu + iA_\mu+ i\omega_\mu \tau^z + iB_\mu \mu^z\,.
\end{equation}
This Lagrangian is invariant under
\begin{eqnarray}
\psi\rightarrow e^{-i\theta}\psi\,,  & A_\mu \rightarrow A_\mu + \partial_\mu \theta & \\
\psi\rightarrow e^{-i\alpha \mu^z}\psi\, , & B_\mu \rightarrow B_\mu + \partial_\mu \alpha & \\
\psi\rightarrow e^{-i\varphi \tau^z}\psi\, , & \omega_\mu \rightarrow \omega_\mu + \partial_\mu \varphi\,, \nonumber \\
 & e_m^n \rightarrow e^n_l [R(2\varphi)]_{lm}.
\end{eqnarray}
The gauge fields for charge, translation and valley are thus respectively $A,B$ and $\omega$. Note that $\omega$ is an SO(2)-restricted version of the spin connection, and $e^n_m$ can be thought of as a restricted vielbein. Hermiticity of the Hamiltonian requires that
\begin{equation}\label{hermconstr}
\partial_n e^n_{m} = 2e^n_l \epsilon^{lm} \omega_n\,.
\end{equation}
This equation can be solved by taking
\begin{equation}
\omega_n = \frac{1}{4}\epsilon^{lm}e_{lp} \partial_n e^p_{m}\,.
\end{equation}
Under time reversal, the gauge fields transform as
\begin{eqnarray}
(A_0,A_x,A_y) & \rightarrow & (A_0,-A_x,-A_y)\\
(B_0,B_x,B_y) & \rightarrow & (-B_0,B_x,B_y) \\
(\omega_0,\omega_x,\omega_y) & \rightarrow & (-\omega_0,\omega_x,\omega_y)\,.
\end{eqnarray}

\subsubsection{CTI$_1$ case}
Let us now add the time reversal symmetric mass term $ M\psi^\dagger\tau^z\mu^z \psi$ to the Lagrangian in Eq. \eqref{Lagr} and integrate out the fermions. The resulting response action contains following mixed Chern-Simons term:
\begin{equation}\label{BdA}
\text{sgn}(M)\times\frac{1}{2\pi} A\mathrm{d}B.
\end{equation}
Note that this term is time-reversal symmetric because the $A$ and $B$ gauge fields transform oppositely under time reversal. The response term in Eq. \eqref{BdA} has previously also been found in Ref.~\cite{Ryu2009} for mono-layer graphene gapped with a staggered sublattice potential. To establish the connection to this work, note that in our notation the low-energy Dirac Hamiltonian of graphene is written as
\begin{equation}
H_{\text{MLG}} = k_x \tau^x \mu^z + k_y \tau^y + M \tau^z\,,
\end{equation}
where $M\tau^z$ is the mass term generated by the sublattice potential. Time-reversal symmetry acts as $\mathcal{T}_{\text{MLG}} = \mu^x \mathcal{K}$. If we perform a basis transformation with $\tau^x$ in (mini-)valley $\mu^z = -1$, we obtain
\begin{equation}
H'_{\text{MLG}} = (k_x \tau^x + k_y\tau^y + M\tau^z)\mu^z\,,
\end{equation}
and $\mathcal{T}'_{\text{MLG}} = \tau^x \mu^x\mathcal{K}$. The basis transformation thus maps both the low-energy Dirac Hamiltonian and the time-reversal operator of graphene to those of the CTI$_1$. And as graphene gapped with a sublattice potential goes to a trivial atomic insulator for $M\rightarrow \infty$, we thus conclude that the response term in Eq. \eqref{BdA} does \emph{not} imply that the CTI$_1$ has non-trivial stable topology, which agrees with our previous conclusion that the CTI$_1$ topology is of the delicate type. Nevertheless, the mixed Chern-Simons term in Eq. \eqref{BdA} does have physical consequences. For example, if we induce $2\k^*$ CDW order by adding $(\mu^x,\mu^y)$ mass terms, then a vortex of the CDW phason trapped by a $\pi$ flux of $B$ carries an electric charge $Q=\pm 1/2$ (for mono-layer graphene the same happens for a vortex in the valence-bond order parameter~\cite{Hou2007,Chamon2008,Chamon2008_2,Ryu2009}). 

Despite the similarity between the low-energy theories of trivially gapped graphene and the CTI$_1$, there are also important differences between the two systems: (1) the Dirac dispersion in the CTI$_1$ case is generated by a U$(1)$ order parameter, and (2) in contrast to graphene, $M\rightarrow \infty$ does not correspond to a trivial atomic limit for the CTI$_1$. This is because the CTI$_1$ mass term changes sign in the Brillouin zone (as it has a different sign in the two mini-valleys). As a result, for $M\gg 1$, the CTI$_1$ state can be thought of as a metallic state with a Fermi surface located at $M(\k)=0$, i.e. where $\varepsilon_+(\k) = \varepsilon_-(\k)$, weakly gapped by the IVC order parameter. 

\subsubsection{TVP case}
In the time-reversal broken case we add the mass term $M \psi^\dagger \tau^z \psi$ to the Lagrangian in Eq. \eqref{Lagr}. Integrating out the fermions now produces following terms with quantized coefficients in the response action:
\begin{equation}\label{responseTVP}
\text{sgn}(M)\times\left[\frac{1}{4\pi} A\mathrm{d}A - \frac{1}{4\pi}\omega \mathrm{d}\omega\right].
\end{equation}
The first term tells us that the system is a quantum anomalous Hall state. The second term is a descendent of the gravitational Chern-Simons term. In the SM (Sec.~\ref{derivationCS}) we provide an explicit perturbative derivation of the gravitational Chern-Simons term in order to ensure that Eq. \eqref{responseTVP} contains the correct level with our normalization convention for the fields.

The Chern-Simons term for $\omega$ implies that the TVP state realizes the second entry in Kitaev's 16-fold way~\cite{Kitaev2006}. To see this, imagine that the valley U(1) symmetry is gauged. The IVC order is a valley charge-2 condensate, and induces a Higgs phase with a surviving deconfined $\mathbb{Z}_2$ gauge field. The valley $\pi$-fluxes, which are screened by a $2\pi$ vortex of the IVC order, remain well-defined anyonic quasi-particles. We will call these anyons $a$. The valley Chern-Simons term implies that the topological spin of $a$ is~\cite{deWild1995}
\begin{equation}\label{topospin}
\theta_{a} = e^{-2\pi i/8}\,,
\end{equation}
which is equal to the topological spin of the $\pi$-fluxes in the second entry of the 16-fold way. We do not find the odd entries in the 16-fold way, because these require breaking the charge conservation symmetry (as they host Majorana zero modes). So for IVC states, there is instead an 8-fold way, and the TVP is the first entry of this 8-fold way.

Because the U$(1)$ charge of every local operator is equal to its U$(1)_v$ charge modulo 2, a U$(1)$ and U$(1)_v$ $\pi$ rotation are equivalent, and hence the continuous part of the symmetry group before exciton condensation is (U$(1)\times$U$(1)_v$)$/\mathbb{Z}_2$. If U$(1)_v$ is gauged, the remaining global symmetry group is U$(1)$/$\mathbb{Z}_2$. The modding out by $\mathbb{Z}_2$ reflects the fact that charges of local operators in the gauge theory are quantized in multiples of $2$ (e.g. $c^\dagger_{K}c^\dagger_{K'}$ is uncharged under the gauge group and hence a local operator). Let us now imagine adiabatically threading a thin solenoid of $\pi$ flux through the system. After gauging U$(1)_v$, a U$(1)$ $\pi$ flux is invisible to all local operators, and hence is `pure gauge', i.e.~it can be removed by a large gauge transformation. So the combined operation of adiabatically threading a U$(1)$ $\pi$ flux followed by a large gauge transformation maps between eigenstates of the U$(1)_v$ gauge theory. But due to the $A\mathrm{d}A$ Chern-Simons term, a charge $1/2$ is nucleated during the flux threading. We thus arrive at the conclusion that the flux insertion creates an anyon with electric charge $1/2$. Due to the $\mathbb{Z}_4$ fusion rules ($a^4 = 1$) ~\cite{Kitaev2006}, and the requirement that all local operators have even integer charge, the only anyon which can have charge $1/2$ is $a$.

\subsection{Connection to the ETI}
We finally comment on the connection between the results obtained in this section and the ETI state introduced in Sec.~\ref{sec:2BLLL}. Our low-energy theory for the TVP and CTI$_1$ is derived from a two-band model. The ETI, however, results from symmetry breaking in a set of bands with an Euler obstruction, and hence requires at least four bands. This means that our low-energy theory cannot be applied to the ETI in a straightforward way. Nevertheless, the ETI and the CTI$_{\pm 1}$ are closely related. To see this, consider for concreteness the ETI toy model in Sec.~\ref{sec:C2z_effective}, where the ETI is constructed by hybridizing the lower band of the $\hat{C}_{2z}\hat{\mathcal{T}}$-symmetric non-interacting model $H_\tau(\bm{k})$ in each valley (Eqs.~\ref{eq:fk_H+} and \ref{eq:fk_H-}). Due to the Euler topology, the lower band in each valley is connected to the upper band by two non-interacting Dirac points. These can be gapped by applying a small $\hat{C}_{2z}$-odd sublattice mass $\sigma^z$, generating Chern bands. If the sublattice mass takes the same positive value in both valleys, then the lower bands of $H_\tau(\bm{k})$ have Chern number $C=\tau$, and hybridizing them with IVC would lead to a CTI$_1$. On the other hand, a negative coefficient for $\sigma^z$ would generate lower bands with $C=-\tau$, giving rise to a CTI$_{-1}$ upon inducing IVC. 
The ETI can hence be interpreted as the $\hat{C}_{2z}$-symmetric boundary between the CTI$_1$ and CTI$_{-1}$. Given that the coefficient of the $B\mathrm{d}A$ term has a different sign for the CTI$_{-1}$ than for the CTI$_1$, we thus conclude that the $B\mathrm{d}A$ term is odd under $\hat{C}_{2z}$, and hence vanishes for the ETI. Similarly, the CTI$_{\pm 1}$ host different vortex core states $(1,\mp i)^T$. We therefore expect that neither of these states will appear as midgap states for the ETI.

\section{ETIs in Experiments: IKS Order}\label{sec:IKSasETI}

\begin{figure*}
    \centering
    \includegraphics[width = \linewidth]{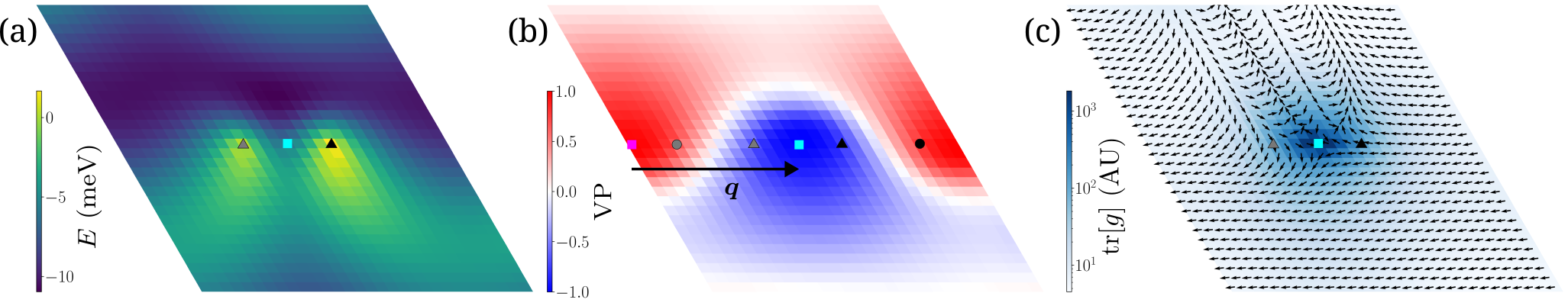}
    \caption{\textbf{Twisted bilayer graphene (TBG) at $\theta = 1.1^\circ$.} Uniaxial heterostrain of strength $\epsilon=0.25\%$ is applied along $\hat{x}$. HF is performed at $\nu=-2$ on a system of size $30\times 30$, with a fixed boost $\bm{q}$ that connects $\Gamma_\text{M}$ to $M_\text{M}$ along the $\hat{x}$ direction, as plotted in b).  The ground state is a spin-unpolarized ETI, referred to as an incommensurate Kekul\'e spiral (IKS) in Ref.~\cite{kwan_kekule_2021}. We project to the central bands, and use the `average' interaction scheme. All plots are shown for one spin sector.  a) The non-interacting dispersion of the valence band in valley $K$. The cyan square labels $\Gamma_M$ and the triangles label the Dirac points. Heterostrain unpins the Dirac points and moves them towards $\Gamma_M$. The dipersion in valley $K^\prime$ is a time-reversed copy of that in valley $K$. 
    b) Momentum-resolved valley polarization of the HF ground state. The magenta square indicates $\Gamma_M$ in valley $K^\prime$ and the non-interacting Dirac points in valley $K'$ are indicated with dots. At some momentum near $\Gamma_M$ of each valley, there is perfect polarization into the opposite valley. c) The color plot shows the trace of the quantum metric of the valley-filtered basis in valley $K$, $\ket{\text{HF}_+}$. We observe diverging quantum metric near $\Gamma_M$, which indicates a singularity. We further expand the valley-filtered basis into the Chern basis $\ket{\text{HF}_+} = \frac{1}{\sqrt{2}}(\ket{C = +1} + e^{-i\theta}\ket{C = - 1})$, and show the angle $\theta$ with arrows. We observe a winding of $4\pi$ at the singularity. In the ETI, the singularity is smoothed out by complete polarization into the opposite valley. 
    }
    \label{fig:TBG_IKS}
\end{figure*}

\label{subsec:TBG}
As promised, after having elucidated the topological and energetic aspects of CTIs and ETIs in model systems, we now turn to exploring their emergence in more realistic systems. In the interests of sharply delineating  \textit{predictions} from the interpretation of existing experiments, we focus here on the latter. We examine two cases where the spiral IVC order anticipated in a textured exciton insulator has already been directly observed in experiment: namely, the IKS states observed via STM in twisted bilayer graphene and mirror-symmetric twisted trilayer graphene (TSTG). In both cases, IKS states emerge in $\hat{C}_{2z}$-symmetric settings so that we identify them as ETIs. In a companion work~\cite{companion}, we use numerical HF studies to propose that CTIs can emerge  in experimentally-accessible parameter regimes in several other $\hat{C}_{2z}$-breaking moiré materials.

\subsection{Twisted bilayer graphene}

\subsubsection{Background}

Magic-angle TBG  is a $\hat{C}_{2z}$-preserving system that is an archetypal example where itineracy, interactions, and topology  are all important in understanding the phase diagram. Near the magic angle $\theta\simeq 1.05^\circ$, each spin and valley possesses two narrow bands at charge neutrality that are connected by Dirac points, and energetically isolated from the remote bands~\cite{Lopez2007,Bistritzer2011}. While  in-plane rotation $\hat{C}_{2z}$ and spinless time-reversal symmetry (TRS) $\hat{\mathcal{T}}$ combine to prevent a non-zero Berry curvature in the single-particle basis, the central bands in each flavor are anomalous and carry a non-zero Euler index.

 The phase diagram of the correlated (often, insulating) states that emerge near integer filling of the central bands at low temperatures  represents one of most basic characterizations of the interacting physics of TBG. It also places important constraints on the theoretical model and influences the physics of proximate metallic  phases.  In the  ``strong-coupling'' framework~\cite{KangVafekPRL,bultinck_ground_2020,TBG4}, correlated insulators can be intuitively constructed by polarizing into Chern bands in a similar vein to quantum Hall ferromagnetism\footnote{This picture works best for small chiral ratios $\kappa$. For larger chiral ratios near the realistic value, the topological heavy fermion framework~\cite{Song2022heavy,Calugaru2023heavy2,Yu2023heavyTTG,Herzog2024heavystrain} is more suitable.}~\cite{Sondhi1993}. While this provides some intuition and has various elegant analytically-tractable limits, it is unable to explain some key experimental facts. Most notable is the empirical finding that despite the reproducible presence of correlated insulators at $\nu=\pm2$, the state at charge neutrality $\nu=0$ is gapless in most devices~\cite{Cao2018,Cao2018b,Cao_2021,Yankowitz_2019,liu2021tuning,Zondiner_2020,uri2020mapping,saito2020independent,Das_2021,saito2021isospin,rozen2021entropic}, precisely where strong-coupling theory would predict the most robust insulators.

 A resolution to this discrepancy was provided in the STM experiment of Ref.~\cite{nuckolls2023quantum}, which obtained high-resolution images of the microscopic graphene-scale ordering over moir\'e length scales. It concluded that in typical TBG devices that had non-negligible amounts of strain, the charge neutrality point is gapless, as pointed out theoretically in Ref.~\cite{parker2020straininduced,Liu2021nematic}, while the normal state across a range of non-zero fillings, including the gapped insulator at $\nu=\pm2$, is consistent with the incommensurate Kekul\'e spiral (IKS) order that was first predicted earlier in Ref.~\cite{kwan_kekule_2021} (see also subsequent theory in Refs.~\cite{wagner_global_2021,wang2022kekule}). The characteristic feature of the IKS, which preserves $\hat{C}_{2z}$ and $\hat{\mathcal{T}}$, is intervalley coherence (IVC) at a finite incommensurate wavevector $\bm{q}$, i.e.~the system hybridizes the two graphene valleys $K$ and $K'$ at finite momentum and forms an intervalley spiral. The IVC generates a symmetry-breaking Kekul\'e pattern on the graphene scale~\cite{Berkeley_STM_Theory,Princeton_STM_Theory}, which modulates slowly on the moir\'e scale according to $\bm{q}$. 
 As noted by Ref.~\cite{kwan_kekule_2021}, the competition between exchange physics and kinetic energy that drives the formation of an intervalley spiral can be understood in terms of the heuristic `lobe principle', which we briefly recapitulated in Sec.~\ref{subsec:intro_picture}. This also provides a quantitative prediction of $\bm{q}$, which can be compared with STM experiments that extract this wavevector by carefully tracking the spiral modulation of the microscopic Kekul\'e pattern across the moir\'e superlattice~\cite{nuckolls2023quantum}. However, the manner in which topology both underpins  the frustration and characterizes the resulting IKS state was previously unknown, and the precise role played by strain has been unclear. As we will argue shortly, these are clarified by the recognition that the IKS state is an ETI. 

Before proceeding, we comment on the real-space texturing of the IVC order parameters \emph{within} each moir\'e unit cell. By comparing to theoretical calculations, Ref.~\cite{nuckolls2023quantum} used characteristic features of the intra-moir\'e-cell IVC modulations seen in STM, in particular the patterns of real-space vortices, as further evidence for IKS order. We caution though that these real-space nodes cannot directly be used to infer CTI or ETI character, because even `trivial IVC' insulators\footnote{By `trivial', we mean that the IVC is not topologically obstructed and is allowed to be non-vanishing across the BZ.} can host such nodes. We show an explicit example for a different moir\'e material in Ref.~\cite{companion}, but provide a simple argument here motivating this possibility. Consider a textured exciton insulator that contains IVC vortices in real-space. As explained in the introduction, the \emph{momentum-space} IVC nodes can in principle be unfrustrated by a small amount of hybridization with remote bands, reflecting the delicate topology of the textured exciton insulators. Such small mixing though does not remove the \emph{real-space} IVC vortices, which are locally stable. Nevertheless, we emphasize that detailed mapping of the intra-moir\'e cell texturing remains invaluable in pinning down the nature of the correlated states~\cite{nuckolls2023quantum}.

\subsubsection{IKS as ETI}
At the non-interacting level, heterostrain unpins the Dirac points from the $K_\text{M},K'_\text{M}$ corners, which migrate towards the mBZ center~\cite{Bi2019} (see the non-interacting band structure in Fig.~\ref{fig:TBG_IKS}a). This tendency of the Dirac points to move towards $\Gamma_\text{M}$ is enhanced by the intrinsic nematic instability, whose origin can be traced to the inhomogeneous Berry curvature of the Chern bands~\cite{Liu2021nematic}. Since strain also broadens the bandwidth, this raises the possibility of IKS states with a finite $\bm{q}$ whose value is determined by the kinetic energy considerations of the lobe principle. Fig.~\ref{fig:TBG_IKS}b,c show an example IKS state for $\nu=-2$, and analogous results can be obtained for $|\nu|=2,3$. HF calculations generally find that the gapped IKS for $|\nu|=2,3$ preserves $\hat{C}_{2z}$. Furthermore, within our spin-collinear calculations, these states are spin-unpolarized at $|\nu|=2$ or spin-polarized at $|\nu|=3$. As argued in Sec.~\ref{sec:2BLLL_ETI}, this necessitates intervalley frustration and the formation of an ETI owing to the $\hat{C}_{2z}\hat{\mathcal{T}}$-protected topology. One signature of this is the perfect valley polarization (which implies vanishing IVC) at certain points in the mBZ as shown in Fig.~\ref{fig:TBG_IKS}b. This perfect valley polarization is necessary to smooth out the singularity in the valley-filtered basis due the non-trivial Euler topology, which manifests in a diverging quantum metric. There is also a non-zero winding in the Chern basis at a singularity close to $\Gamma_M$, as shown in Fig.~\ref{fig:TBG_IKS}c. We notice that the singularity has $4\pi$ winding, which may be contrasted with the two Dirac points with $2\pi$ windings each in the non-interacting dispersion. This results from the hybridization between single particle valence and conduction bands due to energetic reasons.

Beyond clarifying the nature of the non-trivial topology in the IKS, our analysis also emphasizes the importance of the close proximity of the (interaction-renormalized) Dirac points within each valley. Otherwise (as we illustrate using a two-band variant of the LLL model in SM Sec.~\ref{secsupmat:2BLLL_additional}), the IVC state may end up gapless, or spontaneously break $\hat{C}_{2z}$ to neutralize the intervalley frustration. The results of Sec.~\ref{sec:2BLLL} also shed light on the absence of a gapped IKS in TBG for $|\nu|=0,1$. At $\nu=0$, both spin sectors are at charge neutrality with $E_F$ around the Dirac points. Hence, there is minimal kinetic penalty for occupying the valence band of each valley, which precludes kinetically-driven (spiral) IVC which is essential to the formation of any textured exciton insulator. At $|\nu|=1$, moderately strong interactions may induce a spin polarization such that one spin sector is now at odd filling. In this sector, the competition between valley exchange and kinetic energy can then lead to a frustrated intervalley condensate, with a boost $\bm{q}$ chosen according to the lobe principle. However, the other spin sector is still at its charge neutrality point, and therefore remains gapless. This explains the absence of a charge gap in the IKS at $|\nu|=1$~\cite{kwan_kekule_2021}.

We also comment on the $\bm{q}$-independent distinction between the IKS and certain strong-coupling states that also satisfy $\hat{C}_{2z}$ and spinless TRS $\hat{\mathcal{T}}$, focusing on $\nu=-2$ for concreteness\footnote{All strong-coupling states at $|\nu|=3$  are $\hat{\mathcal{T}}$-breaking with $|C|=1$.}. A spin-polarized version of the so-called TIVC insulator can be constructed by occupying $\ket{\psi_+}=\frac{1}{\sqrt{2}}(\ket{KA\uparrow}+\ket{K'B\uparrow})$ and $\ket{\psi_-}=\frac{1}{\sqrt{2}}(\ket{KB\uparrow}+\ket{K'A\uparrow})$ in the Chern basis\footnote{The Chern basis $\ket{\tau\sigma s}$ with Chern number $C=\sigma\tau$ can be obtained by diagonalizing the microscopic sublattice operator $\sigma^z$ within the central bands.}, which exhibits non-vanishing IVC across the mBZ. This state is able to evade IVC frustration because the spin polarization means that there is never a single isolated band below or above the gap within each spin sector, invalidating the argument for the ETI outlined in Sec.~\ref{sec:2BLLL_ETI}. $\ket{\psi_+}$ and $\ket{\psi_-}$ involve hybridization between bands of the same Chern number, which is not topologically obstructed, and are mapped into each other under $\hat{\mathcal{T}}$. A quantum spin Hall variant of the TIVC that still preserves $U(1)_\text{s}$ consists of  $\ket{\psi_+}=\frac{1}{\sqrt{2}}(\ket{KA\uparrow}+\ket{K'B\uparrow})$ and $\ket{\psi_-}=\frac{1}{\sqrt{2}}(\ket{KB\downarrow}+\ket{K'A\downarrow})$, but this state only satisfies a \emph{spinful} TRS, and not spinless $\hat{\mathcal{T}}$-symmetry. This therefore invalidates the Euler obstruction which requires $\hat{C}_{2z}\hat{\mathcal{T}}$ symmetry. The TIVC can be stabilized in low-strain HF calculations including electron-phonon coupling~\cite{blason2022local,kwan2023electronphonon,shi2024moire} and may explain the observed IVC in the ultra-low strain sample of Ref.~\cite{nuckolls2023quantum}.

We discuss the role of $\hat{C}_{3z}$-breaking in enabling a gapped ETI in TBG. If $\hat{C}_{3z}$ is not explicitly or spontaneously broken, the spiral wavevector $\bm{q}$ can only take values $\Gamma_\text{M},K_\text{M}$ or $K'_\text{M}$ for which the Kekul\'e pattern is moir\'e-commensurate . Any of these choices leads to at least one momentum in the mBZ which contains one Dirac point from both valleys. At the Dirac point, the states with conjugate $\hat{C}_{3z}$ eigenvalues are degenerate since they are mapped into each other under $\hat{C}_{2z}\hat{\mathcal{T}}$. 
Hence, it is not possible, even in the presence of $\hat{C}_{2z}\hat{\mathcal{T}}$-preserving IVC, to energetically isolate a single band above or below $E_F$ in each spin sector without breaking $\hat{C}_{3z}$. Our analysis thus highlights why strain plays such a crucial role in stabilizing IKS order in TBG/TSTG, by providing the requisite $\hat{C}_{3z}$-breaking, an aspect that has until now been somewhat mysterious.

Finally, the $\hat{C}_{2z}$-symmetry can be explicitly broken by aligning to the hBN substrate, whose effect can be modelled with a sublattice mass $\Delta_\text{hBN} \sigma^z$~\cite{Serlin900,Sharpe_2019,Zhang2019anomalous,Kwan2021domain}. This imbues the non-interacting bands with non-zero Chern numbers, and has been invoked to explain the QAH effect observed at $\nu=+3$~\cite{Bultinck2019mechanism,Zhang2019anomalous,Cea2020hBN}. While the IKS in Ref.~\cite{kwan_kekule_2021} persists for a finite range of $\Delta_\text{hBN}$, the ETI designation is no longer valid due to the lack of $\hat{C}_{2z}$. By studying the valley-filtered basis, we find in our HF calculations that the IKS becomes a CTI$_{\pm1}$ in the presence of a small $\Delta_\text{hBN}$. 

\subsection{Mirror-symmetric twisted trilayer graphene}\label{sec:IKSinTSTG}

Mirror-symmetric twisted trilayer graphene (TSTG) shares many similarities with TBG, since its Hilbert space can be decomposed into a mirror-even TBG sector and a mirror-odd monolayer graphene sector at zero interlayer potential~\cite{khalaf_magic_2019,calugaru2021TSTG1}. Soon after the experimental identification of the IKS in TBG, similar hallmark signatures were reported in mirror-symmetric twisted trilayer graphene (TSTG) around $\nu=\pm2$~\cite{kim2023imaging}. Transport measurements also yield results consistent with the IKS~\cite{zhou2024doubledome}.

In previous work, we explored the HF phase diagram of TSTG under varying heterostrain and displacement field~\cite{wang_kekule_2024}.  These studies indeed identify the usual IKS at finite strain; in the absence of a displacement field $\Delta V$, this is an ETI with similar properties to the IKS state in TBG.

 In addition, Ref.~\cite{wang_kekule_2024} also identified a distinct $\hat{C}_{2z}$-breaking Kekul\'e spiral phase at large displacement field, which survives to the limit of zero strain where the wavevector $\bm{q}$ becomes commensurate with the moir\'e lattice. Since $\hat{C}_{2z}$ is spontaneously broken, this phase cannot be an ETI. 
 As discussed in Ref.~\onlinecite{companion}, the question of whether this is a CTI is more subtle due to the presence of multiple bands near $E_F$, and the lack of clear spectral gaps that are present in the other platforms studied above. Our analysis suggests that the $\hat{C}_{2z}$-breaking Kekul\'e spiral is also not a CTI~\cite{companion}.

\section{Discussion}\label{sec:discussion}

Textured exciton insulators belong in the regime where interactions, band dispersion, and electronic topology are all significant. In contrast, the physics at weak-coupling is centered around the Fermi surface, and nesting, which is not generic, is necessary to open a full insulating gap. In the case of intervalley order, the spiral wavevector $\bm{q}$ is a purely kinetic property set by the details of the Fermi surfaces that are local in the BZ. On the other hand, when interactions dominate, a global view of the BZ is important. For trivial single-particle bands, conventional intervalley spiral order can arise from a localized Mott-like picture. For topological bands, ferromagnetic and topological insulators are expected, akin to quantum Hall ferromagnetism.  
When interactions and kinetic energy are comparable, both global properties of the single-particle wavefunctions across the whole BZ, such as the band topology, and local properties, such as the details of the kinetic dispersion, come together to stabilize and shape the textured exciton insulators. We emphasize that the physics here cannot be fully understood through the lens of either the weak- or strong-coupling limits.

For  moir\'e systems, the intermediate-coupling regime can be accessed by appropriate choices of physical parameters such as the interaction strength, twist angle, displacement field, and strain. These conspire to stabilize the IKS state in magic-angle bi- and trilayer graphene, which as we have argued above is an example of an ETI. As we show in Ref.~\cite{companion},  several $\hat{\mathcal{C}}_{2z}$-breaking moir\'e materials favor the formation of CTIs in physically plausible parameter regimes.  The presence of band topology can often be experimentally verified through observation of correlated (fractional) Chern insulators, including with a perpendicular magnetic field. From the perspective of strongly-interacting topological physics, it is typically desirable for interactions to be dominant relative to the band dispersion. In this context, we believe that textured exciton insulators are important for several reasons. As evidenced in the HF phase diagrams at integer fillings, they are often the correlated phases that are adjacent to more conventional flavor-polarized (topological) insulators. An understanding of the stability of such strong-coupling states thus requires careful consideration of the energetic competition with textured exciton insulators. Furthermore, factors such as interaction-induced band renormalization~\cite{Guinea2018,rademaker2019smoothening,cea2019pinning,goodwin2020hartree,kang2021cascades}, and experimentally relevant variables like twist angle inhomogeneity~\cite{uri2020mapping} and strain~\cite{Kerelsky2019,Choi2019,Xie2019stm}, may complicate our assessment of the relative strength of interactions, and favor the emergence of intermediate-coupling orders. A prime example of this is the ubiquity of IKS order in TBG and TSTG devices examined by STM~\cite{nuckolls2023quantum,kim2023imaging}, which arises due to unintentional strains whose introduction is often unavoidable when preparing twisted samples.

For  the models considered here, as well as TBG, TSTG and most of the materials in Ref.~\onlinecite{companion}, although several single-particle bands may participate non-trivially in the formation of the textured exciton insulator, the intervalley frustration can be straightforwardly diagnosed since the relevant interacting bands are isolated. A natural question that arises is how the classification of CTIs and ETIs generalizes to more complicated multiband situations. For instance, the Euler index $e_2$ relevant to the ETI is a property of a $\hat{C}_{2z}\hat{\mathcal{T}}$-symmetric set of two bands, but there are other possible multi-gap topologies characterized by different symmetries or greater numbers of bands. However, involvement of other bands can complicate the analysis if there are no clear spectral gaps to higher remote bands. 

We have shown that textured exciton insulators inherit a delicate topology that obstructs an adiabatic connection to an atomically localized description. One way to understand this is that the strong or fragile topology of the constituent symmetric bands forces the valley pseudospin to form a complex texture in momentum space. The latter implies a non-vanishing quantum geometry for the textured exciton insulator, as we show in Appendix~\ref{secapp:QGbound} by deriving a lower bound for the integrated trace of the quantum metric~\cite{Herzog2022superfluidbound}. The Chern or Euler topology of the underlying bands also leads to certain conditions on the components of their quantum geometric tensor~\cite{Xie2020bounded,Kwon2024quantumEuler}. It would be interesting to combine this with the intervalley frustration to develop more refined quantum geometric bounds for the CTI or ETI, and investigate their physical consequences for quantities such as the superfluid weight for superconductivity~\cite{peotta2015superfluidity,liang2017superfluid,Xie2020bounded,Herzog2022superfluidbound,torma2022superconductivity,Yu2023EulerTBG}. 

Another important consideration is the impact of corrections beyond mean-field theory. It may be possible to establish the stability of the various textured exciton insulator phases using more sophisticated techniques. For instance, the presence of IKS order in strained TBG has been theoretically corroborated at $|\nu|=3$ in the density-matrix renormalization group study of Ref.~\cite{wang2022kekule}. Since the competing spin-valley polarized phases that appear at strong interaction strengths are likely to be close to Slater determinant states, we expect that inclusion of quantum fluctuations should relatively favor the textured exciton insulators. This is a question that should be considered both in highly-controllable settings such as the simplified models introduced in this work, but also for more realistic ones that capture the more involved features of moiré materials.

The numerical phase diagrams in this work are based on HF calculations assuming a generalized translation symmetry. The latter is generated by $\hat{T}_{\bm{R}}e^{i\bm{q}\cdot\bm{R}\tau^z/2}$, where $\bm{q}$ is the IVC spiral wavevector, and $\hat{T}_{\bm{R}}$ is a translation operator by a moir\'e lattice vector $\bm{R}$. This guarantees that valley-diagonal observables, such as the moir\'e-scale charge density, remain moir\'e-periodic. However, alternative symmetry-breaking orders such as CDWs can also exploit the non-negligible dispersion, as proposed in Ref.~\cite{pierce2021unconventional} for TBG. In the flat-band limit, Refs.~\cite{Chen2012FQHtorusFTI,Mukherjee2019FQHsphereFTI,yang2023phase} have shown that phase separation can be energetically favorable for the LLL model.
More detailed calculations are required to fully flesh out the various phase diagrams.

Consider the situation where the non-trivial interacting physics predominantly occurs within one pair of TR-related Chern bands $c^\dagger_{\bm{k},\tau}$ with $C=\tau n$. At half-filling, the CTI$_n$ is the unique gapped mean-field phase that has $\hat{\mathcal{T}}$ and generalized translation symmetries. Beyond mean-field level, it is still meaningful to detect the topological frustration of a CTI$_n$ based on correlators of $c^\dagger_{\bm{k},+}c_{\bm{k}+\bm{q},-}$. However, there is the possibility of other genuinely many-body incompressible phases that still preserve translation and time-reversal symmetries, especially for strong interactions. One exotic example for $n=1$ is a fractional topological insulator~\cite{bernevigQSHE2006,Levin_Stern,Neupert2011FTI,Stern2015review,Neupert_2015,Levin2012classification} adiabatically connected to the product of two time-reversed lattice analogs of the Moore-Read state~\cite{moore1991nonabelions}.

 The fluctuations of the textured exciton insulators, which have a gapless branch due to broken $U(1)_\text{v}$-symmetry, impact the finite-temperature responses, and may help seed other proximate phases such as superconductivity. An avenue for future work is to study whether there are universal features in the collective modes arising from the topological intervalley frustration, which may constrain their coupling to the low-energy electronic degrees of freedom. Furthermore, the range of stability of the CTIs and ETIs to finite doping should be established, as has been done theoretically for the IKS in TBG~\cite{wagner_global_2021}.

In a companion paper~\cite{companion}, we have investigated the HF phase diagrams for several $\hat{\mathcal{C}}_{2z}$-breaking moir\'e systems, and we leave a detailed discussion of the feasibility of realizing CTIs in specific settings to that work. Looking beyond moir\'e, other platforms that contain the minimal ingredients --- topological bands in the intermediate coupling regime with a $U(1)$ index that is flipped under time-reversal symmetry --- are prime candidates for hosting textured exciton insulators.

Topological phases such as (fractional) Chern insulators in 2d materials are usually experimentally identified by measuring quantized electrical Hall response in transport, or tracking how the incompressible state evolves in a magnetic field according to the Streda formula with various probes. 
Such signatures are absent in textured exciton insulators. They are hence challenging to detect directly, and require comparatively indirect evidence to verify their presence. Properties of the integer phase diagram can help narrow down the possibilities. For instance, the presence of an insulating phase and unbroken time-reversal symmetry (e.g.~absence of anomalous Hall effect) at odd integer fillings implies either density wave order or intervalley coherence if the more exotic possibility of topological order is not invoked. If there is independent evidence of well-isolated topological bands, the scenario of intervalley coherence then likely implies a CTI or ETI phase. In graphene systems, IVC can be probed directly by imaging the Kekul\'e distortion on the graphene lattice scale. The intervalley wavevector $\bm{q}$ has been extracted experimentally by analyzing the long-wavelength spiral modulation in moir\'e superlattices~\cite{nuckolls2023quantum,kim2023imaging}. In Ref.~\cite{nuckolls2023quantum}, the identification of vortices in the spatial profile of the local valley and sublattice order within the moir\'e unit cell was further used to support the presence of IKS order in TBG. An interesting question is whether textured exciton insulators more generally impose non-trivial constraints on the real-space patterns of such vortices. 
Finally, as derived in Sec.~\ref{sec:field_theory}, order parameter defects can induce CDW order, which may be detectable with local imaging techniques.

\begin{acknowledgements}
We thank Jonah Herzog-Arbeitmann, Andrei Bernevig, Jiabin Yu and Frank Schindler for insightful discussions, and are especially grateful to Andrei Bernevig for a compact but illuminating conversation about non-compactness. This work was supported by a Leverhulme Trust International Professorship (Grant Number LIP-202-014, ZW), by a University of Zurich postdoc grant (FK-23-134, GW),  by the European Research Council under the European Union Horizon 2020 Research and Innovation Programme via Grant Agreements No. 804213-TMCS (SAP) and No. 101076597-SIESS (NB), and by EPSRC Grants EP/S020527/1 and EP/X030881/1 (SHS). YHK is supported by a postdoctoral research fellowship
at the Princeton Center for Theoretical Science.

\end{acknowledgements}

\begin{appendix}
\section{CTI and topology}\label{secapp:CTI_localization}
The simplest setting of a CTI involves an energetically isolated pair of Chern bands which are interchanged by TRS and labelled with a `valley' $U(1)_v$ label $\tau=\pm$. The $U(1)_v$ symmetry prevents these bands from hybridizing with each other in the absence of interactions. Due to the non-trivial stable Chern topology, there is evidently an obstruction to $U(1)_v$-symmetric exponential Wannier localization, let alone an atomically trivial description. As detailed in Sec.~\ref{sec:1BLLL} of the main text, a CTI can form at half-filling when interactions spontaneously break the valley conservation, leading to a charge gap. Since CTIs preserve TRS, the resulting filled band has $C=0$ and hence can be exponentially localized. Despite the removal of the Wannier obstruction, one can still ask whether the CTI can be treated in a Mott or atomic limit picture, where the ground state is a Slater determinant of real-space fully localized valley moments.  In  this appendix, we use a toy model calculation to illustrate that this is not possible---the Wannier obstruction of the $U(1)_v$-symmetric bands translates to an obstruction against atomic localization for the CTI. Certain one-body correlation functions are required to have nonvanishing non-onsite elements, and the local charge density has non-zero fluctuations. We then demonstrate the obstruction against atomic localization is `delicate', in that mixing with remote bands allows for deformation to the atomic limit without closing the gap. Finally, we discuss whether the Wannier functions of the CTI can be made compact.

\subsection{Toy two-band model and local moment obstruction}\label{secapp:toy_local}
As a motivating example, consider a system with two orbitals $\sigma=A,B$ and two valleys $\tau=\pm$. Importantly, here we consider the orbitals to be trivial atomically localized orbitals. For any many-body state, we define the filling to be $\nu=-2$ when all orbitals are empty, and $\nu=+2$ when all orbitals are occupied. The single-particle $U(1)_\text{v}$-preserving Hamiltonian is
\begin{equation}\label{appeq:CTIlocal_H}
    H=\begin{pmatrix}
        \bm{d}(\bm{k})\cdot\bm{\sigma} & 0 \\
        0 & -\bm{d}(\bm{k})\cdot\bm{\sigma}
    \end{pmatrix}
\end{equation}
where TRS $\hat{\mathcal{T}}=\tau^x\mathcal{K}$ constrains $d_x(\bm{k}),d_z(\bm{k})$ to be odd functions, and $d_y(\bm{k})$ to be an even function. We assume that the $2\times 2$ Hamiltonian $\bm{d}(\bm{k})\cdot\bm{\sigma}$ is gapped, and the valence band has Chern number $C=+1$, while the conduction band has $C=-1$. We let $\alpha_{v,\sigma}(\bm{k})$ and $\alpha_{c,\sigma}(\bm{k})$ denote the negative and positive eigenvalue eigenvector of $\bm{d}(\bm{k})\cdot\bm{\sigma}$ respectively. In components, we have
\begin{align}
\alpha_v(\bm{k}) &= \left(\begin{matrix} \cos\theta_\k e^{i\varphi_1(\k)} \\ \sin \theta_\k e^{i\varphi_2(\k)} \end{matrix}\right),\nonumber\\
\alpha_c(\bm{k}) &= i\sigma^y\alpha_v^*(\bm{k})=\left(\begin{matrix} \sin\theta_\k e^{-i\varphi_2(\k)} \\ -\cos \theta_\k e^{-i\varphi_1(\k)} \end{matrix}\right).
\end{align}
Crucially, the non-zero Chern number implies that $\bm{d}(\bm{k})/|\bm{d}(\bm{k})|$ covers the unit sphere once, such that $\cos\theta_{\bm{k}}$ and $\sin\theta_{\bm{k}}$ each vanish at least somewhere in the BZ. For simplicity, we assume that this happens once in the BZ for each of $\cos\theta_{\bm{k}}$ and $\sin\theta_{\bm{k}}$. We consider a smooth but non-periodic gauge, such that $\varphi_1(\bm{k})$ winds by $2\pi$ around where $\cos\theta_{\bm{k}}$, and $\varphi_2(\bm{k})$ winds by $2\pi$ around where $\sin\theta_{\bm{k}}$ vanishes.

Let 
\begin{equation}
    \ket{u^{\tau}_n(\bm{k})}=\sum_{\sigma}u^\tau_{n,\sigma}(\bm{k})\ket{\bm{k},\tau,\sigma}
\end{equation}
denote the eigenstates of $H$, where $n=c,v$ indexes the conduction or valence band. We have the inverse relation
\begin{equation}
   \ket{\bm{k},\tau,\sigma}=\sum_{n}[u^\tau_{n,\sigma}(\bm{k})]^*\ket{u^{\tau}_n(\bm{k})}.    
\end{equation}
Note that the Fourier transform
\begin{align}\label{appeq:rFT}
\ket{\bm{r},\tau,\sigma}&=\frac{1}{\sqrt{N}}\sum_{\bm{k}}e^{-i\bm{k}\cdot\bm{r}}\ket{\bm{k},\tau,\sigma}\nonumber\\
    &=\frac{1}{\sqrt{N}}\sum_{\bm{k} n}e^{-i\bm{k}\cdot\bm{r}}[u^\tau_{n,\sigma}(\bm{k})]^*\ket{u^{\tau}_n(\bm{k})},
\end{align}
where $N$ is the number of unit cells, yields atomically localized basis orbitals $\ket{\bm{r},\tau,\sigma}$.

From Eq.~\ref{appeq:CTIlocal_H}, we see that $H$ is designed such that the valence band of $\tau=+$ has the same A/B-orbital Bloch state as the conduction band of $\tau=-$, such that
\begin{align}
    u^+_{v,\sigma}(\bm{k})=\alpha_{v,\sigma}(\bm{k}),&\quad u^+_{c,\sigma}(\bm{k})=\alpha_{c,\sigma}(\bm{k}),\nonumber\\ u^-_{v,\sigma}(\bm{k})=\alpha_{c,\sigma}(\bm{k}),&\quad u^-_{c,\sigma}(\bm{k})=\alpha_{v,\sigma}(\bm{k}).
\end{align}

A single Chern band within a valley does not admit an exponentially-localized Wannier representation. 
However, the two Chern bands within a valley carry opposite Chern numbers and hence can be easily recombined to yield atomically localized orbitals (see Eq.~\ref{appeq:rFT}). Ignoring any energetic considerations for the moment, consider building a new band $\ket{\psi_{+,A}(\vec{k})}$ by invoking inter-Chern coherence between the conduction and valence bands in the $\tau=+$  valley:
\begin{align}\label{appeq:psi_Adiag}
    \ket{\psi_{+,A}(\bm{k})}&=\sum_{n}[u^+_{n,A}(\bm{k})]^*\ket{u^{+}_n(\bm{k})}\nonumber\\
    &=\sum_{n\sigma}[u^+_{n,A}(\bm{k})]^*u^+_{n,\sigma}(\bm{k})\ket{\bm{k},+,\sigma}\nonumber\\
    &=\sum_{\sigma}\left([\alpha_{v,A}(\bm{k})]^*\alpha_{v,\sigma}(\bm{k})\ket{\bm{k},+,\sigma}\right.\nonumber\\&\quad\quad\quad\left.+[\alpha_{c,A}(\bm{k})]^*\alpha_{c,\sigma}(\bm{k})\ket{\bm{k},+,\sigma}\right)\nonumber\\
    &=\ket{\bm{k},+,A}.
\end{align} 
If we now build a $\nu=-1$ state by fully occupying this new band, then all one-body correlation functions
\begin{align}
    P^{\tau\sigma,\tau'\sigma'}(\bm{r},\bm{r}')&=\langle c^\dagger_{\bm{r},\tau,\sigma} c_{\bm{r}',\tau',\sigma'}\rangle\nonumber\\&=\frac{1}{N}\sum_{\bm{k}}e^{-i\bm{k}\cdot(\bm{r}-\bm{r}')}P^{\tau\sigma,\tau'\sigma'}(\bm{k})    
\end{align}
would be purely on-site $\bm{r}=\bm{r}'$ and localized on $\tau=+$ and $\sigma=A$. 

If the conduction band is much higher in energy than the valence band, then the physics can be projected onto the valence bands, The situation is then analogous to the LLL model in Sec.~\ref{sec:1BLLL}, and a CTI can be stabilized at $\nu=-1$ by breaking $U(1)_v$ due to interactions (note that the valence band projection amounts to energetically forbidding the single-valley inter-Chern state discussed above, so the only route to inter-Chern coherence is  to also require inter-valley coherence). We consider the occupied band of the following specific CTI:
\begin{align}\label{appeq:psi_CTI}
    \ket{\psi_\text{CTI}(\bm{k})}&=\sum_{\tau}[u^\tau_{v,A}(\bm{k})]^*\ket{u^{\tau}_v(\bm{k})}\nonumber\\
    &=\sum_{\tau\sigma}[u^\tau_{v,A}(\bm{k})]^*u^\tau_{v,\sigma}(\bm{k})\ket{\bm{k},\tau,\sigma}\nonumber\\
    &=\sum_{\sigma}\left([\alpha_{v,A}(\bm{k})]^*\alpha_{v,\sigma}(\bm{k})\ket{\bm{k},+,\sigma}\right.\nonumber\\&\quad\quad\quad \left.+
    [\alpha_{c,A}(\bm{k})]^*\alpha_{c,\sigma}(\bm{k})\ket{\bm{k},-,\sigma}\right).
\end{align}
The construction of $\ket{\psi_\text{CTI}}$ above is superficially very similar to that of $\ket{\psi_{+,A}}$ in Eq.~\ref{appeq:psi_Adiag}. In both cases, we are combining a $C=-1$ and a $C=+1$ band to create a $C=0$ band, such that there is no obstruction in the latter to constructing exponentially-localized Wannier functions. Indeed, even the expansion coefficients are similar, which has been deliberately done in an attempt to maximize the localization. The only difference is that $\ket{\psi_\text{CTI}}$ mixes the valleys: yet this difference is crucial to the topological structure of the CTI, as we now elucidate. 

The key point is that $\ket{\psi_\text{CTI}}$ does not admit a fully localized representation, and possesses non-onsite correlations. To see this, we first write down the projector in momentum space
\begin{equation}
    P_\text{CTI}^{\tau\sigma,\tau'\sigma'}(\bm{k})=\left(u^\tau_{v,A}(\bm{k})[u^\tau_{v,\sigma}(\bm{k})]^*\right)\left([u^{\tau'}_{v,A}(\bm{k})]^*u^{\tau'}_{v,\sigma'}(\bm{k})\right).
\end{equation}
Now consider the following summed one-body correlation function
\begin{align}\label{appeq:summed_1body_corr}
    \left\langle\left(\sum_\tau c^\dagger_{\tau\sigma}(\bm{r})\right)\left(\sum_{\tau'}c_{\tau',\sigma'}(\bm{r}')\right)\right\rangle\quad\quad\quad\nonumber\\\quad\quad\quad\quad=\frac{1}{N}\sum_{\bm{k},\tau\tau'}e^{-i\bm{k}\cdot(\bm{r}-\bm{r}')}P^{\tau\sigma,\tau'\sigma'}(\bm{k}).
\end{align}
For the CTI, the above evaluates to $\delta_{\bm{r},\bm{r}'}\delta_{\sigma,\sigma',A}$ by exploiting orthonormality. While this is purely onsite, this does \emph{not} correspond to the expected value for a band $\ket{\psi_{\tau_x,A}}$ of fully localized in-plane valley moments localized on the $A$ orbitals
\begin{equation}
    \ket{\psi_{\tau_x,A}}=\prod_{\bm{r}}\frac{1}{\sqrt{2}}\left(\ket{\bm{r},+,A}+\ket{\bm{r},-,A}\right),
\end{equation}
which instead yields $2\delta_{\bm{r},\bm{r}'}\delta_{\sigma,\sigma',A}$ for Eq.~\ref{appeq:summed_1body_corr}. In fact, any one-body correlator for $\ket{\psi_{\tau_x,A}}$ is purely onsite. 
On the other hand, observables for $\ket{\psi_\text{CTI}}$ that carry definite valley charge give non-onsite contributions. For instance, the correlator
\begin{align}
    \langle c^\dagger_{+\sigma}(\bm{r})c_{-,\sigma'}(\bm{r}')\rangle&=\frac{1}{N}\sum_{\bm{k}}e^{-i\bm{k}\cdot(\bm{r}-\bm{r}')}u^+_{v,A}(\bm{k})[u^+_{v,\sigma}(\bm{k})]^*\nonumber\\&\quad\quad\quad\quad\quad\times([u^{-}_{v,A}(\bm{k})]^*u^{-}_{v,\sigma'}(\bm{k})
\end{align}
cannot be strictly onsite, because the product of $u$'s necessarily vanishes at certain points in the BZ for Chern bands. We thus find that any operator with definite valley charge retains a memory of the non-trivial topology of the $U(1)_v$-symmetric bands, which is manifested as an obstruction to a trivial atomic limit with strictly onsite expectation values.

The non-locality of correlation functions in $\ket{\psi_{\text{CTI}}}$ also leads to onsite charge fluctuations. Consider the onsite number operator $n(\bm{r})=\sum_{\tau\sigma}c^\dagger_{\tau,\sigma}(\bm{r})c_{\tau,\sigma}(\bm{r})$ and its variance
\begin{align}
    \text{Var}[n(\bm{r})]&=\langle n(\bm{r})^2\rangle-\langle n(\bm{r})\rangle^2\nonumber\\&=\langle n(\bm{r})\rangle -\sum_{\tau\tau'\sigma\sigma'}|\langle c^\dagger_{\tau,\sigma}(\bm{r})c_{\tau',\sigma'}(\bm{r})\rangle|^2.
\end{align}
For both $\ket{\psi_{+,A}}$ and $\ket{\psi_{\tau_x,A}}$, we have $\text{Var}[n(\bm{r})]=0$ as expected. For $\ket{\psi_{\text{CTI}}}$, we have
\begin{equation}
    \text{Var}_{\text{CTI}}[n(\bm{r})]=1-\frac{1}{N^2}\sum_{\bm{k}\bm{k}'}\text{Tr}[P_\text{CTI}(\bm{k})P_\text{CTI}(\bm{k}')]
\end{equation}
where $\text{Tr}$ is a trace over all orbitals labeled by $\sigma$ and $\tau$. Since $P_\text{CTI}(\bm{k})$ necessarily varies in the BZ for a CTI, we thus find finite onsite charge fluctuations.

\subsection{Additional bands and delicate topology}\label{secapp:CTI_remote}

In this subsection, we first illustrate generally that a CTI in a model with one band per valley can be deformed infinitesimally into a trivial IVC state without closing the charge gap by introducing remote degrees of freedom. By trivial IVC, we mean that the IVC is non-vanishing throughout the BZ, and hence the valley-filtered bands have zero Chern invariant. We consider a CTI$_n$ state with $n=1$ for simplicity. We therefore conclude that the obstruction to an atomic limit for a CTI is a form of delicate topology inherited from the strong topology of the $U(1)_v$-symmetric system.

Begin with a general one-band CTI constructed from $\ket{u_\tau(\bm{k})}$ (the cell-periodic part of the Bloch function), where $\tau=\pm$ indicates the valley, which is locked to the Chern number $C=\tau$. We choose a smooth gauge for these bands, such that $\partial_{k_a}\ket{u_\tau(\bm{k})}$ is finite. This can always be done if we allow the gauge to be non-periodic. 

The general form of the filled band of the CTI is
\begin{equation}
    \ket{u_\text{CTI}(\bm{k})}=\sum_{\tau}c_{\tau}(\bm{k})\ket{u_\tau(\bm{k})}.
\end{equation}
We assume that the CTI is fully-polarized towards $\tau=-$ at $\bm{k}_0$. This means that $c_-(\bm{k})\sim 1$ near $\bm{k}_0$. At the same time, we must have a vortex in $c_+(\bm{k})$, which reflects the fact that the overall valley pseudospin has a meron centered at $\bm{k}_0$ when plotted in a smooth gauge. With appropriate choice of smooth gauge, we can choose $c_+(\bm{k})\sim (k_x-k_{0x})+i(k_y-k_{0y})$.
Hence the CTI wavefunction near $\bm{k}_0$ can be parameterized as
\begin{align}
    \ket{u_\text{CTI}(\bm{k})}&=\left((k_x-k_{0x})+i(k_y-k_{0y})\right)\ket{u_+(\bm{k})}\nonumber\\&\quad\quad\quad\quad+\sqrt{1-|\bm{k}-\bm{k}_0|^2}\ket{u_-(\bm{k})}.
\end{align}

We now introduce some remote set of states $\ket{u_{+,\text{rem}}(\bm{k})}$ in valley $+$. This could be a band defined over the BZ, but we will see shortly that this is not necessary for the present argument. We simply require $\ket{u_{+,\text{rem}}(\bm{k})}$ to be gauge-fixed such that it is smooth in the vicinity of $\bm{k}_0$. We would like to mix in an infinitesimal amplitude $\delta(\bm{k})$ of $\ket{u_{+,\text{rem}}(\bm{k})}$ near $\bm{k}_0$. Hence, we choose $\delta(\bm{k})$ to be an infinitesimal and smooth function that decays rapidly away from $\bm{k}_0$. Having $\delta(\bm{k})$ decay for large $\bm{k}-\bm{k}_0$ avoids potential issues about topological constraints due to any non-trivial topology of $\ket{u_+(\bm{k})}$ and $\ket{u_{+,\text{rem}}(\bm{k})}$. We now write down the filled band corresponding to some new IVC state
\begin{align}\label{eq:trivial}
    \ket{u_\text{trivial}(\bm{k})}&=\left((k_x-k_{0x})+i(k_y-k_{0y})\right)\ket{u_+(\bm{k})}\nonumber\\&\quad\quad+\delta(\bm{k})\ket{u_{+,\text{rem}}(\bm{k})}\\&\quad\quad+\sqrt{1-|\bm{k}-\bm{k}_0|^2-\delta(\bm{k})^2}\ket{u_-(\bm{k})}.\nonumber
\end{align}
Note that we can perform an analogous deformation using $\ket{u_{-,\text{rem}}(\bm{k})}$ at the other lobe at $-\bm{k}_0$ where the CTI is fully-polarized towards $\ket{u_+(\bm{k})}$. The short range of $\delta(\bm{k})$ prevents the deformations from interfering with each other.

$\ket{u_\text{trivial}(\bm{k})}$ is evidently an infinitesimal deformation of $\ket{u_\text{CTI}(\bm{k})}$. Furthermore, $\ket{u_\text{trivial}(\bm{k})}$ is trivial because the IVC never vanishes. If the charge gap of the CTI was finite, then the charge gap of $\ket{u_\text{trivial}(\bm{k})}$ remains finite. 

We address whether $\ket{u_\text{trivial}(\bm{k})}$ corresponds to a `physically smooth' state. The criterion we use here is that the quantum geometry of $\ket{u_\text{trivial}(\bm{k})}$ is finite and does not diverge. The quantum geometric tensor of some band $\ket{u(\bm{k})}$ is
\begin{equation}
    Q^{ab}_{\bm{k}}=\braket{\partial_{k_a}u({\bm{k}})|Q_{\bm{k}}|\partial_{k_b}u({\bm{k}})}
\end{equation}
where $Q_{\bm{k}}=1-\ket{u(\bm{k})}\bra{u(\bm{k})}$. For the quantum geometric tensor to be not divergent, a sufficient condition is that $\partial_{k_a}\ket{u({\bm{k}})}$ does not diverge. Since the Bloch states in Eq.~\ref{eq:trivial} are smooth, it is clear that $\partial_{k_a}\ket{u_\text{trivial}(\bm{k})}$ does not diverge. 

At the same time, the `valley-filtered' (VF) basis does diverge. We can write down the normalized VF basis $\ket{u_{\text{VF},+}(\bm{k})}$ for $\ket{u_\text{trivial}(\bm{k})}$ in valley $\tau=+$ near $\bm{k}_0$
\begin{align}
    \ket{u_{\text{VF},+}(\bm{k})}&=\frac{(k_x-k_{0x})+i(k_y-k_{0y})}{\sqrt{|\bm{k}-\bm{k}_0|^2+\delta(\bm{k})^2}}\ket{u_+(\bm{k})}\\& \quad\quad+\frac{\delta(\bm{k})}{\sqrt{|\bm{k}-\bm{k}_0|^2+\delta(\bm{k})^2}}\ket{u_\text{rem}(\bm{k})}.\nonumber
\end{align}
This certainly has a diverging quantum geometry, since it rapidly flips from $\ket{u_\text{rem}(\bm{k})}$ to $\ket{u_+(\bm{k})}$ as we move infinitesimally away from $\bm{k}_0$ for infinitesimal $\delta(\bm{k})$. However, the key is that the valley polarization near $\bm{k}_0$ in $\ket{u_\text{trivial}(\bm{k})}$ cancels out this divergence. 

We now show that the initial CTI can be further deformed to a trivial atomic limit using the remote states. To do so, we now specify that $\ket{u_{\tau,\text{rem}}(\bm{k})}$ comes from a valleyful and atomically-trivial symmetric set of orbitals that satisfies TRS. We can then deform $\delta(\bm{k})$ to be $1/\sqrt{2}$ across the entire BZ without encountering singularities, leading to
\begin{equation}
    \ket{u_\text{atomic}(\bm{k})}=\frac{1}{\sqrt{2}}(\ket{u_{+,\text{rem}}(\bm{k})}+\ket{u_{-,\text{rem}}(\bm{k})}),
\end{equation}
which clearly represents a limit of atomically localized moments. 

\subsection{Non-compactness of translation-invariant CTI with intra-valley $\hat{C}_{2n}$ symmetry}\label{secapp:compactness}

In App.~\ref{secapp:CTI_localization} it was shown that the CTI exhibits a form of delicate topology which obstructs an atomic limit. Here we show the that in the presence of an intra-valley $\hat{C}_{2n}$ symmetry, i.e.~a $2n$-fold rotation symmetry along an axis perpendicular to the plane, the CTI Wannier functions cannot even be made compact if strict translation symmetry is imposed --- i.e. with these symmetries the CTI is a noncompact atomic insulator~\cite{SchindlerBernevigNOnCompact2021}. We also show that if the CTI admits compact Wannier functions in the absence of an intra-valley $\hat{C}_{2n}$ symmetry, such CTIs are extremely rare.

We start from the filled CTI band, which we write as
\begin{equation}
|u_{\text{CTI}}(\k)\rangle = \sum_{\tau}c_\tau(\k)|v_\tau(\k)\rangle|\tau\rangle\,,
\end{equation}
where we have used the notation $|u_\tau(\k)\rangle = |v_\tau(\k)\rangle|\tau\rangle$ to make explicit that the two Chern bands with Chern number $C=
\tau$ live in different valleys and hence are mutually orthogonal. The corresponding CTI Wannier state centered at position $\R$ is then
\begin{equation}
|W_\R(\r)\rangle = \frac{1}{N}\sum_{\k} e^{i\k\cdot(\r-\R)}\sum_{\tau}c_\tau(\k)|v_\tau(\k)\rangle|\tau\rangle
\end{equation}
Let us now consider the part of the Wannier function which is supported in valley $\tau$
\begin{equation}\label{Wtau}
\langle\tau|W_\R(\r)\rangle = \frac{1}{N}\sum_{\k} e^{i\k\cdot(\r-\R)}c_\tau(\k)|v_\tau(\k)\rangle.
\end{equation}
Even though it is constructed from states in a single Chern band, this function is exponentially localized because $|c_\tau(\k)|$ varies in the Brillouin zone. Note that if $|c_\tau(\k)|$ were fixed to 1, then \eqref{Wtau} would be a Wannier function for a Chern band, which cannot be exponentially localized. The crucial property that enables exponential Wannier localization is that the Chern-number-enforced singularity in $|v_\tau(\k)\rangle$ is removed by taking $c_\tau(\k)$ to vanish at the location of the singularity. Let us now assume that it is possible to choose $c_\tau(\k)$ such that $\langle\tau|W_\R(\r)\rangle$ is compact. Combining these functions for different $\R$, we would then have set of compact functions that \emph{almost} spans the complete Chern band. Almost, because in $\k$-space we are missing the state at the location of the singularity. To get the missing state, we can consider a different set of functions
\begin{equation}\label{Wtauprime}
\langle\tau|W'_\R(\r)\rangle = \frac{1}{N}\sum_{\k} e^{i\k\cdot(\r-\R)}c'_\tau(\k)|v_\tau(\k)\rangle\,,
\end{equation}
where the phase of $c'_\tau(\k)$ is related to that of $c_\tau(\k)$ by a singular gauge transformation, which moves the singularity to a different location in $\k$-space, and $|c'_\tau(\k)|$ vanishes at the new location of the singularity. Eq.~\eqref{Wtauprime} provides us with a different set of exponentially localized functions, which span the Chern band except for the momentum state at the new location of the singularity. But \eqref{Wtau} and \eqref{Wtauprime} together form an overcomplete basis for the Chern band. In Sec.~III G of Ref.~\cite{DubailRead} it was proven that there exists no (overcomplete) set of compact functions that can span a topologically non-trivial band. This implies that if $\langle\tau|W_\R(\r)\rangle$ is compact, then $\langle\tau|W'_\R(\r)\rangle$ cannot be compact (but both can be exponentially localized). Put differently, if a $c_\tau(\k)$ exists such that $\langle \tau|W_\R(\r)\rangle$ is compact, then $c_\tau(\k)$ comes with a unique, special location in momentum space where the zero of $c_\tau(\k)$ has to be. Any other $c'_\tau(\k)$ with a zero that is not at this special point cannot give rise to compact Wannier functions for the CTI. This means that given a set of time-reversal related opposite Chern bands, the CTI made from those Chern bands can only admit compact Wannier functions if the nodes in the IVC order parameter are \emph{exactly} at this special point. If the Chern bands have a $\hat{C}_{n}$ symmetry with $n=2,3,\dots$, then this special point has to be a high-symmetry point. As discussed in the main text, the CTI must have its order parameter nodes, which correspond to the zeros of $c_{\pm}(\k)$, at two momenta which are interchanged by time-reversal symmetry. In the presence of both intra-valley $\hat{C}_{2n}$ symmetry and strict translation symmetry (which forbids a non-zero spiral wavevector $\q$ for the CTI), these nodes are therefore necessarily away from the high-symmetry point where they would have to be to allow for a compact Wannier representation. 

It is also interesting to note that the result of Ref.~\cite{Chen}, i.e.~that a compact tight-binding Hamiltonian cannot produce an exactly flat Chern band, follows as a corollary. To see this, consider the tight-binding Hamiltonian $H(\k)$ and assume it has an exact flat band with energy $E$ and corresponding eigenvectors $|u(\k)\rangle$, and non-flat bands $E_n(\k)$. We start by writing the flat band projector as
\begin{equation}
|u(\k)\rangle \langle u(\k)| = \oint \frac{\mathrm{d}z}{2\pi i} \frac{1}{z-H(\k)}\,,
\end{equation}
where the integral is along a circle in the complex plane centered at $E$, and with a radius smaller than the gap to the neighbouring bands. We can rewrite this as
\begin{align}
\oint \frac{\mathrm{d}z}{2\pi i} \frac{1}{z-H(\k)} & =  \oint \frac{\mathrm{d}z}{2\pi i} \frac{1}{\text{det}(z-H(\k))}\text{Adj}(z-H(\k))\nonumber\\ 
& =  \frac{1}{\prod_n (E-E_n(\k))}\text{Adj}(E-H(\k))\,,
\end{align}
where Adj is the adjugate matrix. If the Hamiltonian is compact, then H$(\k)$ consists of polynomials in $e^{i\mathbf{a}_1\cdot \k}$ and $e^{i\mathbf{a}_2\cdot\k}$, where $\mathbf{a}_1$ and $\mathbf{a}_2$ form a basis for the Bravais lattice. The same is then true for Adj$(E-H(\k))$, and hence for $\prod_n (E-E_n(\k))|u(\k)\rangle\langle u(\k)|$. The Fourier transforms of $\prod_n (E-E_n(\k)) u^*_a(\k)|u(\k)\rangle$ would then be compact, for any choice of $a$. Since the $u_a(\k)$ cannot all vanish at the same point in momentum space, we have thus obtained an overcomplete set of compact functions that spans the band. It must therefore be trivial.

\section{Quantum geometric bounds for textured exciton insulators}\label{secapp:QGbound}

In this appendix, we consider lower bounds on the integrated trace of the quantum metric for CTIs and ETIs. The derived lower bound only uses the fact that the valley pseudospin is forced to point along opposite poles somewhere in the BZ for textured exciton insulators, and does not incorporate possible refinements to the bound from spatial symmetries or the topology of the $U(1)_v$-symmetric bands. Our discussion closely follows the formalism and derivations of Ref.~\cite{Herzog2022superfluidbound}. For simplicity, we consider a square real-space unit cell with basis lattice vectors $\bm{a}_1=a\hat{x}$ and $\bm{a}_2=a\hat{y}$, so that a general lattice vector is $\bm{R}=r_1\bm{a}_1+r_2\bm{a}_2$ with integer $r_1,r_2$. The corresponding basis RLVs are $\bm{b}_1=\frac{2\pi}{a}\hat{x}$ and $\bm{b}_2=\frac{2\pi}{a}\hat{y}$, with a general RLV being $\bm{G}=g_1\bm{b}_1+g_2\bm{b}_2$.

We first define the Abelian quantum geometric tensor
\begin{equation}\label{eqapp:QGT}
\text{Tr}\mathcal{G}_{ij}=\text{Tr}\left[P(\partial_i P)(\partial_j P)\right]
\end{equation}
where $P$ is the gauge-invariant Hermitian projector onto the $N_\text{occ}$ occupied bands (we consider $N_\text{occ}=1$ appropriate for CTIs and ETIs), $\partial_i$ indicates a momentum derivative along $k_i$ with $i=x,y$, and the trace $\text{Tr}$ is taken over the space of all orbitals (note that this subsumes all possible degrees of freedom including valley). The momentum argument $\bm{k}$ has been suppressed above. For simplicity, we neglect the `embedding' of orbitals in the unit cell so that we can choose $P(\bm{k})=P(\bm{k}+\bm{G})$. Eq.~\ref{eqapp:QGT} can be split into the symmetric quantum metric and anti-symmetric Berry curvature
\begin{equation}
\text{Tr}\mathcal{G}_{ij}=g_{ij}-\frac{i}{2}f_{ij}.
\end{equation}
We will focus on the  quantum metric, which takes the form 
\begin{equation}
g_{ij}=\frac{1}{2}\text{Tr}\left[(\partial_i P)(\partial_j P)\right],
\end{equation}
and is positive semi-definite. We define the integrated trace of the quantum metric (a dimensionless scalar)
\begin{equation}
G=\frac{1}{2}\int_\text{BZ} \frac{d^2\bm{k}}{(2\pi)^2} \sum_i \text{Tr}\left[(\partial_i P)(\partial_i P)\right]
\end{equation}
where the integral is taken over the BZ. Our goal will be to derive a finite lower bound on $G$. We first expand in a Fourier series
\begin{align}
    P(\bm{k})&=\sum_{\bm{R}}e^{-i\bm{R}\cdot\bm{k}}p(\bm{R}),\nonumber\\ p(\bm{R})&=\frac{a^2}{(2\pi)^2}\int_\text{BZ} d^2\bm{k}e^{i\bm{R}\cdot\bm{k}}P(\bm{k}).
\end{align}
From Hermiticity, we have $p^\dagger(\bm{R})=p(-\bm{R})$. Since there is one occupied band, we also have $\sum_{\bm{R}}||p(\bm{R})||^2=1$, where $||A||^2=\text{Tr}A^\dagger A$ is the Frobenius norm. $G$ can be expressed in dual $R$-space as
\begin{equation}\label{appeq:G_Rspace}
G=\sum_{\bm{R}}\frac{|\bm{R}|^2}{2a^2}||p(\bm{R})||^2,
\end{equation}
which is a sum of positive terms (except for $\bm{R}=0$ which vanishes).

For a textured exciton insulator, we know that the IVC has to vanish and point along opposite poles at (at least) two distinct points $\bm{k}=\pm \bm{k}^*$ in the BZ. We allow $\bm{k}^*$ to lie anywhere (except at time-reversal invariant momenta) to minimize the bound on $G$. The difference of the projectors at these nodal points is
\begin{equation}
P(\bm{k}^*)-P(-\bm{k}^*)=\sum_{\bm{R}}-2i\sin(\bm{k}^*\cdot\bm{R})p(\bm{R}).
\end{equation}
Taking the Frobenius norm and using the triangle inequality leads to
\begin{align}
    ||P(\bm{k}^*)-P(-\bm{k}^*)||&\leq \sum_{\bm{R}}2|\sin(\bm{k}^*\cdot\bm{R})|\times||p(\bm{R})||\nonumber\\&\leq \sum_{\bm{R}\neq \bm{0}}2||p(\bm{R})||.
\end{align}
The LHS of the above equation gives $||P(\bm{k}^*)-P(-\bm{k}^*)||=\sqrt{2}$, because $P(\pm\bm{k}^*)$ orient along opposite directions in valley space, and are hence orthogonal. We are then left with the problem of minimizing Eq.~\ref{appeq:G_Rspace} subject to the constraints
\begin{equation}
    \frac{1}{\sqrt{2}}\leq \sum_{\bm{R}\neq \bm{0}}||p(\bm{R})||\quad\text{and}\quad\sum_{\bm{R}}||p(\bm{R})||^2=1.
\end{equation}
Intuitively, we expect a lower bound on $G$ to arise from putting as much of the weight  of $||p(\bm{R})||$ as is consistent with the normalization condition on the harmonics $\bm{R}$ with the smallest sizes, as long as this saturates the first inequality above. This intuition is formalized in the `concentration lemma'~\cite{Herzog2022superfluidbound}. For the case of the square lattice, we consider $||p(\bm{a}_1)||=||p(-\bm{a}_1)||=\alpha$ and $||p(\bm{a}_2)||=||p(-\bm{a}_2)||=\beta$ such that that $\alpha+\beta=\frac{1}{2\sqrt{2}}$ and the rest of the weight is placed at $\bm{R}=0$. Eq.~\ref{appeq:G_Rspace} is minimized for $\alpha=\beta=\frac{1}{4\sqrt{2}}$, leading to 
\begin{equation}
G\geq\frac{1}{16}.
\end{equation}

The above result can be generalized to different BZ geometries. Furthermore, we anticipate that the bound on $G$ may be tightened by accounting for the quantum geometry intrinsic to the valley-symmetric bands which is finitely bounded from below due to their non-trivial topology. In addition, there may be refinements if the positions of the IVC nodes are constrained, which could arise due to spatial symmetries such as $\hat{C}_{3z}$. These questions are somewhat detail-dependent and hence beyond the scope of our analysis here, so we leave their resolution to future work.

\end{appendix}

\newpage
\clearpage

\setcounter{section}{0}
\setcounter{figure}{0}
\let\oldthefigure\thefigure
\renewcommand{\thefigure}{S\oldthefigure}

\setcounter{table}{0}
\renewcommand{\thetable}{S\arabic{table}}

\renewcommand{\thesection}{S\arabic{section}}
\renewcommand{\thesubsection}{\thesection.\arabic{subsection}}
\renewcommand{\thesubsubsection}{\thesubsection.\arabic{subsubsection}}

\onecolumngrid
	\begin{center}\textbf{\large --- Supplementary Material ---}

\end{center}

\section{Additional details of the LLL model}\label{secsupmat:1BLLL_additional}

In this section, we present additional details of the LLL model that were omitted or only briefly covered in the main text in Sec.~\ref{sec:1BLLL}.

\subsection{Form factors and density operator}
\label{secsupmat:1BLLL_form}

We have the following relations between the magnetic Bloch operator $d^\dagger_{\bm{k},\tau}$ and the Landau gauge operator $c^\dagger_{k,\tau}$
\begin{gather}
d^\dagger_{\bm{k},\tau}=\frac{1}{\sqrt{N_x}}\sum_{n}^{}e^{i\tau k_x(k_y+nQ)}c^\dagger_{k_y+nQ,\tau}\\
c^\dagger_{k_y+nQ,\tau}=\frac{1}{\sqrt{N_x}}\sum_{k_x} e^{-i\tau k_x(k_y+nQ)}d^\dagger_{\bm{k},\tau}.
\end{gather}
The position operator $\psi^\dagger_{\tau}(\bm{r})$ is expressed as
\begin{align}
\hat{\psi}^\dagger_\tau(\bm{r})=&\frac{1}{\sqrt{L_y\pi^{\frac{1}{2}}}}\sum_k e^{-iky}e^{-\frac{1}{2}(x-\tau k)^2}c^\dagger_{k\tau}\\
=&\frac{1}{\sqrt{N_xL_y\pi^{\frac{1}{2}}}}\sum_{n,\bm{k}} e^{-i(k_y+nQ)y}e^{-\frac{1}{2}(x-\tau (k_y+nQ))^2}e^{-i\tau k_x(k_y+nQ)}d^\dagger_{\bm{k}\tau}.
\end{align}

We now consider the density operator in momentum space
\begin{align}
\rho_\tau(\bm{q})&\equiv\int d\bm{r}\,e^{-i\bm{q}\bm{r}}\psi_\tau^\dagger(\bm{r})\psi_\tau(\bm{r})\\
&=\int d\bm{r}\,e^{-i\bm{q}\bm{r}}\frac{1}{L_y\sqrt{\pi}}\sum_{k,k'}e^{i(k'-k)y-\frac{1}{2}(x-\tau k)^2-\frac{1}{2}(x-\tau k')^2}c^\dagger_{k\tau}c_{k'\tau}\\
&=\int dx\,e^{-iq_xx}\frac{1}{\sqrt{\pi}}\sum_ke^{-\frac{1}{2}(x-\tau k)^2-\frac{1}{2}(x-\tau(k+q_y))^2}c^\dagger_{k\tau}c_{k+q_y,\tau}\\
&=e^{-\frac{\bm{q}^2}{4}}\sum_ke^{-iq_x\tau(k+\frac{q_y}{2})}c_{k\tau}^\dagger c_{k+q_y,\tau}\\
&=e^{-\frac{\bm{q}^2}{4}}\frac{1}{N_x}\sum_{\lfloor{k}\rfloor,n_k,k_x,k_x'}e^{-iq_x\tau(\lfloor{k}\rfloor+n_kQ+\frac{q_y}{2})}e^{-i\tau k_x(\lfloor{k}\rfloor+n_kQ)}e^{i\tau k_x'(\lfloor{k}\rfloor+n_kQ+q_y)}d^\dagger_{(k_x,\lfloor{k}\rfloor)\tau}d_{(k_x',\lfloor{k+q_y}\rfloor)\tau}\\
&=e^{-\frac{\bm{q}^2}{4}}\sum_{\lfloor{k}\rfloor,k_x}e^{-iq_x\tau(\lfloor{k}\rfloor+\frac{q_y}{2})}e^{-i\tau k_x\lfloor{k}\rfloor}e^{i\tau\lfloor{q_x+k_x}\rfloor(\lfloor{k}\rfloor+q_y)}d^\dagger_{(k_x,\lfloor{k}\rfloor)\tau}d_{(\lfloor{k_x+q_x}\rfloor,\lfloor{k+q_y}\rfloor)\tau}\\
	 &=e^{-\frac{\bm{q}^2}{4}}\sum_{\bm{k}\in\text{1BZ}}e^{i\tau\big(-q_x(k_y+\frac{q_y}{2})-k_xk_y+\lfloor{k_x+q_x}\rfloor(k_y+q_y)\big)}d^\dagger_{\bm{k}\tau}d_{\lfloor{\bm{k}+\bm{q}}\rfloor\tau}\\
  &\equiv \sum_{\bm{k}\in \text{BZ}}\lambda_\tau(\bm{k},\bm{q})d^\dagger_{\bm{k}\tau}d_{\lfloor{\bm{k}+\bm{q}}\rfloor\tau}
\end{align}
where we have recast the final result in terms of operators in the first BZ using the $\lfloor\bm{k}\rfloor$ notation that maps momenta $\bm{k}$ onto $[0,Q)\times [0,Q)$. With this prescription, we are in effect working in a \emph{periodic gauge}, with a branch cut at $k_x=Q$. This is ideally suited for HF numerics that use a consistent gauge for all $\bm{k}$ that map to the same $\lfloor \bm{k}\rfloor$. In the \emph{smooth} gauge (as presented in the main text), we have instead
\begin{align}
\rho_\tau(\bm{q})&=   e^{-\frac{\bm{q}^2}{4}}\sum_{\bm{k}\in\text{BZ}}e^{i\tau\big(-q_x(k_y+\frac{q_y}{2})-k_xk_y+(k_x+q_x)(k_y+q_y)\big)}d^\dagger_{\bm{k}\tau}d_{\bm{k}+\bm{q},\tau}\\
&=e^{-\frac{\bm{q}^2}{4}}\sum_{\bm{k}\in\text{BZ}}e^{i\tau(k_x+\frac{q_x}{2})q_y}d^\dagger_{\bm{k}\tau}d_{\bm{k}+\bm{q},\tau}\equiv \sum_{\bm{k}\in\text{BZ}}\Lambda_\tau (\bm{k},\bm{q})d^\dagger_{\bm{k}\tau}d_{\bm{k}+\bm{q},\tau}
\end{align}
where $\Lambda(\bm{k},\bm{q})$ is the form factor defined in a smooth gauge.

\subsection{Exciton vortex lattice order parameter}\label{secsupmat:1BLLL_EVL}

Our goal in this section is to construct the exciton vortex lattice (EVL) order parameter~\cite{Bultinck2019mechanism}. We first construct a trial IVC order parameter in real space\begin{equation}
    \Delta_\text{EVL}(\bm{r})=\langle c^\dagger_{\bm{r},+}c_{\bm{r},-}\rangle = \sum_k C_k e^{iky}e^{-(x+\frac{k}{2})^2},
\end{equation}
with some to be determined coefficients $C_k$. Since the IVC (excitonic) channel experiences a doubled magnetic field, we use a basis of Landau gauge LLL wavefunctions with magnetic length $\xi=\frac{1}{\sqrt{2}}$. The chirality of position-momentum locking is determined by the fact that $c^\dagger_{\bm{r},+}c_{\bm{r},-}$ probes the the exciton with electron in $\tau=-$ and hole in $\tau=+$. 

To fix the coefficients $C_k$, we need to impose the appropriate magnetic translation symmetries. Since the exciton experiences a doubled field, its primitive magnetic unit cell has half the area of the underlying electrons. Consider the magnetic unit cell spanned by $\bm{a}^\text{exc}_1=a\hat{y}$ and $\bm{a}^\text{exc}_2=\frac{1}{2}a\hat{x}+\frac{1}{2}a\hat{y}$. Consider the choice of corresponding commuting magnetic translation operators
\begin{equation}
    \tilde{T}^\text{exc}_1=e^{iap_y},\quad \tilde{T}^\text{exc}_2=e^{\frac{i\pi s}{2}}e^{-iQy}e^{ia\frac{p_x+p_y}{2}}
\end{equation}
where $s$ can take values $\pm1$ (this affects the overall sign of $\tilde{T}_2^\text{exc}$). These operators satisfy $(\tilde{T}^\text{exc}_2)^2 (\tilde{T}^\text{exc}_1)^{-1}=e^{2iQy}{T}_x$. The latter fact explains the choice of phases---we recover the gauge choice for the magnetic Bloch basis in Eq.~\ref{eq:MTO}. 

We now impose symmetry under magnetic translations. We first consider $\tilde{T}_1^\text{exc}$ and eigenphase $e^{ib_1 a}$
\begin{gather}
    \tilde{T}_1^\text{exc}\Delta_\text{EVL}(\bm{r})=\sum_k C_k e^{ik(y+a)}e^{-(x+\frac{k}{2})^2}\stackrel{!}{=}e^{ib_1 a}\Delta_\text{EVL}(\bm{r})
    \\
    \rightarrow k=b_1+jQ\text{ for }j\in\mathbb{Z}\\
    \rightarrow \Delta_\text{EVL}(\bm{r})=\sum_{j=-\infty}^{\infty}C_je^{i(jQ+b_1)y}e^{-(x+\frac{jQ+b_1}{2})^2}.
\end{gather}

We now consider $\tilde{T}_2^\text{exc}$ and eigenphase $e^{ib_2 a}$
\begin{align}
    \tilde{T}_1^\text{exc}\Delta_\text{EVL}(\bm{r})&=\sum_j C_j e^{\frac{i\pi s}{2}}e^{iQy}e^{i(jQ+b_1)(y+\frac{a}{2})}e^{-(x+\frac{a}{2}+\frac{jQ+b_1}{2})^2}\\
    &=\sum_j C_{j} e^{\frac{i\pi s}{2}}e^{i\pi j }e^{\frac{ib_1 a}{2}}e^{i((j+1)Q+b_1)y}e^{-(x+\frac{(j+1)Q+b_1}{2})^2}\stackrel{!}{=}e^{ib_2 a}\Delta_\text{EVL}(\bm{r})
    \\
    \rightarrow C_{j+1}&=C_j e^{\frac{i\pi s}{2}}e^{i\pi j}e^{i(\frac{b_1}{2}-b_2)a}\\
    \rightarrow \Delta_\text{EVL}(\bm{r})&=C_0\sum_j e^{\frac{i\pi  sj^2}{2}}e^{i(\frac{b_1}{2}-b_2)ja}e^{i(jQ+b_1)y}e^{-(x+\frac{jQ+b_1}{2})^2}. 
\end{align}

We now project this into the LLL magnetic Bloch basis
\begin{align}
    \Delta_\text{EVL}(\bm{k},\bm{k}')&=\langle d^\dagger_{\bm{k},+}d_{\bm{k}',-}\rangle\\
    &\sim \int d\bm{r}\,\Delta_\text{EVL}(\bm{r})\phi_{\bm{k},+}(\bm{r})\phi_{\bm{k}',-}^*(\bm{r})\\
    &\sim\sum_{j,n}\int dx\,e^{\frac{i\pi \sigma j^2}{2}}e^{i(\frac{b_1}{2}-b_2)ja}e^{i(k_x+k_x')k_y}e^{i(k_x+k_x')nQ}e^{i(jQ+b_1)k_x'}\\
    &\quad\quad\quad\quad \times e^{-(x+\frac{jQ+b_1}{2})^2}e^{-\frac{1}{2}(x-(k_y+nQ))^2}e^{-\frac{1}{2}(x+(k_y+nQ+jQ+b_1))^2}
\end{align}
where in the last line we have integrated over $y$ to enforce $k_y'=k_y+b_1$. We can perform the Gaussian integration in the last line to yield
\begin{align}
\Delta_\text{EVL}(\bm{k},\bm{k}')&\sim \sum_{j,n}e^{\frac{i\pi  sj^2}{2}}e^{i(\frac{b_1}{2}-b_2)ja}e^{i(k_x+k_x')k_y}e^{i(k_x+k_x')nQ}e^{i(jQ+b_1)k_x'}\\
    &\quad\quad\quad\quad \times e^{-k_y^2}e^{-(2nQ+jQ+b_1)k_y}e^{-(nQ+\frac{jQ+b_1}{2})^2}\\
&\sim \sum_{j,n}e^{\frac{i\pi  sj^2}{2}}e^{i(\frac{b_1}{2}-b_2)ja}e^{i(k_x+k_x')k_y}e^{i(k_x-k_x'-b_1+2b_2)nQ}e^{i(jQ+b_1)k_x'}e^{-k_y^2}e^{-(jQ+b_1)k_y}e^{-(\frac{jQ+b_1}{2})^2}\\
&\sim \sum_j e^{\frac{i\pi  sj^2}{2}}e^{i(\frac{b_1}{2}-b_2)ja}e^{i(2k_x-b_1+2b_2)k_y}e^{i(jQ+b_1)(k_x-b_1+2b_2)}e^{-k_y^2}e^{-(jQ+b_1)k_y}e^{-(\frac{jQ+b_1}{2})^2}\\
&\sim \sum_j e^{\frac{i\pi  sj^2}{2}}e^{i(k_x-\frac{b_1}{2}+b_2)(2k_y+jQ+b_1)}e^{-\frac{1}{4}(2k_y+jQ+b_1)^2}\\
&\sim \sum_j e^{\frac{(i s - 1)\pi j^2}{2}}e^{-(k_y+\frac{b_1}{2})^2}e^{2i(k_x-\frac{b_1}{2}+b_2)(k_y+\frac{b_1}{2})}e^{jQ(i(k_x-\frac{b_1}{2}+b_2)-(k_y+\frac{b_1}{2}))}
\end{align}
where in the second line we took $j\rightarrow j-2n$, and in the third line we performed the sum over $n$ to constrain $k_x'=k_x-b_1+2b_2$. Finally, we simplify in terms of the Jacobi theta function of the third kind\footnote{In the normalization used here, we have the quasiperiodicity
\begin{equation}
    \theta_3(z+1,\alpha)=\theta_3(z,\alpha),\quad \theta_3(z+\alpha,\alpha)=e^{-i\pi\alpha}e^{-2\pi iz}\theta_3(z,\alpha).
\end{equation}}
\begin{equation}
    \theta_3(z,\alpha)=\sum_{n=-\infty}^\infty e^{\pi i n^2 \alpha}e^{2\pi inz}=\theta_3(-z,\alpha)
\end{equation}
leading to
\begin{gather}
    \Delta_\text{EVL}(\bm{k},\bm{k}')=\delta_{k_x'=k_x-b_1+2b_2}\delta_{k_y'=k_y+b_1}\times \Delta_0 e^{(p_x+ip_y)^2-p_x^2}\times\theta_3\left(\frac{p_x+ip_y}{Q},\frac{s+i}{2}\right)\\
    p_x=k_x-\frac{b_1}{2}+b_2,\quad p_y=k_y+\frac{b_1}{2}.
\end{gather}
$\theta_3(z,\alpha)$ has nodes at $z=m\alpha + n + \frac{\alpha + 1}{2}$ where $m,n\in\mathbb{Z}$. Hence the exciton order parameter has nodes at $\pm(Q/4,3Q/4)+(\frac{b_1}{2}-b_2,-\frac{b_1}{2})$ for $s=1$ and $\pm(Q/4,Q/4)+(\frac{b_1}{2}-b_2,-\frac{b_1}{2})$ for $s=-1$.

To make the physics more transparent, we rewrite the above in a more symmetric manner in terms of the momentum boost $\bm{q}=\bm{k}-\bm{k}'$ of valley $\tau=-$ relative to valley $\tau=+$
\begin{equation}\label{eq:exciton_lattice}
    \Delta_{\text{EVL},\bm{q}}(\bm{k}-\frac{\bm{q}}{2})=\langle d^\dagger_{\bm{k}-\frac{\bm{q}}{2},+}d_{\bm{k}+\frac{\bm{q}}{2},-} \rangle=\Delta_0 e^{2ik_xk_y-k_y^2}\theta_3\left(\frac{k_x+ik_y}{Q},\frac{s+i}{2}\right)=\Delta_{\text{EVL},\bm{q}}(-\bm{k}-\frac{\bm{q}}{2}).
\end{equation}
So $\Delta_{\text{EVL},\bm{q}}(\bm{k}-\frac{\bm{q}}{2})$ either has vortices at $\pm(Q/4,3Q/4)$ for $s=1$ and $\pm(Q/4,Q/4)$ for $s=-1$. It will be useful to write it out explicitly when $\bm{q}=0$
\begin{gather}
    \Delta_{\text{EVL},\bm{q}=\bm{0}}(\bm{k})=\sum_j e^{\frac{(i s - 1)\pi j^2}{2}}e^{-k_y^2}e^{2ik_xk_y}e^{jQ(ik_x-k_y)}=\Delta_{\text{EVL},\bm{q}=\bm{0}}(-\bm{k})\\
    \Delta_{\text{EVL},\bm{q}=\bm{0}}(\bm{k}+\bm{G})=e^{2ik_yG_x}\Delta_{\text{EVL},\bm{q}=\bm{0}}(\bm{k})
\end{gather}
where $\bm{G}$ is a reciprocal lattice vector.

\begin{figure}
    \centering
    \includegraphics[width = 0.5\linewidth]{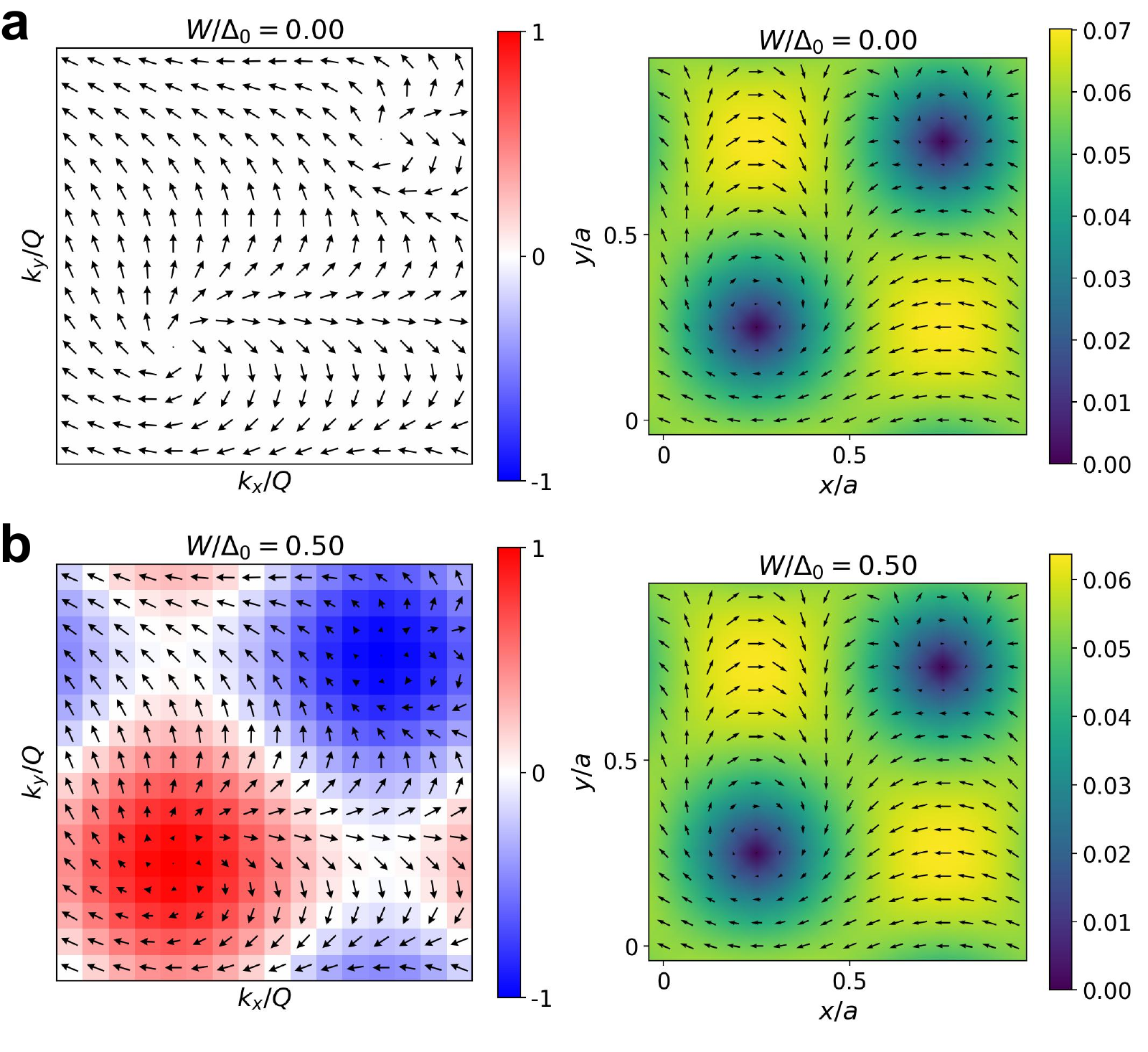}
    \caption{\textbf{Properties of the CTI$_1$ constructed from the EVL ansatz with $\bm{q}=0$ and $s=-1$ in the LLL model.} a) Results for $W/\Delta_0=0$. Left shows the in-plane pseudospin (arrows) and out-of-plane valley polarization (color) in the BZ. Right shows the in-plane pseudospin in the real-space unit cell, with color indicating the strength of IVC. b) Same as a) but for $W/\Delta_0=0.5$.} 
    \label{figapp:EVL_BML_ansatz}
\end{figure}

To recap, we have constructed a family of EVL order parameters $\Delta_{\text{EVL},\bm{q}}$ parameterized by the intervalley boost vector $\bm{q}$. For a given $\bm{q}$, the node positions are fixed to one of two options. This may seem surprising given that the underlying Landau levels have continuous translation invariance, so it would seem that one should be able to shift the nodes continuously. However, the resolution is that the opposite winding of the LLLs and the gauge choice when defining the vector potential and magnetic Bloch basis fix the positions. For instance, in the Landau gauge $\bm{A}=-\tau Bx\hat{y}$ and with a consistent identification of the BZ in the two valleys, the positions $x=\frac{na}{2}$ are special for $\bm{q}=0$. This is because both valleys possess a Landau gauge eigenfunction that is peaked at $x$ and has a $k_y$-momentum that folds to the same momentum in the BZ. For different $\bm{q}$, the special $x$-positions change accordingly, thus generating EVLs with different origins in real space. 

In the main text, we constructed a CTI$_1$ ansatz using the EVL order parameter and showed its valley pseudospin order in momentum space. The ansatz is generated by solving the following mean-field Hamitonian
\begin{equation}
\begin{gathered}
    H^\text{MF}=\sum_{\bm{k},\tau,\tau'}d^\dagger_{\bm{k},\tau}h_{\tau,\tau'}(\bm{k})d_{\bm{k},\tau'}\\
    h(\bm{k})=\begin{pmatrix}
        \epsilon_+(\bm{k}) & [\Delta_{\text{EVL}}(\bm{k})]^*\\
        \Delta_{\text{EVL}}(\bm{k}) & \epsilon_-(\bm{k})
    \end{pmatrix},
\end{gathered}
\end{equation}
where $\epsilon_\tau(\bm{k})$ is the dispersion in each valley whose band width is controlled by $W$.
In Fig.~\ref{figapp:EVL_BML_ansatz}, we also show the corresponding real-space properties of the EVL and the CTI$_1$. The local valley polarization in real space always vanishes, consistent with TRS. The IVC in the CTI$_1$ state is overall slightly weaker than the EVL, but both have two vortices with the same winding. Interestingly, the real-space profile of the local IVC appears similar for different values of $W/\Delta_0$. We comment that the real-space vortices can be understood as a direct consequence of the effective twisted boundary conditions on the unit cell arising from the magnetic translation symmetry, since the IVC fermion bilinear experiences a net magnetic field. This forces the IVC in real space to vanish at least once in each unit cell to accommodate the winding. 

\subsection{Phase diagram with lowest and second harmonics in dispersion}\label{secsupmat:1BLLL_nnn}

\begin{figure}
    \centering
    \includegraphics[width = 0.6\linewidth]{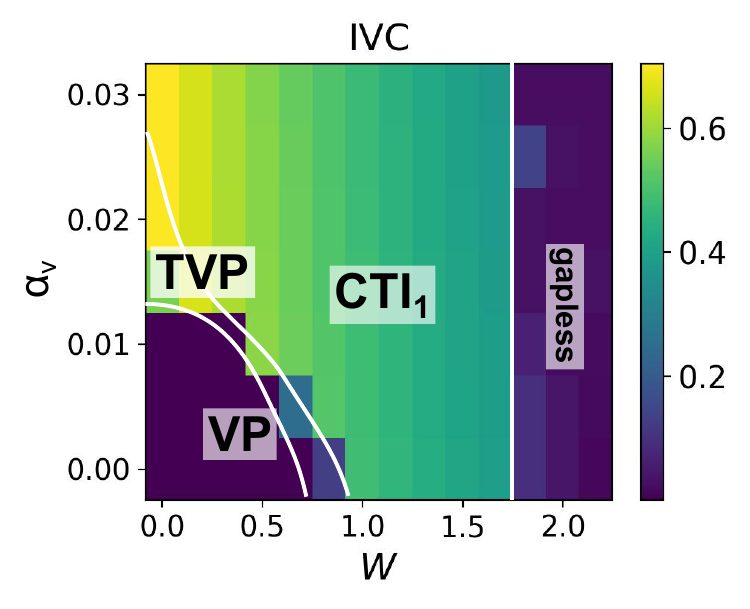}
    \caption{\textbf{HF phase diagram of the LLL model with second-harmonic potential terms at  half-filling.} $\alpha_\text{v}$ is the valley interaction anisotropy, and $W$ controls the scale of the dispersion $\epsilon_{\tau}(\bm{k})=-\frac{W}{4}(\cos k_xa + \cos k_y a+0.2\cos 2k_xa + \cos 2k_ya)$. The gate distance $d_\text{sc}=6a$, and the interaction scale $U=1$. Color indicates the magnitude of intervalley coherence (IVC). White lines indicate approximate phase boundaries. System size is $24\times 24$, and the valley boost is fixed to either $\bm{q}=(0,0)$ or $(Q/2,Q/2)$. [VP: valley-polarized; TVP: tilted valley-polarized; CTI$_1$: Chern texture insulator]}
    \label{figapp:1BLLL_phase_24x24_nnn}
\end{figure}

In Fig.~\ref{figapp:1BLLL_phase_24x24_nnn}, we perform identical calculations as in Fig.~\ref{fig:1BLLL_phase_24x24}, except we include a second-harmonic component to the single-particle dispersion
\begin{equation}
    \epsilon_{\tau}(\bm{k})=-\frac{W}{4}(\cos k_xa + \cos k_y a+0.2\cos k_xa + 0.2\cos k_y a).
\end{equation}
This breaks the perfect intervalley nesting of the Fermi surfaces for half-filling.

The phase diagram for weak and intermediate values of $W$ is largely unchanged. For larger $W$ (weaker interactions), we find that the CTI$_1$ gives way to a gapless metallic phase.

\subsection{More general dispersions}\label{secsupmat:1BLLL_moredispersion}

\begin{figure}
    \centering
    \includegraphics[width = 0.65\linewidth]{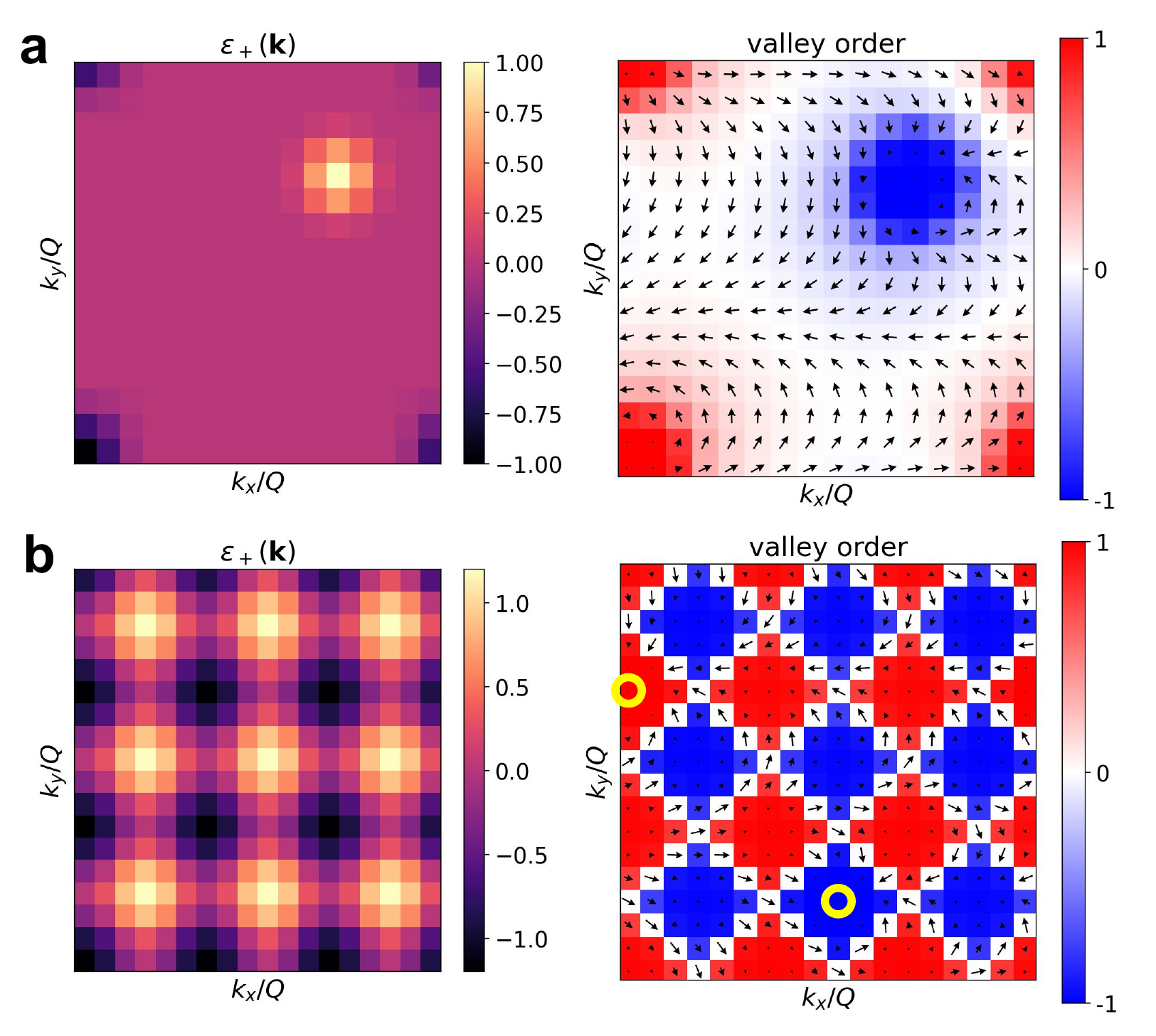}
\caption{\textbf{CTI$_1$ states in the LLL model for different single-particle dispersions.} The gate distance $d_\text{sc}=6a$ and valley interaction anisotropy $\alpha_\text{v}=0$ for all cases. a) Left shows the single-particle dispersion in valley $\tau=+$, with minimum at $(0,0)$ and maximum at $(11Q/16,11Q/16)$. The dispersion in valley $\tau=-$ is related by TRS. Right shows the valley order of the HF ground state, which is a CTI$_1$ with boost $\bm{q}=(5Q/16,5Q/16)$. b) Similar as a), except the single-particle dispersion is composed of third harmonics, and the CTI$_1$ has $\bm{q}=(Q/2,Q/6)$. The vortices in the IVC are indicated with yellow circles.}
    \label{fig:1BLLL_warped_harmonic}
\end{figure}

In Fig.~\ref{fig:1BLLL_warped_harmonic}a, we design a single-particle dispersion for $\tau=+$ such that the minimum is at $(0,0)$ and the maximum is at $(11Q/16,11Q/16)$. Due to TRS, the maximum in $\tau=-$ is at $(5Q/16,5Q/16)$. The lobe principle predicts an ideal boost wavevector $\bm{q}=(5Q/16,5Q/16)$ in order to overlap the $\tau=+$ minimum with the $\tau=-$ maximum\footnote{Due to TRS, this automatically also overlaps the $\tau=-$ minimum with the $\tau=+$ maximum.}. Indeed, Fig.~\ref{fig:1BLLL_warped_harmonic}a shows that the lowest HF state is a CTI$_1$ with this boost $\bm{q}$, where the merons are shifted so that they are centered at the extrema of the dispersion. By repeating the calculation for different spacings of the extrema in $\epsilon_\tau(\bm{k})$, we find that the HF energy is minimized when the peak and trough are furthest apart in momentum space. This is expected because bringing the merons together incurs an exchange penalty as the valley pseudospin needs to rotate rapidly. But it is clear that the CTI$_1$ phase is flexible enough to accommodate deformations to the single-particle dispersion, and does not rely on nesting that is often required for weak-coupling instabilities.

In Fig.~\ref{fig:1BLLL_warped_harmonic}b, we further generalize to a larger number of band extrema by considering a third-harmonic potential. Several choices of $\bm{q}$ are compatible with the lobe principle, and yield degenerate CTI$_1$ states. While the pseudospin in the BZ (Fig.~\ref{fig:1BLLL_warped_harmonic}b) has sizable valley polarization in many regions due to the single-particle dispersion, there are still only two merons which are highlighted by yellow circles. They can be identified by tracking the winding of the IVC order, or checking where the pseudospin orients along the poles (which is required at the meron cores). The merons are spaced as far apart as possible in the BZ in order to reduce the exchange cost. This underscores the difference between the CTI, and a more trivial IVC state between topologically trivial bands. While the kinetic dispersion leaves a noticeable fingerprint on the momentum-space valley order in both phases, the latter does not have a non-zero topological winding imposed on the IVC order parameter. In particular, it can be connected to the limit of fully localized valley moments in real-space forming an in-plane (spiral) pattern, which in momentum-space corresponds to uniform IVC across the BZ. The same cannot be done for the CTI, which must have vortices/merons.

\section{Additional details of toy models for the ETI}\label{secsupmat:additional_ETI}

\subsection{Additional results on ETI toy model with projection}\label{secsupmat:additional_ETI_k1k2}

\begin{figure}
    \centering
    \includegraphics[width = 0.5\linewidth]{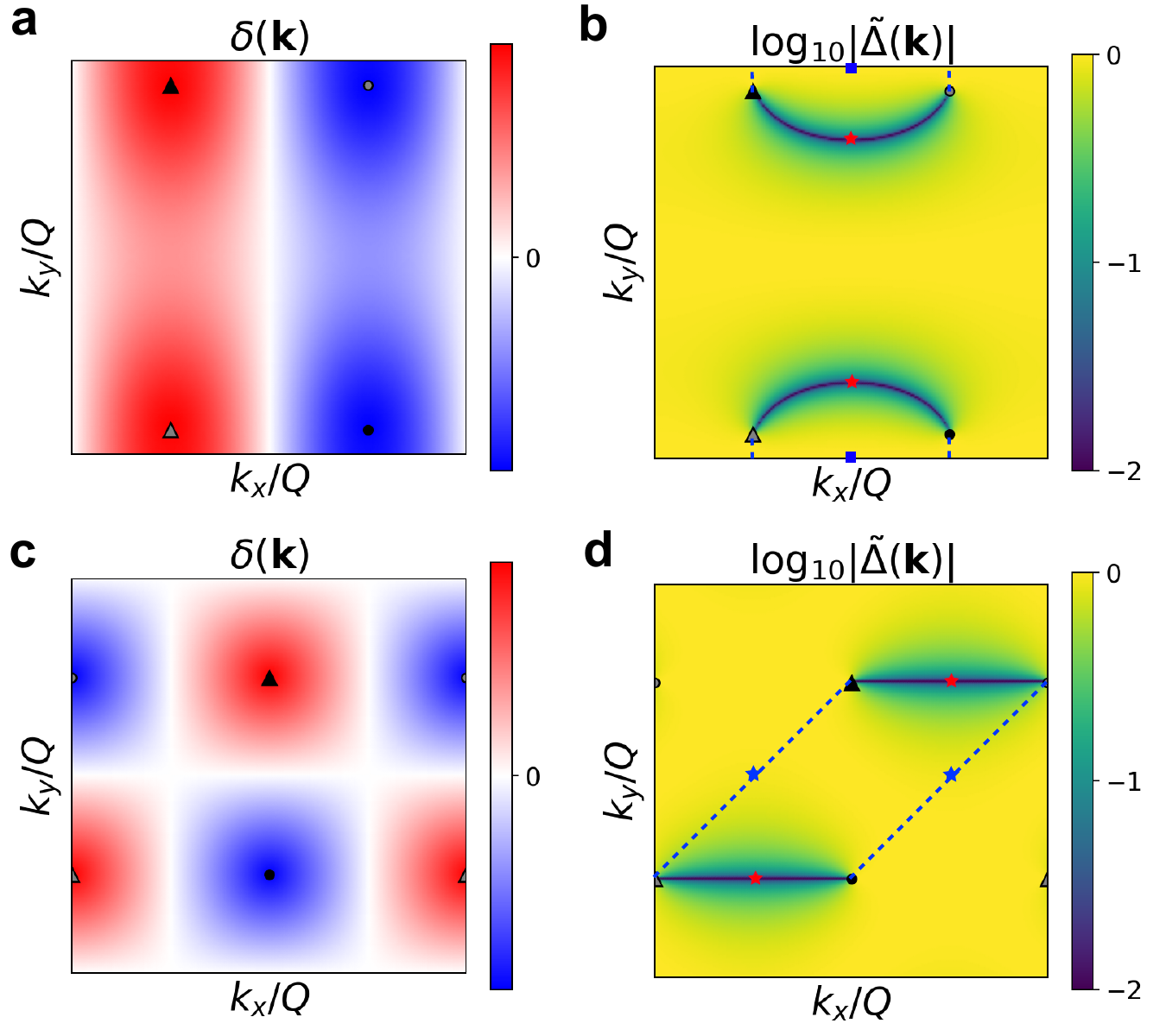}
\caption{\textbf{Toy model for ETI in the BZ.} 
The model (see Eq.~\ref{eq:fk_H+} for definition in the absence of IVC) is defined over the entire BZ. Any boost $\bm{q}$ has been absorbed into the momentum origin such that TRS takes $\bm{k}\rightarrow -\bm{k}$.
a) Difference in valence band valley dispersion $\delta(\bm{k})$. Dirac points in valley $\tau=+$ are at $\bm{k}_1=(Q/4,-Q/16)$ [black triangle] and $\bm{k}_2=(Q/4,Q/16)$ [grey triangle].
TRS-related Dirac points in valley $\tau=-$ are indicated with dots.
b) Projected IVC matrix element $\tilde{\Delta}(\bm{k})$ for a TIVC mass of strength $\Delta_\text{TIVC}=1$ (Eq.~\ref{eq:TIVCmass}). Note that the lower limit of the color scale has been clamped. Red stars indicate residual Dirac points in the projected theory. By additional application of other $\bm{k}$-dependent IVC terms, the nodal lines can be rewired to the dashed blue lines, and the residual Dirac points annihilate at the blue square. c,d) Same as a,b) except $\tau=+$ Dirac points are at $\bm{k}_1=(Q/2,3Q/4)$ and $\bm{k}_2=(0,Q/4)$. Note that even after the nodal lines are rewired, there are still gapless points (blue stars). 
}
    \label{figapp:2BLLL_fk_model}
\end{figure}

In this subsection, we discuss the effect of changing the non-interacting Dirac point positions $\bm{k}_1,\bm{k}_2$ in the kinetic term of the ETI toy model introduced in Sec.~\ref{sec:C2z_effective} of the main text. We also explain why a uniform TIVC mass leads to nodal lines in the projected IVC $\tilde{\Delta}(\bm{k})$ that connect non-interacting Dirac points from opposite valleys.

\subsubsection{Non-interacting Dirac point positions}

Figs.~\ref{figapp:2BLLL_fk_model}a,b show $\delta(\bm{k})$ and $\tilde{\Delta}(\bm{k})$ for a system where the non-interacting Dirac points within a valley are close to each other. Recall that these quantities are defined via projection into the lower bands of the non-interacting model. Note that $\delta(\bm{k})$ vanishes along two lines in the BZ, which cover all the time-reversal invariant momenta. The projected theory with a uniform TIVC term has two residual Dirac points, but by applying other IVC perturbations, these can be annihilated (blue square) to yield a direct gap at $\nu=-1$. The resulting rewired nodal lines that connect the Dirac points within a valley (dashed blue lines) are significantly shorter, which suggests a favorable ground state energy as the IVC perturbation is able to hybridize the bands over a larger region of the BZ.

In contrast, $\bm{k}_1,\bm{k}_2$ are chosen in Figs.~\ref{figapp:2BLLL_fk_model}c,d to model the situation in the two-band LLL model (see Sec.~\ref{secsupmat:2BLLL_additional}), where the Dirac points in one valley are as far apart as possible. The nodal lines can again be rewired (dashed blue lines) to connect the non-interacting Dirac points within a valley, but this leads to two issues. Firstly, the rewired nodal lines are longer than the original ones, suggesting that such a configuration is not energetically favorable. Secondly, it is not possible to open a direct gap in this way, because the nodal lines unavoidably cross the locus of $\delta(\bm{k})=0$ (blue stars in Fig.~\ref{figapp:2BLLL_fk_model}d). This explains the result in Sec.~\ref{secsupmat:2BLLL_IVC} that the two-band and $\hat{C}_{2z}$-symmetric variant of the LLL model yields a gapless state with two Dirac points when $\hat{C}_{2z}$ and $\hat{\mathcal{T}}$ are preserved.

In a similar vein, we expect the exchange energetics to be favored by having the non-interacting Dirac points in one valley far away (in the $\bm{q}$-boosted frame) from those in the other valley. If the Dirac points of the two valleys are close together in the BZ, then the valley pseudospin would be required to rotate rapidly in momentum space between them.

\subsubsection{Nodal lines for uniform TIVC term}

We show that in the presence of a uniform TIVC term, the projected IVC $\tilde{\Delta}(\bm{k})$ of Eq.~\ref{eq:toy_model_Htilde} has nodal lines that connect non-interacting Dirac points from opposite valleys. Since the lower bands arise from a $\hat{C}_{2z}\hat{\mathcal{T}}$-symmetric two-band model in each valley, they can be described at each $\bm{k}$ in the BZ by an in-plane sublattice angle $\phi_\tau(\bm{k})$ (the Bloch states cannot have any non-zero sublattice polarization due to symmetry). In a smooth gauge, $\phi_\tau(\bm{k})$ is smooth except for singularities at the Dirac points at valley $\tau$ where the sublattice angle winds by $2\pi\tau$, assuming the scenario where there are two Dirac points of identical winding in each valley for simplicity. Consider the uniform TIVC term $\sim \tau^x\sigma^x$\footnote{More generally for this argument, we could have considered an arbitrary direction in valley space $\cos(\theta)\tau^x\sigma^x+\sin(\theta)\tau^y\sigma^x$. We could have also considered a momentum-dependent TIVC angle $\theta(\bm{k})$, as long as it is smooth and has no vortices.}. When projected to the lower bands, this vanishes when $\phi_+(\bm{k})+\phi_-(\bm{k})=\pi$. Now consider a contractible loop in the BZ that encircles $\bm{k}_1$ and $\bm{k}_2$, but not $-\bm{k}_1$ or $-\bm{k}_2$. Around this loop, $\phi_+(\bm{k})$ winds by $4\pi$ but $\phi_-(\bm{k})$ has no net winding. Hence, there must be at least two points along the loop where $\tilde{\Delta}(\bm{k})$ vanishes. By deforming this loop, we can trace nodal lines by invoking smoothness away from the non-interacting Dirac points. These nodal lines can only terminate at non-interacting Dirac points where smoothness breaks down. This argument can be repeated for a loop encircling only the non-interacting Dirac points in valley $\tau=-$. Consistency of the above statements requires nodal lines that connect the non-interacting Dirac points in different valleys. TRS further constrains the relative positions of the nodal lines.

\subsubsection{Nodal lines for TIVC and inter-Chern terms}

Now consider the inclusion of an inter-Chern IVC term, such that in the sublattice basis, the inter-valley part of the (unprojected) Hamiltonian is given by
\begin{equation}
    H_{+-}(\bm{k}) = \begin{pmatrix}
        g(\bm{k}) & 1 \\
        1 & g^*(\bm{k})
        
    \end{pmatrix},
\end{equation}
where for simplicity we have chosen the TIVC term to be constant with magnitude 1. After projection to the lower bands, we find
\begin{equation}
    \tilde{\Delta}(\bm{k}) = e^{i(\phi_+(\bm{k}) - \phi_-(\bm{k}))}h(\bm{k}) + h^*(\bm{k}),
\end{equation}
where $h(\bm{k}) = e^{i\phi_-(\bm{k})} + g(\bm{k})$. We have $\tilde{\Delta}(\bm{k}) = 0$ if 
\begin{equation}\label{eq:delta_zero}
    \phi_+(\bm{k}) - \phi_-(\bm{k}) + 2\arg(h(\bm{k})) \equiv \pi \mod 2\pi.
\end{equation}
Traversing a closed loop surrounding a single non-interacting Dirac point, say, in valley $\tau = +$, we find that $\phi_+(\bm{k}) - \phi_-(\bm{k})$ winds by $2\pi$, while $2\arg(h(\bm{k}))$ winds by a multiple of $4\pi$, so that somewhere on the loop Eq.~\ref{eq:delta_zero} must be satisfied. As such, we find that nodal lines start and end at non-interacting Dirac points. However, for a loop surrounding two non-interacting Dirac points, even if they are from the same valley, we can no longer guarantee that it cuts through a nodal line, due to the possible winding of $h(\bm{k})$ that compensates for the winding of non-interacting Dirac points. This means that nodal lines no longer have to connect non-interacting Dirac points of opposite valleys.
\subsection{ETI toy model without projection}\label{secsupmat:noprojection}

We consider the  toy model defined in Sec.~\ref{sec:C2z_effective} of the main text, but without projecting into the lower bands.

\begin{figure}[h]
    \centering
    \includegraphics[width = 0.4\linewidth]{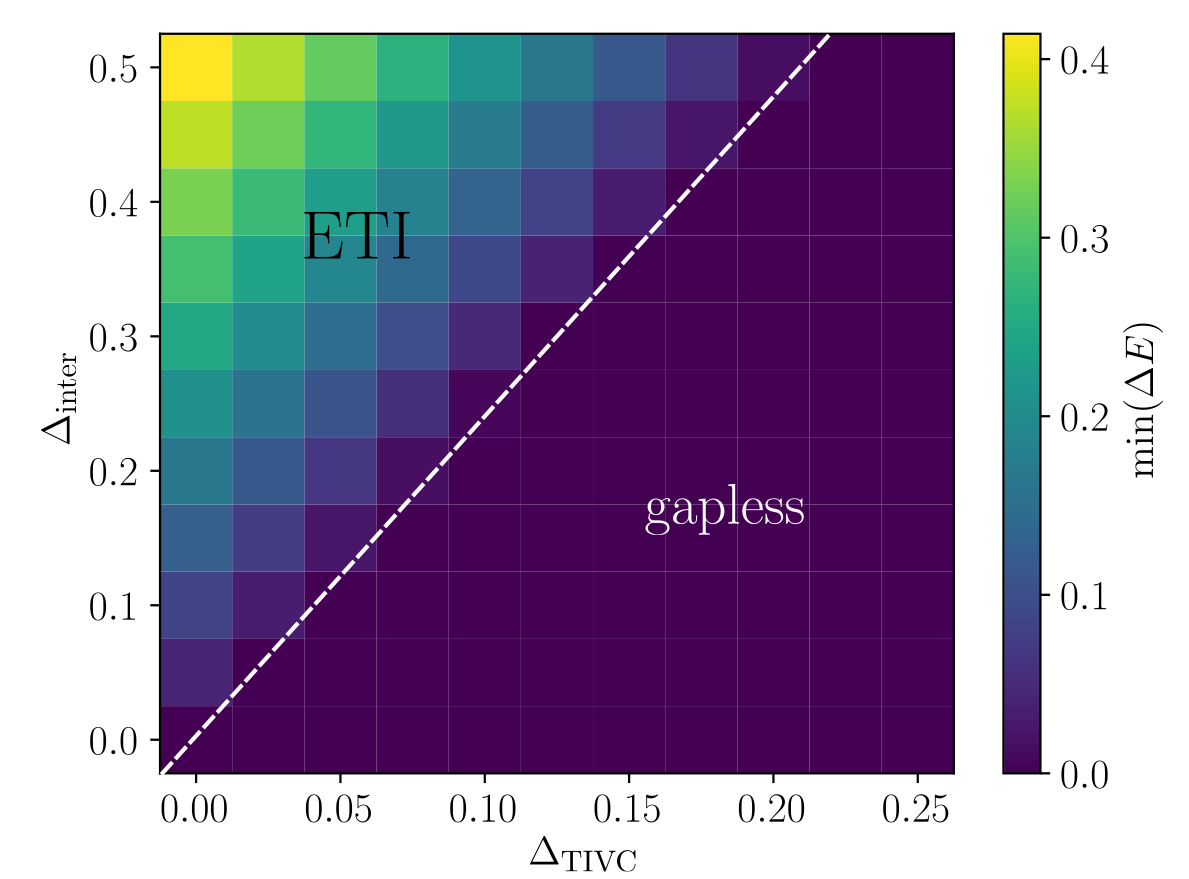}
    \caption{The minimum direct charge gap in the BZ for the unprojected ETI toy model. We observe that large $\Delta_{\text{TIVC}}$ favors gapless states while large $\Delta_{\text{inter}}$ favors ETI.}
    \label{fig:ETIvsGapless}
\end{figure}

For the following discussion, we will use $\bm{k}_1 = (3Q/4, 9Q/16)$, $\bm{k}_2 = (3Q/4, 7Q/16)$ and $\bm{k}_{\text{inter}} = (3Q/4, Q/2)$. The ground state (at $\nu = -1$) is either a gapless state or a gapped ETI, which depends on the relative strength of $\Delta_{\text{TIVC}}$ and $\Delta_{\text{inter}}$, as shown in Fig.~\ref{fig:ETIvsGapless}.

\begin{figure}[h]
    \centering
    \includegraphics[width = 0.8\linewidth]{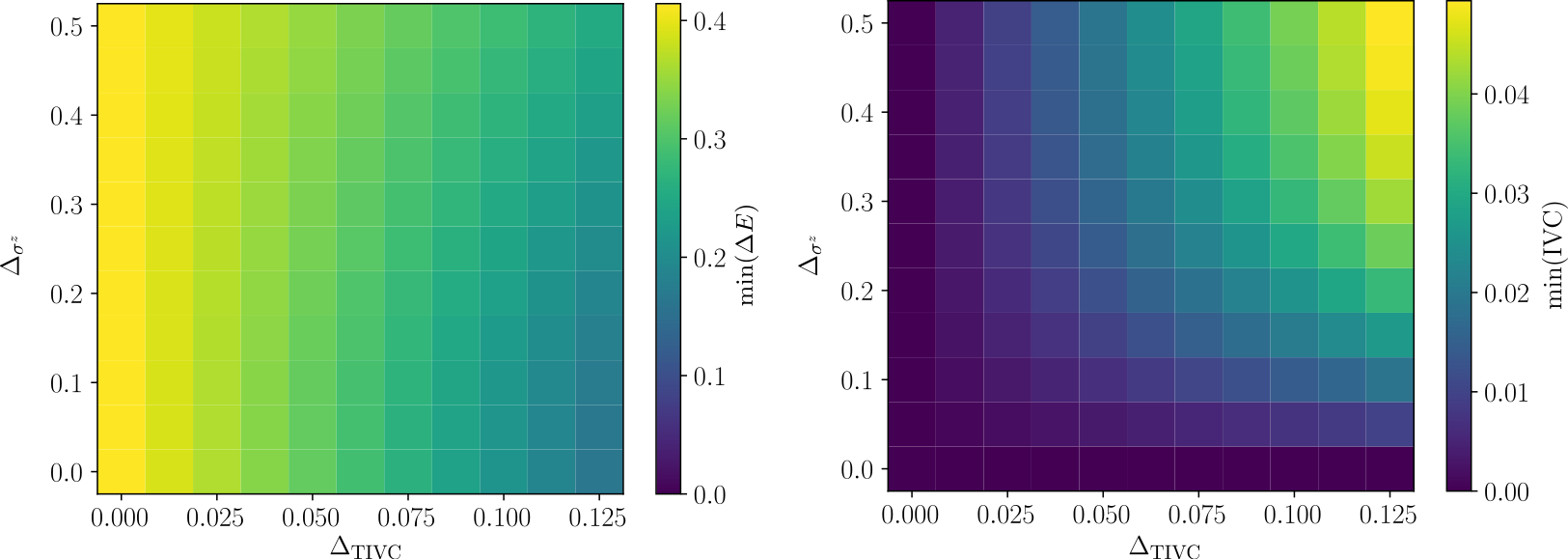}
    \caption{The minimum direct gap and minimum IVC strength in the BZ for the toy model. For ETI ($\hat{C}_{2z}\hat{\mathcal{T}}$-symmetric) and CTI (no $\hat{C}_{2z}\hat{\mathcal{T}}$-symmetry), the minimum IVC strength in the BZ is zero. In this phase diagram, the state is ETI on the horizontal axis, CTI on the vertical axis, and trivial IVC everywhere else.}
    \label{fig:CTI_ETI_trivial}
\end{figure}

We also consider a $\hat{C}_{2z}\hat{\mathcal{T}}$-breaking term given by
\begin{equation}
    H_{\sigma^z} = \Delta_{\sigma^z}\sigma^z
\end{equation}
which preserves TRS. This $\hat{C}_{2z}\hat{\mathcal{T}}$-breaking term can either convert ETI into a CTI or a trivial IVC state, as shown in Fig.~\ref{fig:CTI_ETI_trivial}. We note that the transition between ETI, CTI and trivial IVC does not involve the closing of the charge gap.

\subsection{Continuum $k\cdot p$ model}
In this subsection, we consider a continuum $k\cdot p$ version of the ETI toy model introduced in Sec.~\ref{sec:C2z_effective} in the main text. While the $k\cdot p$ model can describe the physics within a local patch of momentum space, it does not capture the constraints imposed by the combination of a finite BZ and associated topological winding of inter-Chern hybridization.

We consider two bands in each valley that are described by $2\times2$ matrices $H_+(\bm{k})$ and $H_-(\bm{k})$ in sublattice space. We first take the limit of a continuum $\bm{k}\cdot\bm{p}$ expansion about $\bm{k}=0$, and study the following model
\begin{gather}\label{eq:kp_H+}
    H_+(\bm{k})=\begin{pmatrix}
        0 & (k-k_1)(k-k_2)\\
        (k-k_1)^*(k-k_2)^* & 0
    \end{pmatrix}
\end{gather}
with $k=k_x+ik_y$, and $H_-(\bm{k})=H_+^*(-\bm{k})$. The system satisfies TRS and two-fold rotation which act on valley and sublattice as $\hat{\mathcal{T}}=\tau^x\mathcal{K}$ and $\hat{C}_{2z}=\tau^x\sigma^x$. As in the two-band LLL model discussed in App.~\ref{secsupmat:2BLLL_additional}, the sublattice basis should be thought of as having Chern number $C=\tau\sigma$. The Dirac points at charge neutrality are located at $\tau\bm{k}_1$ and $\tau\bm{k}_2$ and have opposite winding in the two valleys. Since we will focus on filling $\nu=-1$, we will project the problem onto the lower bands
\begin{gather}
    \ket{\text{lower},+,\bm{k}}=\ket{+}\otimes\frac{1}{\sqrt{2}}\begin{pmatrix}
        1\\-e^{-i\theta(\bm{k})}
    \end{pmatrix}\\
    \ket{\text{lower},-,\bm{k}}=\ket{-}\otimes\frac{1}{\sqrt{2}}\begin{pmatrix}
        1\\-e^{i\varphi(\bm{k})}
    \end{pmatrix}
\end{gather}
where
\begin{gather}
    \theta(\bm{k})=\text{arg}[(k-k_1)(k-k_2)]\\
    \varphi(\bm{k})=\text{arg}[(k+k_1)(k+k_2)].
\end{gather}
As the goal is to induce a $\hat{\mathcal{T}}$- and $\hat{C}_{2z}$-symmetric IVC state, we add a uniform mean-field $U(1)_\text{v}$-breaking term
\begin{equation}\label{appeq:TIVCmass}
    H_\text{TIVC}=\Delta\tau^x\sigma^x
\end{equation}
which could arise from interaction effects.
Within the lower band subspace of the two valleys, the Hamiltonian can be parameterized as
\begin{gather}\label{appeq:toy_model_Htilde}
    \tilde{H}(\bm{k})=\begin{pmatrix}
        \epsilon(\bm{k})+{\delta(\bm{k})} & \tilde{\Delta}(\bm{k}) \\
        [\tilde{\Delta}(\bm{k})]^*& \epsilon(\bm{k})-{\delta(\bm{k})}
    \end{pmatrix}
\end{gather}
with the model-specific quantities
\begin{gather}
    \epsilon(\bm{k})=-\frac{|(k-k_1)(k-k_2)|+|(k+k_1)(k+k_2)|}{2}\\
    \delta(\bm{k})=\frac{|(k+k_1)(k+k_2)|-|(k-k_1)(k-k_2)|}{2}\label{appeq:toy_model_dk}.
\end{gather}
For the uniform TIVC mass of, we have
\begin{gather}
    \tilde{\Delta}(\bm{k})=-\frac{\Delta}{2}\left(e^{i\theta(\bm{k})}+e^{i\varphi(\bm{k})}\right)\label{appeq:toy_model_Delta}.
\end{gather}

\begin{figure}
    \centering
    \includegraphics[width = 0.6\linewidth]{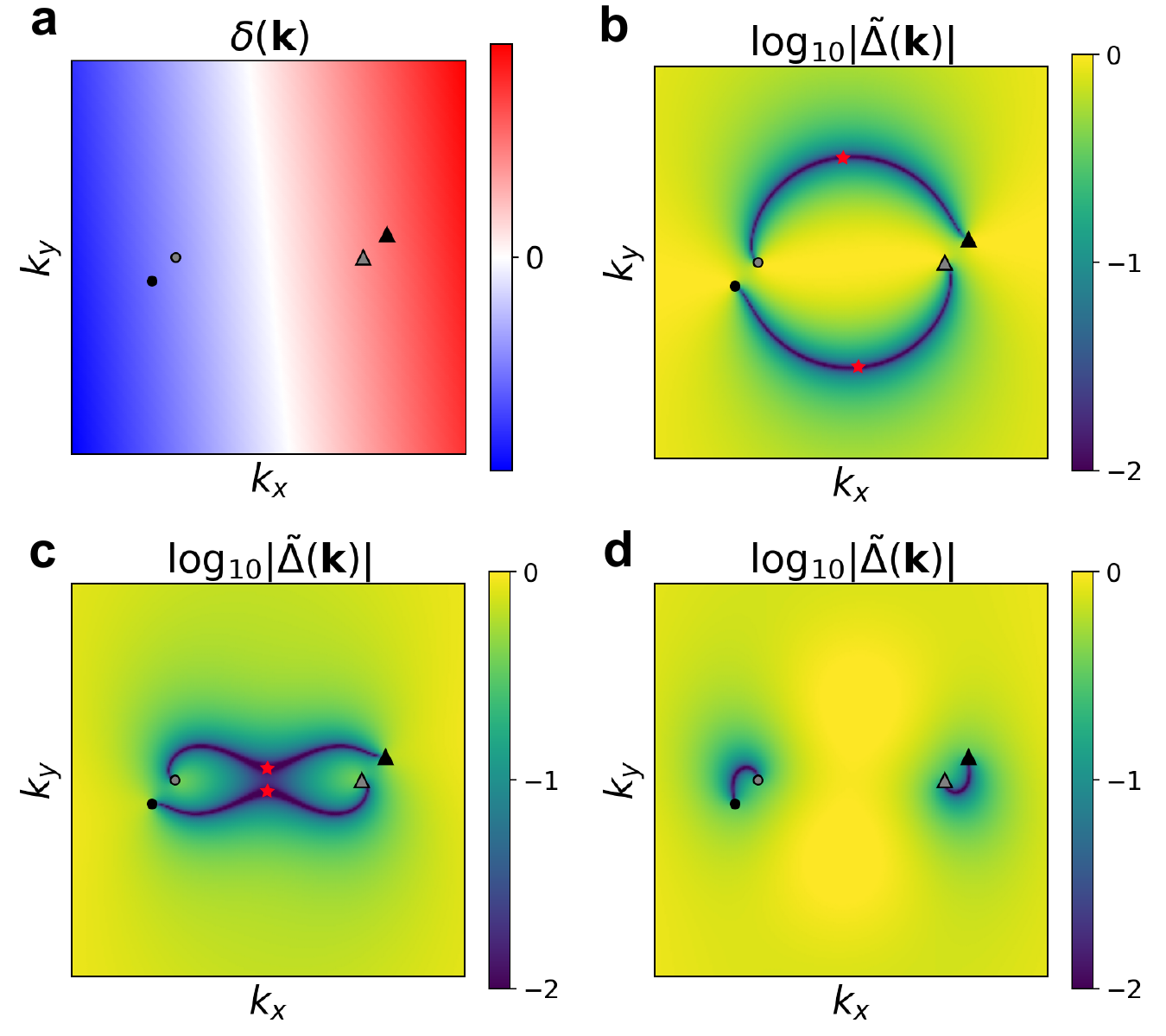}
\caption{\textbf{Toy $\bm{k}\cdot\bm{p}$ model for ETI.} The model (see Eq.~\ref{eq:kp_H+} for definition in the absence of IVC) is unbounded for large momentum. Any boost $\bm{q}$ has been absorbed into the momentum origin such that TRS takes $\bm{k}\rightarrow -\bm{k}$. Dirac points in valley $\tau=+$ are at $\bm{k}_1=(1.25,0.25)$ [black triangle] and $\bm{k}_2=(1.00,0.00)$ [grey triangle]. TRS-related non-interacting Dirac points in valley $\tau=-$ are indicated with dots.
a) Difference in lower band valley dispersion $\delta(\bm{k})$. b,c,d) Projected IVC matrix element $\tilde{\Delta}(\bm{k})$ for a TIVC term of strength $\Delta=1$ and a $\tau^x$ perturbation with momentum-dependent strength $A_{\tau^x}e^{-|k|^2/2}$. b,c,d) correspond to $A_{\tau^x}=0,1,2$ respectively. Note that the lower limit of the color scale has been clamped. Red stars indicate residual Dirac points in the projected Hamiltonian. The residual Dirac points have annihilated in d), leading to the formation of an ETI.
}
    \label{fig:2BLLL_toy_model}
\end{figure}

In order to open a direct gap at $\nu=-1$ corresponding to half-filling of $\tilde{H}(\bm{k})$, $\sqrt{|\tilde{\Delta}(\bm{k})|^2+\delta(\bm{k})^2}$ must be non-vanishing for all momenta $\bm{k}$. In Fig.~\ref{fig:2BLLL_toy_model}a, we show a representative plot of the valley dispersion difference $\delta(\bm{k})$. This takes values of opposite signs near the non-interacting Dirac points of the two valleys, since the non-interacting Dirac points are associated with high-energy features of the lower bands. In between, there is a line of $\delta(\bm{k})=0$, which crosses the time-reversal invariant momentum $\bm{k}=0$ due to TRS.
In Fig.~\ref{fig:2BLLL_toy_model}b, we plot the off-diagonal term $|\tilde{\Delta}(\bm{k})|$, which exhibits two nodal lines that connect non-interacting Dirac points in opposite valleys at $\bm{k}_1$ and $-\bm{k}_2$, as well as $\bm{k}_2$ and $-\bm{k}_1$. There are two residual Dirac points (red stars) where $\delta(\bm{k})=\tilde{\Delta}(\bm{k})=0$ with opposite winding number. Generally for any $\bm{k}_1,\bm{k}_2$, there will be residual Dirac points since the nodal lines connect non-interacting Dirac points in opposite valleys, and hence intersect the locus of $\delta(\bm{k})=0$.

To bring the residual Dirac points together and annihilate them, we project an additional IVC term\footnote{More generally, the total set of possible intervalley terms that preserve $\hat{C}_{2z}$ and $\hat{\mathcal{T}}$ are $\tau_x\sigma_x,\tau_x,\tau_y\sigma_z$ for momentum-even functions and $\tau_x\sigma_y$ for momentum-odd functions. Out of these, only $\tau_x\sigma_x$ and $\tau_x\sigma_y$ commute with $C=\tau_z\sigma_z$ and hence hybridize bands within the Chern sectors.} $\sim \tau^x$ to $\tilde{H}(\bm{k})$, which preserves the $\hat{C}_{2z}$ and $\hat{\mathcal{T}}$ symmetries. We consider a momentum-dependent perturbation localized near $\bm{k}=0$, which avoids issues with global obstructions arising from the fact that $\tau^x$ couples sublattice bands with opposite Chern number. As shown in Figs.~\ref{fig:2BLLL_toy_model}c,d, a sufficiently strong $\tau^x$ perturbation `rewires' the nodal lines of the projected IVC matrix element $|\tilde{\Delta}(\bm{k})|$, so that they connect the non-interacting Dirac points within each valley. As a result, the residual Dirac points annihilate and a direct gap opens at $\nu=-1$, leading to an ETI.

\section{Two-band LLL model}\label{secsupmat:2BLLL_additional}

In this section, we present the two-band LLL model, which is an explicit $\hat{C}_{2z}\hat{\mathcal{T}}$-symmetric interacting LLL-based model whose valley-resolved Hilbert space has a non-trival Euler class $|e_2|=1$. This generalizes the LLL model introduced in Sec.~\ref{sec:1BLLL}, which only has one band per valley. In addition to the valley index $\tau=\pm$, the two-band LLL model also has a sublattice degree of freedom with $\sigma=+$ corresponding to sublattice $A$, and $\sigma=-$ corresponding to sublattice $B$. The sublattice and valley-dependent magnetic field is $\bm{B}=-\tau\sigma B\hat{z}$ leading to Chern numbers $C=\tau\sigma$. The action of two-fold rotation and time-reversal symmetry on the magnetic Bloch creation operators is
\begin{equation}
        \hat{C}_{2z}d^\dagger_{\bm{k},\tau\sigma}\hat{C}_{2z}^{-1}=d^\dagger_{-\bm{k},\bar{\tau}\bar{\sigma}},\quad
    \hat{\mathcal{T}}d^\dagger_{\bm{k},\tau\sigma}\hat{\mathcal{T}}^{-1}=d^\dagger_{-\bm{k},\bar{\tau}\sigma},
\end{equation}
where $\bar{\sigma}=-\sigma$. $\hat{C}_{2z}\hat{\mathcal{T}}$ maps the Chern bands onto each other within a given valley, demonstrating the non-trivial Euler class.

\subsection{Interaction term}\label{secsupmat:2BLLL_interaction}
We take the interaction to be density-density in sublattice and valley space
\begin{equation}
    \hat{H}_\text{int}=\frac{1}{2A}\sum_{\bm{q}\in\text{all},\tau\tau'\sigma\sigma'}\tilde{U}_{\tau\sigma;\tau'\sigma'}(\bm{q})\rho_{\tau\sigma}(\bm{q})\rho_{\tau'\sigma'}(-\bm{q})
\end{equation}
where we have allowed for interaction anisotropies to influence the competition between various phases. For example, having a weaker inter-sublattice versus intra-sublattice interaction would tend to disfavor sublattice-polarized phases. We follow a similar prescription as the LLL model in Sec.~\ref{subsec:1BLLL_model}, and consider a dual gate screened interaction $u_0(\bm{q})=2\pi U\frac{\tanh qd_\text{sc}}{q}$ with effective dielectric constants depending on the valley-sublattice flavor. In particular, we have
\begin{equation}\label{eqapp:2BLLL_anis}
    \tilde{U}_{\tau\sigma;\tau'\sigma'}(\bm{q})=[1-(u_\text{v}-1)\delta_{\tau,\tau'}][1-(u_\text{sub}-1)\delta_{\sigma,\sigma'}]u_0(\bm{q})
\end{equation}
with anisotropy factors $u_\text{v},u_\text{sub}$, where $u_\text{v}<1$ ($u_\text{v}>1$) corresponds to intravalley interactions being stronger (weaker) than intervalley interactions. For a completely isotropic interaction potential $u_\text{v}=u_\text{sub}=1$, $\hat{H}_\text{int}$ has a $U(2)\times U(2)$ symmetry corresponding to rotations within each pair of bands with identical Chern numbers $C=\tau\sigma$ (i.e.~a Chern sector). In other words, the form factors themselves satisfy $U(2)\times U(2)$.

\subsection{Kinetic term}
We now turn to the valley-diagonal kinetic term. Since we have single-valley $\hat{C}_{2z}\mathcal{\hat{T}}$ symmetry and opposite Chern numbers in the two sublattices, we expect that the resulting band structure in each valley will have Dirac nodes with a net $4\pi$ winding, as is the case for TBG. This $4\pi$ winding encodes the non-trivial Euler index $|e_2|=1$ carried by the two bands within a given valley. In fact, we have the tools at our disposal to explicitly construct a Hamiltonian that possesses these topological features---we can use the EVL order parameter $\Delta_{\text{EVL}}(\bm{k})$ from Sec.~\ref{subsec:1BLLL_EVL}, which motivates the following single-particle Hamiltonian
\begin{align}\label{eq:2BLLL_SP}
\begin{split}
    \hat{H}^\text{SP}=&\sum_{\bm{k}\in \text{BZ}}\bigg[\Delta_{\text{EVL}}^*(\bm{k})d^\dagger_{\bm{k},+A}d_{\bm{k},+B}+\Delta_{\text{EVL}}(\bm{k})d^\dagger_{\bm{k},+B}d_{\bm{k},+A}\\
    &\quad \quad\,+\Delta_{\text{EVL}}^*(\bm{k})d^\dagger_{\bm{k},-B}d_{\bm{k},-A}+\Delta_{\text{EVL}}(\bm{k})d^\dagger_{\bm{k},-A}d_{\bm{k},-B}\bigg],
\end{split}
\end{align}
which preserves $\hat{C}_{2z}$ and $\hat{\mathcal{T}}$. The dispersion is plotted in Fig.~\ref{fig:2BLLL_sym_nosym}a with $\bm{q}=0$ and $s=-1$ in the EVL order parameter (Eq.~\ref{eq:1BLLL_EVL_OP}). 
There are two Dirac points in each valley with identical winding, which are maximally separated in momentum space.

We note that the kinetic term $\hat{H}^\text{SP}$ (Eq.~\ref{eq:2BLLL_SP}) satisfies a $U(2)$ subgroup of the $U(2)\times U(2)$ form factor symmetry, corresponding to identical rotations in the two Chern sectors. 

\subsection{Phase diagram at $\nu=-1$}

\begin{figure}
    \centering
    \includegraphics[width = 0.6\linewidth]{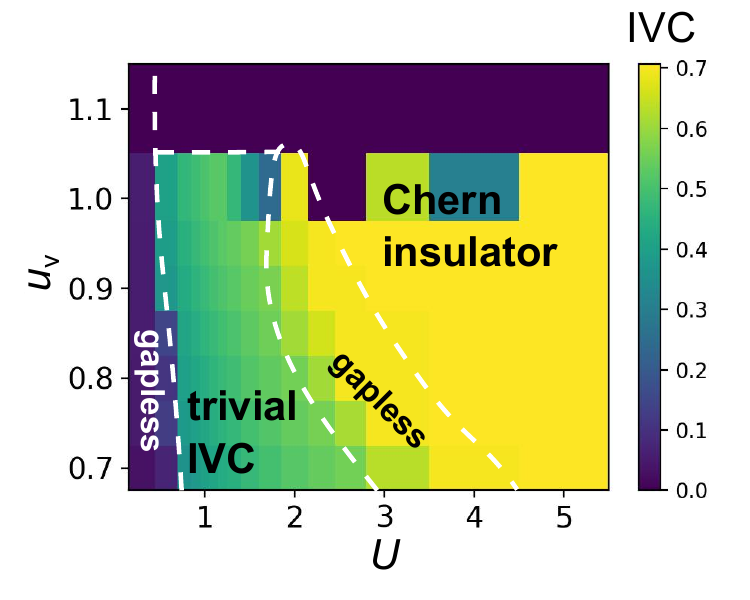}
    \caption{\textbf{HF phase diagram of the two-band LLL model at $\nu=-1$.} $u_\text{v}$ is the valley interaction anisotropy factor and $U$ controls the scale of the interaction (see Eq.~\ref{eqapp:2BLLL_anis}).  Color indicates the magnitude of intervalley coherence (IVC). White lines indicate approximate phase boundaries. The valley boost is fixed to either $\bm{q}=(0,0)$ or $(Q/2,Q/2)$. The gate distance $d_\text{sc}=6a$, sublattice interaction anisotropy $u_\text{sub}=1$, and system size is $16\times 16$.}
    \label{figapp:2BLLL_nu1_phase}
\end{figure}

Fig.~\ref{figapp:2BLLL_nu1_phase} shows the phase diagram of the two-band LLL model at $\nu=-1$ as a function of the valley interaction anisotropy $u_\text{v}$ and interaction scale $U$. The interactions have been chosen to be isotropic in sublattice space $u_\text{sub}=1$. For strong interactions, we find a $|C|=1$ Chern insulator phase. For large $u_\text{v}$, this is simply a valley- and sublattice-polarized insulator. For small $u_\text{v}$, the Chern insulator becomes intervalley coherent. This IVC is not obstructed, as the IVC is predominaintly between bands of the same Chern number $C=\tau\sigma$. For intermediate $U\sim 1$, we find a time-reversal symmetric IVC phase. However, we find that $\hat{C}_{2z}$ is broken (see Fig.~\ref{fig:2BLLL_sym_nosym}b for an example). Furthermore, inspection of the valley-filtered basis reveals that the intervalley coherence is not frustrated. This will be explained in more detail in the next subsection.

We comment briefly on the phase diagram at $\nu=0$. Because the non-interacting band structure consists of just Dirac points at $E_F$, there is no `lobe principle' that would motivate kinetically-driven IVC. Indeed, we do not find any appreciable regions of such phases in the phase diagram.

\subsection{IVC states at $\nu=-1$}\label{secsupmat:2BLLL_IVC}
\begin{figure*}
    \centering
    \includegraphics[width = 0.9\linewidth]{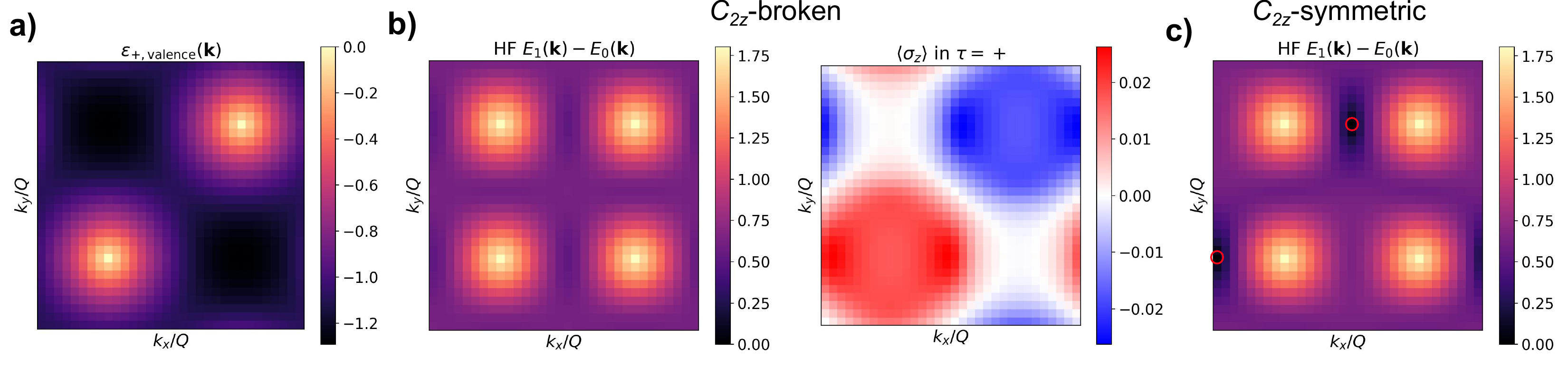}
\caption{\textbf{Two-band LLL model and intervalley-coherent states.} a) Non-interacting valence band dispersion in valley $\tau=+$. The single-particle band structure is constructed from Eq.~\ref{eq:2BLLL_SP} using $\bm{q}=0$ and $s=-1$ in the EVL ansatz of Eq.~\ref{eq:1BLLL_EVL_OP}. The conduction band dispersion can be obtained using particle-hole symmetry, and there are Dirac points at $\bm{k}=\pm(Q/4,Q/4)$. The dispersion in valley $\tau=-$ is identical. b) Properties of a $\hat{C}_{2z}$-broken IVC state at $\nu=-1$ with $\bm{q}=(Q/2,0)$. Left: Difference between the energy of the second-lowest HF band $E_1(\bm{k})$ and the lowest HF band $E_0(\bm{k})$. The latter is fully occupied. Right: Momentum-resolved sublattice polarization of the HF state in valley $\tau=+$. Parameters are $U=u_\text{v}=u_\text{sub}=1,d_\text{sc}=6a$, and system size is $24\times 24$. c) Same as b) left, except where $\hat{C}_{2z}$ has been enforced in the HF calculation. Red circles indicate residual Dirac points.}
    \label{fig:2BLLL_sym_nosym}
\end{figure*}

In the non-interacting problem at $\nu=-1$, $E_F$ lies within the valence bands, which possess two high-energy peaks (i.e.~the Dirac points) at $\bm{k}=\pm(Q/4,Q/4)$ and two low-energy troughs at $\bm{k}=\pm(Q/4,3Q/4)$ in each valley [see Fig.~\ref{fig:2BLLL_sym_nosym}a]. This suggests a `lobe' construction that boosts valley $\tau=-$ by $\bm{q}=(Q/2,0)$, and results in four pairs of coincident peak/trough features. Note that this is in contrast to the lobe construction for the IKS in TBG, where there is a single high-energy region and low-energy trough in each valley, leading to two pairs of coincident peak/trough features after boosting~\cite{kwan_kekule_2021}. For weak interactions, the large non-interacting Fermi surfaces and lack of nesting will lead to a gapless state. For sufficiently strong interactions, the system is expected to simply polarize into flavor and sublattice space and form a `strong-coupling' Chern insulator. For intermediate interaction strengths though, we anticipate that the system will exploit the above `lobe' construction and form a gapped IVC spiral state.

As shown in Fig.~\ref{fig:2BLLL_sym_nosym}b for $U=1$, we indeed find a gapped IVC phase at $\nu=-1$ with $\bm{q}=(Q/2,0)$ that preserves TRS. The occupied HF band has dominant weight on the single-particle valence bands, and modulates its valley pseudospin to adapt to the kinetic energy. However, while there is no net sublattice polarization, the system breaks $\hat{C}_{2z}$, as can be deduced by the momentum-resolved sublattice polarization which has small opposite values around the positions of the non-interacting Dirac points. There is also no indication of any topological frustration, since the IVC does not vanish anywhere in the BZ, unlike for the CTI$_n$.

Another way to see the lack of frustration is to consider `unfolding' the single occupied intervalley-coherent HF band
\begin{equation}
    \ket{\text{HF},\bm{k}}=\alpha(\bm{k})\ket{\text{HF}_+,\bm{k}}+\beta(\bm{k})\ket{\text{HF}_-,\bm{k}}
\end{equation}
which implicitly defines a pair of time-reversal-related valley-diagonal bands $\{\ket{\text{HF}_+,\bm{k}}\}$ and $\{\ket{\text{HF}_-,\bm{k}}\}$, dubbed the valley-filtered bands. In Fig.~\ref{fig:2BLLL_sym_nosym}b, we find these valley-filtered bands $\{\ket{\text{HF}_\tau,\bm{k}}\}$ are topologically trivial with $C=0$. Since there is no topological obstruction to hybridization between $C=0$ bands, the IVC is therefore of a `trivial' nature.

As argued in the main text, if the system was gapped and preserved $\hat{C}_{2z}$ and $\hat{\mathcal{T}}$, the resulting state would be an ETI. The interpretation of the `trivial' IVC state in Fig.~\ref{fig:2BLLL_sym_nosym}b is therefore that the system removes the topological obstruction by spontaneously breaking $\hat{C}_{2z}$ and sublattice-polarizing locally in momentum space. Unlike in the one-band LLL model, the presence of multiple bands within each valley allows for such `unfrustration' of the IVC order parameter. 
We note that an alternative $\hat{C}_{2z}$-breaking scenario to Fig.~\ref{fig:2BLLL_sym_nosym}b consists of inducing identical sublattice masses at the two Dirac points for each valley. The resulting IVC state would instead be a CTI$_1$ since the unfolded bands have $|C|=1$. This is energetically disfavored because it does not resolve the topological obstruction to IVC.

In Fig.~\ref{fig:2BLLL_sym_nosym}c, we attempt to generate an ETI by enforcing $\hat{C}_{2z}$ in the HF calculation. However, the resulting state remains gapless with two residual Dirac points at $E_F$ that are not at any of the non-interacting Dirac point positions. This means that while the two-band LLL model has the same topological features as the strained TBG Hamiltonian, the non-topological details prevent stabilization of an ETI. One key difference between the two models is the positions of the (renormalized) Dirac points in the BZ before IVC is induced. In TBG, due to the combination of the intrinsic nematic instability and the effect of external uniaxial heterostrain, the Dirac points migrate towards $\Gamma_\text{M}$ and become close to each other within each valley. Hence, the sign-changing sublattice polarization involved in the trivial IVC of Fig.~\ref{fig:2BLLL_sym_nosym}b is unlikely, since the rapidly changing sublattice order would lead to a large exchange penalty, and the result is a topologically frustrated IKS. On the other hand, the Dirac points in the two-band LLL model are spaced as far as possible in the BZ, and there is no analogous mechanism that brings them together. Therefore, the system is more susceptible to the sublattice texturing shown in Fig.~\ref{fig:2BLLL_sym_nosym} that alleviates the topological frustration to IVC. Furthermore, even when $\hat{C}_{2z}$ is imposed, there are still residual Dirac points which remain far apart.

\section{Perturbative derivation of valley/gravitational Chern-Simons term}\label{derivationCS}
\begin{figure}
\begin{center}
\includegraphics[scale=0.7]{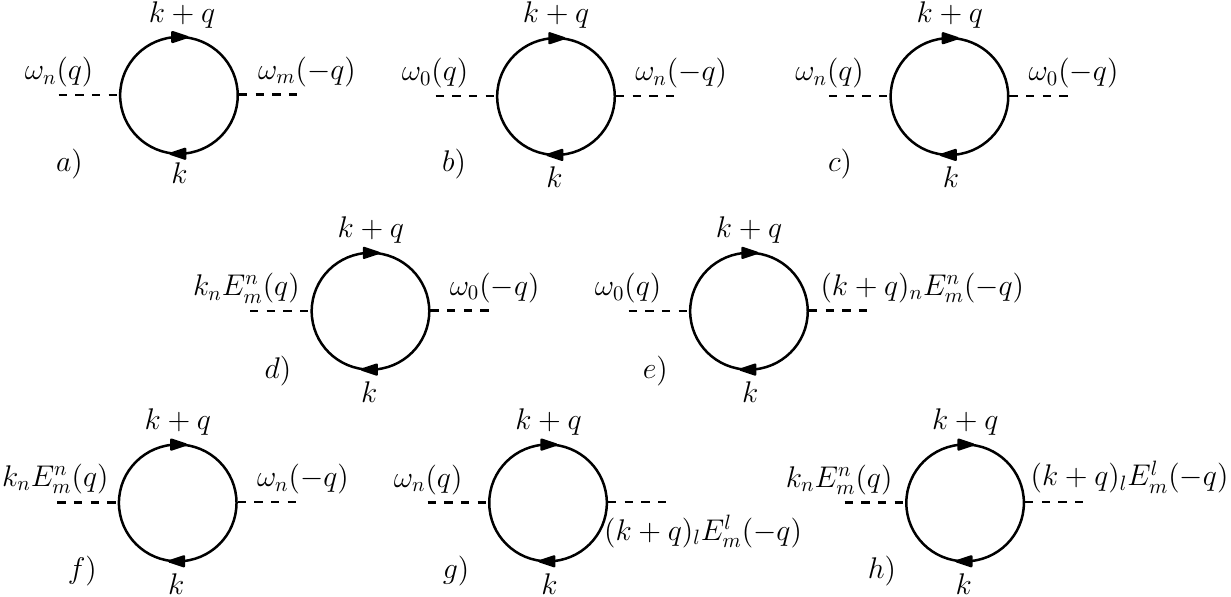}
\end{center}
\caption{Feynman diagrams which can contribute to the gravitational Chern-Simons term.}\label{diags}
\end{figure}

Consider a single massive Dirac fermion coupled to the $\omega$ (valley) gauge field defined in the main text (Sec.~\ref{sec:field_theory}). Its action is given by
\begin{equation}
\mathcal{L} = \psi^\dagger(i\partial_t -\omega_0 \tau^z)\psi - \psi^\dagger e^n_m \tau^m(i\partial_n - \omega_n \tau^z)\psi - M\psi^\dagger \tau^z \psi.
\end{equation}
We write
\begin{equation}
e^n_m = \delta^n_m + E^n_m\,,
\end{equation}
where we consider $E^n_m$ to be small, and we will work to lowest order in $E^n_m$. To lowest order, Eq. \eqref{hermconstr} becomes
\begin{equation}\label{LO}
\partial_n E^n_m = 2\epsilon^{lm}\omega_l.
\end{equation} 
As a first step, we rewrite the Lagrangian as
\begin{equation}
\mathcal{L} = \psi^\dagger(i\partial_t -\tau^n i \partial_n - M\tau^z)\psi - \omega_0\psi^\dagger\tau^z\psi  - \frac{1}{2}\left[\psi^\dagger E^n_m \tau^m i\partial_n \psi - i\partial_n \psi^\dagger E^n_m \tau^m  \psi  \right].
\end{equation}
Going to momentum space, this becomes 
\begin{eqnarray}
S & = &  \int\frac{d^3 k}{(2\pi)^3} \psi^\dagger(k)(k_0 - \tau^n k_n - M \tau^z)\psi(k) \\
&& - \int\frac{d^3 k}{(2\pi)^3}\int\frac{d^3 q}{(2\pi)^3}\left[\omega_0(q) \psi^\dagger(k+q)\tau^z \psi(k) + \frac{1}{2}(2k_n + q_n)E^n_m(q) \psi^\dagger(k+q) \tau^m q_n \psi(k)  \right].\nonumber
\end{eqnarray}
Using Eq. \eqref{LO} we can write this as
\begin{eqnarray}
S & = &  \int\frac{d^3 k}{(2\pi)^3} \psi^\dagger(k)(k_0 - \tau^n k_n - M \tau^z)\psi(k) \\
&& - \int\frac{d^3 k}{(2\pi)^3}\int\frac{d^3 q}{(2\pi)^3}\bigg\{\omega_0(q) \psi^\dagger(k+q)\tau^z \psi(k) + \left[k_nE^n_m(q)- i\epsilon^{nm}\omega_n(q)\right] \psi^\dagger(k+q) \tau^m \psi(k)  \bigg\}. \nonumber
\end{eqnarray}
We will now perturbatively integrate out the Dirac fermions, using the propagator
\begin{equation}
\includegraphics[scale=0.7]{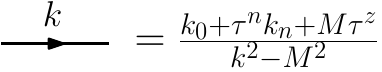}
\end{equation}
where $k^2 = k_0^2 - \bm{k}^2$.  The gravitational Chern-Simons term in the resulting effective action can receive contributions from the diagrams shown in Fig. \ref{diags}.

\begin{itemize}
\item Diagram (a) in Fig. \ref{diags} evaluates to
\begin{align}
&(-1)\omega_n(q)\omega_m(-q)(-\epsilon^{nl}\epsilon^{ms})\int \frac{d^3k}{(2\pi)^3}\text{tr}\left[\frac{k_0 +k_a \tau^a + M\tau^z }{k^2-M^2} \tau^l \frac{k_0 + q_0 +(k_b+q_b) \tau^b + M\tau^z }{(k+q)^2-M^2} \tau^s  \right]\nonumber \\
&= \omega_n(q)\omega_m(-q)\epsilon^{nl}\epsilon^{ms}\int \frac{d^3k}{(2\pi)^3} \frac{k_0}{k^2-M^2}\frac{M}{(k+q)^2-M^2}\text{tr}(\tau^l\tau^z\tau^s) \nonumber \\
& \hspace{5 cm}+ \frac{M}{k^2-M^2}\frac{k_0+q_0}{(k+q)^2-M^2}\text{tr}(\tau^z\tau^l\tau^s) +\cdots \nonumber\\
&= \omega_n(q)\omega_m(-q)(-\epsilon^{nl}\epsilon^{ms})( q_0 i\epsilon^{ls})\times 2M\int\frac{d^3k}{(2\pi)^3}\frac{1}{k^2-M^2}\frac{1}{(k+q)^2-M^2}+\cdots
\end{align}
The integral is now identical to the one that appears in the perturbative calculation of the Chern-Simons term for the charge gauge field. We thus find that the first diagram gives
\begin{equation}
\omega_n(q)\omega_m(-q)(-i\epsilon^{nm})q_0 \frac{\text{sgn}(M)}{8\pi}+\cdots \,,
\end{equation}
where we have also performed a derivative expansion and dropped higher powers of $q$ coming from the integral.

\item Diagram (b) gives
\begin{align}
&(-1)\omega_0(q)\omega_n(-q) i\epsilon^{nm} \int \frac{d^3 k}{(2\pi)^3}\text{tr}\left[\frac{k_0 +k_a \tau^a + M\tau^z }{k^2-M^2} \tau^z \frac{k_0 + q_0 +(k_b+q_b) \tau^b + M\tau^z }{(k+q)^2-M^2} \tau^m  \right]\nonumber \\
&= -\omega_0(q)\omega_n(-q) i\epsilon^{nm} \times 2M  \int \frac{d^3 k}{(2\pi)^3}\left[\frac{1 }{k^2-M^2} \frac{ (k_m+q_m) }{(k+q)^2-M^2}  + \frac{k_m  }{k^2-M^2}  \frac{ 1 }{(k+q)^2-M^2}   \right] +  \cdots \nonumber\\
&= -\omega_0(q)\omega_n(-q) i\epsilon^{nm} \times 2M \int \frac{d^3 k}{(2\pi)^3} \frac{2k_m}{[(k-q/2)^2-M^2][(k+q/2)^2-M^2]}   +  \cdots\nonumber \\
& = 0 +\cdots\,,
\end{align}
where the last equality follows because the integrand is odd under $k\rightarrow -k$. Note that in the above derivation we have only kept track of the part which can contribute to the Chern-Simons term---we have not shown that diagram (b) is identically zero.  By symmetry it follows that diagram (c) also does not contribute to the Chern-Simons term. 

\item Evaluating diagram (d) we find
\begin{align}
&(-1)E^n_m(q)\omega_0(q)\int\frac{d^3k}{(2\pi)^3}k_n \text{tr}\left(\frac{k_0+k_a\tau^a + M\tau^z}{k^2-M^2}\tau^m\frac{k_0+q_0+(k_b+q_b)\tau^b + M\tau^z}{(k+q)^2-M^2}\tau^z \right)\nonumber \\
&= -E^n_m(q)\omega_0(q)\times 2M\int\frac{d^3k}{(2\pi)^3} \frac{1}{(k-q/2)^2-M^2}\frac{1}{(k+q/2)^2-M^2}2k_nk_m+\cdots
\end{align}
This integral is UV divergent and needs to be regularized.  Let us write the result of evaluating the regularized integral as $\text{sgn}(M) q_nq_m f(q_0/M,\bm{q}/M,q_0/\Lambda,\bm{q}/\Lambda)$, where $f$ is a dimensionless function and $\Lambda$ is a UV scale needed for regularization.  For the Chern-Simons term we only need to keep the constant part of $f$, which for future use we write as $X/16\pi$. Below we will see that $X$ can be fixed by gauge invariance. The final result for diagram (d) is then
\begin{equation}
-\frac{1}{2}E^n_m(q)\omega_0(q) q_nq_m\frac{\text{sgn}(M)}{8\pi}X+\cdots = \omega_n(q)i\epsilon^{nm}q_m\omega_0(-q)\frac{\text{sgn}(M)}{8\pi}X+\cdots
\end{equation}
By symmetry we now have that diagram (e) contributes
\begin{equation}
\omega_0(q)i\epsilon^{nm}q_m\omega_n(-q)\frac{\text{sgn}(M)}{8\pi}X.
\end{equation}

\item Evaluating diagram (f) gives
\begin{align}
& E^n_m(q) \omega_l(-q)(i\epsilon^{ls})(i\epsilon^{ms})q_0\times 2M\int\frac{d^3k}{(2\pi)^3}\frac{1}{k^2-M^2}\frac{1}{(k+q)^2-M^2}k_n + \cdots\nonumber\\
& = E^n_m(q) \omega_l(-q)(i\epsilon^{ls})(i\epsilon^{ms})q_0\times 2M\int\frac{d^3k}{(2\pi)^3}\frac{1}{(k-q/2)^2-M^2}\frac{1}{(k+q/2)^2-M^2}(k_n-\frac{q_n}{2})+ \cdots\nonumber\\
& =-\frac{q_n}{2} E^n_m(q) \omega_l(-q)(i\epsilon^{ls})(i\epsilon^{ms})q_0\frac{\text{sgn}(M)}{8\pi}+\cdots\nonumber\\
& = -i\epsilon^{lm}\omega_l(q)\omega_m(-q) q_0 \frac{\text{sgn}(M)}{8\pi} +\cdots
\end{align}
From symmetry it follows that (g) produces an identical contribution.

\item And finally from diagram (h) we get
\begin{align}
& -i\epsilon^{mt}q_0 E^n_m(q)E^l_t(-q)\times 2M\int\frac{d^3k}{(2\pi)^3}\frac{1}{k^2-M^2}\frac{1}{(k+q)^2-M^2}k_n(k_l+q_l)+\cdots \nonumber\\
& = (1+X) i\epsilon^{nl}\omega_n(q)\omega_l(-q)q_0 \frac{\text{sgn}(M)}{8\pi)} + \cdots
\end{align}

\end{itemize}
Putting everything together we obtain
\begin{align}\label{C}
& \frac{\text{sgn}(M)}{8\pi}\left[ (2-X)\omega_n(q)\omega_m(-q)(-i\epsilon^{nm}q_0) + X \omega_0(q)\omega_n(-q)(i\epsilon^{nm}q_m) + X \omega_n(q) \omega_0(-q) (i\epsilon^{nm}q_m) \right].
\end{align}
Gauge invariance imposes on the quadratic part of the effective action $\omega_\mu(q)\Pi^{\mu\nu}(q)\omega_\nu(-q)$ that it should satisfy $q_\mu\Pi^{\mu\nu}(q) = 0$.  This implies that we can write $\Pi^{\mu\nu} = \Pi_1(q)(q^2\eta^{\mu\nu} - q^\mu q^\nu) + \Pi_2(q)\epsilon^{\mu\nu\lambda}ip_\nu$, where $\eta^{\mu\nu} = \text{diag}(1,-1,-1)$ and $p^\mu = \eta^{\mu\nu}p_\nu$.  The only value of $X$ which is compatible with this form is $X=1$, in which case Eq. \eqref{C} becomes following Chern-Simons term in real space:
\begin{equation}\label{CS8}
-\frac{\text{sgn}(M)}{8\pi}\epsilon^{\mu\nu\lambda}\omega_\mu \partial_\nu \omega_\lambda.
\end{equation}
Adding up the contributions of two gapped Dirac fermions with the same sign for the mass term we arrive at the properly quantized Chern-Simons term $\omega\mathrm{d}\omega/4\pi$ given in the main text.

\section{CTI and TVP edge physics}\label{secsupmat:edgephysics}

In this appendix we study the edge physics of the CTI and TVP states using the low-energy theory of Sec.~\ref{sec:field_theory}. Consider a boundary with normal vector $\n$ ($\n^2=1$). In order to ensure that the Dirac operator on a region $M$ with boundary $\partial M$ is Hermitian we have to impose following boundary conditions (BCs):
\begin{equation}\label{BC}
\int_{\partial M} \langle \psi'(x)|\n\cdot\boldsymbol{\sigma}|\psi(x)\rangle = 0\,,
\end{equation}
for any two states $|\psi\rangle$ and $|\psi'\rangle$ in the Hilbert space.  As an example, let us take a boundary at $x=0$ with boundary normal $\n=(-1,0)$. One obvious way to satisfy the BCs in Eq.~\eqref{BC} is by taking the wavefunctions to vanish at the boundary. However, we can also satisfy the BCs by allowing states which are non-vanishing at the boundary provided that they are oriented (on the Bloch sphere) along a particular direction in the $y-z$ plane. 

Let us now consider the Dirac operator in the presence of this boundary:
\begin{equation}
H_{B} = (i\partial_x - k_x^*)\tau^x + (i\partial_y - k_y^*)\tau^y + M\tau^z\,,
\end{equation} 
where we again assume that $M>0$. This operator has solutions exponentially localized at the boundary which satisfy
\begin{equation}
H_B |\psi(k_y)\rangle = -(k_y-k_y^*) |\psi(k_y)\rangle\,,
\end{equation}
with 
\begin{equation}
|\psi(k_y)\rangle = e^{-ik_yy}e^{-ik^*_x x}\left( \begin{matrix} 1 \\ -i \end{matrix}\right)\times e^{-Mx}.
\end{equation}
To allow these linearly dispersing chiral edge states in our Hilbert space we thus have to choose the BCs $|\psi\rangle \propto (1,-i)^T$ at $x=0$. If we change the sign of the mass term, then the exponentially localized boundary states are proportional to $(1,i)^T$ and hence are not in our Hilbert space. This means that for a particular choice of BC, and hence a particular choice of Hilbert space, there are chiral edge states for one sign of the mass, and not for the other. This is of course just a reflection of the fact that a 2D Dirac fermion is the critical point between a trivial phase and a quantum Hall phase.

So far we have focused on one of the two mini-valleys. Fixing the boundary condition in one mini-valley to be $(1,-i)^T$ forces us to impose the same BCs in the other mini-valley because of the time-reversal symmetry. In this case, taking $M>0$ in first mini-valley and $M<0$ in the second (corresponding to the CTI$_1$ state) leads to two counter-propagating chiral edge states at momenta $\pm k_y^*$. If we change the sign of the mass in one of the two mini-valleys we remove one branch of edge states and we are left with a chiral edge --- this case applies to the TVP state. Depending on which mass is changed, we either end up with a single left-moving mode at $k_y = k_y^*$, or a single right-moving mode at $k_y = -k_y^*$. 

For a general straight edge with normal vector $\n$ there are two choices of BC: $|\psi\rangle \propto (1,\eta ie^{i\theta})$, where $e^{i\theta} = n_x+in_y$ and $\eta  = \pm 1$. If we take masses $(M,-M)$ in the mini-valleys $\mu^z=(1,-1)$, then there are two counter-propagating edge modes if $\eta = \text{sgn}(M)$. The left-moving mode sits at a momentum parallel to the edge given by
\begin{equation}\label{kpara}
k^*_\parallel = \epsilon^{ij}k^*_i n_j\, ,
\end{equation}
and the right-moving mode sits at $-k^*_\parallel$ ($\epsilon^{ij}=-\epsilon^{ji}$ and $\epsilon^{xy}=1$). If we take $\eta = -\text{sgn}(M)$ there are no edge modes. For the TVP with masses $(M,M)$ there is a chiral edge mode, with the chirality determined by $\text{sgn}(M)$, located at $\text{sgn}(M)\eta k^*_\parallel$.

Let us discuss to what extent edges of the CTI$_1$ state with two counter-propagating modes are non-trivial. If these modes located at the same edge cross at a particular $k_\parallel$, they can hybridize without breaking translation along the edge. But note that the hybridization induces (decaying) charge oscillations with wavevector $2k^*_{\perp} =2 \n \cdot \k^*$ perpendicular to the edge. So gapping out edges of the CTI$_1$ state with counter-propagating modes induces a region of local CDW order, similar to the halo of CDW order at the IVC vortex core discussed in Sec. \ref{vortexCTI}.

\section{Numerical determination of quantum metric}
The differential Fubini-Study metric is given by
\begin{equation}
    d s^2 = 1 - |\braket{\psi|\phi}|^2
\end{equation}
where $\ket{\phi} = \ket{\psi} + \ket{\delta \psi}$ is  infinitesimally close to $\ket{\psi}$ in the Hilbert space, and both $\ket{\psi}$ and $\ket{\phi}$ are normalized. For a set of wavefunctions parameterised by Bloch momentum $\bm{k}$, we can define a metric tensor, such that
\begin{equation}\label{eq:metric_tensor}
    ds^2= g_{ab}q_a q_b
\end{equation}
for some infinitesimal vector $\bm{q}$. The $\bm{k}$ dependence of $ds^2$ and $g_{ab}$ is implicitly understood. For a set of wavefunctions $\ket{u(m, n)}$ defined on some discrete $\bm{k} = (m/N)\bm{G}_1 + (n/N)\bm{G}_2$, where $\bm{G}_1 = k_\theta(\sqrt{3}, 0)$ and $\bm{G}_2 = k_\theta(-\frac{\sqrt{3}}{2}, \frac{3}{2})$, we can define, for each $m$ and $n$,
\begin{align}
    \delta s_A^2 = & 1 - |\braket{u(m + 1, n)|u(m, n)}|^2 \\
    \delta s_B^2 = & 1 - |\braket{u(m, n + 1)|u(m, n)}|^2 \\
    \delta s_C^2 = & 1 - |\braket{u(m + 1, n + 1)|u(m, n)}|^2.
\end{align}
Solving for the metric tensor using Eq.~\ref{eq:metric_tensor}, we find
\begin{equation}
    \text{tr}[g] := g_{aa} + g_{bb} = \frac{2}{3}(\delta s^2_A + \delta s^2_B + \delta s^2_C)/(\Delta k)^2.
\end{equation}
where $\Delta k = \sqrt{3}k_\theta/N$. For the TBG calculation, we ignore the small change in the RLV $\bm{G}_i$ due to strain when calculating $\text{tr}[g]$.

\end{document}